\documentclass[useAMS,usenatbib]{mn2e}

\usepackage{graphicx}

\RequirePackage{snapshot}

\usepackage{ifpdf}

\usepackage{times}

\usepackage{booktabs}

\usepackage{aas_macros}

\ifpdf
	\usepackage{epstopdf}
\fi

\newcommand{\au}{$\mathrm{au} $}

\usepackage{subfig}

\usepackage{bm}

\title[Disc evolution affected by encounters]{Protoplanetary disc evolution affected by star--disc interactions in young stellar clusters}
\author[G. P. Rosotti et al.]{Giovanni P. Rosotti\thanks{E-mail:
rosotti@usm.lmu.de}$^{1,2,3}$, James E. Dale$^{2,3}$, Maria de Juan Ovelar$^{4}$, David A. Hubber$^{2,3}$,
\newauthor  J. M. Diederik Kruijssen$^{5}$, Barbara Ercolano$^{2,3}$,  Stefanie Walch$^{5}$\\
$^{1}$Max-Planck-Institut f\"ur extraterrestrische Physik, Giessenbachstra\ss{}e, D-85748 Garching, Germany\\
$^{2}$ Excellence Cluster Universe, Boltzmannstr. 2, D-85748 Garching, Germany\\
$^{3}$ Universitats-Sternwarte M\"unchen, Scheinerstra\ss{}e 1, D-81679 M\"unchen, Germany\\
$^{4}$ Leiden Observatory, Leiden University, P.O. Box 9513, 2300RA Leiden, The Netherlands\\
$^{5}$ Max-Planck-Institut f\"ur Astrophysik, Karl-Schwarzschild-Str. 1, D-85748 Garching, Germany}

\begin{document}

\date{Accepted 2014 April 04. Received 2014 April 04; in original form 2013 October 11}

\pagerange{\pageref{firstpage}--\pageref{lastpage}} \pubyear{2013}

\maketitle

\label{firstpage}

\begin{abstract}

Most stars form in a clustered environment. Therefore, it is important to assess how this environment influences the evolution of protoplanetary discs around young stars. In turn, this affects their ability to produce planets and ultimately life. We present here for the first time 3D SPH/N-body simulations that include both the hydrodynamical evolution of the discs around their natal stars, as well as the dynamics of the stars themselves. The discs are viscously evolving, accreting mass onto the central star and spreading. We find penetrating encounters to be very destructive for the discs as in previous studies, although the frequency of such encounters is low. We also find, however, that encounter influence the disc radii more strongly than other disc properties such as the disc mass. The disc sizes are set by the competition between viscous spreading and the disruptive effect of encounters. As discs spread, encounters become more and more important. In the regime of rapid spreading encounters simply truncate the discs, stripping the outer portions. In the opposite regime, we find that the effect of many distant encounters is able to limit the disc size. Finally, we predict from our simulations that disc sizes are limited by encounters at stellar densities exceeding $\sim 2-3 \times 10^3 \ \mathrm{pc}^{-2}$.

\end{abstract}


\section{Introduction}

Stars form in regions of enhanced ambient gas and stellar densities compared to the Galactic field \citep{lada03}. Whether or not these density peaks are long-lived or disperse on a dynamical time (i.e.~whether they become bound stellar clusters or unbound associations) depends crucially on their initial densities and the resulting star formation efficiencies \citep{kruijssen12}. In Milky Way-like galaxies, about $10$\% of all stars are born in bound stellar clusters \citep{bastian08}, but this number increases with the gas surface density to up to $\sim50$\% in high-density starburst environments \citep{goddard10,adamo11,kruijssen12d,silvavilla13}.

The cluster environment leaves a spectacular imprint on the star formation process. Through the collective feedback of young stars such as stellar winds and photoionising radiation, natal gas is ejected and the accretion discs surrounding protostars may be destroyed by external photoevaporation \citep{adams04,pelupessy12,dale13}. Combining the current observational and theoretical understanding of planet-, star-, cluster- and galaxy formation, \citet{longmore14} estimate that some 10\% of all stars in the Universe may have the formation of planets (or lack thereof) in their habitable zones affected by their natal cluster environment. In this paper, we concentrate on the dispersal of gas from protoplanetary discs through encounters with neighbouring stars. It serves as a first step to obtaining a detailed understanding of how the cluster environment affects the evolution of protoplanetary discs.

In isolation, an effective viscosity causes the redistribution of angular momentum within the gaseous disc \citep{lyndenbellpringle}. This leads to disc spreading on the one hand and mass accretion onto the central star on the other hand. While the latter process is routinely observed \citep[e.g.,][]{Gullbring98,Natta2004,2008ApJ...681..594H,2012ApJ...755..154M}, there are only a few observational reports of disc spreading \citep{Isella2009,2011A&A...529A.105G}, as high spatial resolution is needed to resolve the disc size. Within the current limitations, these works show how simple theoretical models are able to reproduce the observed rate of disc spreading. In practice, these works have concentrated on the nearest star-forming regions, namely Taurus-Auriga and Ophiucus, which are characterized by a lower stellar density when compared with more crowded regions, like the Orion Nebula Cluster (ONC). After several Myr of this slow and quiet evolution, it appears that another destructive process kicks in, and the disc is rapidly cleared on a $\sim 10^5 \ \mathrm{yr}$ timescale \citep{2010ApJS..186..111L,2011MNRAS.410..671E,2013MNRAS.428.3327K}. Currently, internal photoevaporation is the best candidate mechanism for such a fast disc dispersal \citep{UVswitch,AlexanderEvol,2009ApJ...705.1237G,Owen1st,Owen11Models}.

Does a clustered environment impact this picture of disc evolution? \citet{2012A&A...546L...1D} found evidence of a dependence of the observed disc sizes on the environmental surface stellar density. In particular, discs in crowded environments, that is, at stellar densities above $10^{3.5}\,\mathrm{pc}^{-2}$, are systematically smaller than their counterparts in less crowded fields. Observationally, it is known that proximity to high mass stars may alter the evolution of protoplanetary discs via external photoevaporation \citep{1998AJ....115..263O,2010ApJ...725..430M,2012ApJ...757...78M}. The high-energy radiation from massive stars can ionize and evaporate the material in the atmosphere of discs even at distances of $\simeq 1 \ \mathrm{pc}$ \citep{Johnstone98,adams04}. Although there are spectacular images of this process in silhouette discs (proplyds) in the Orion Nebula Cluster (ONC), overall this process is not expected to be the main driver of disc evolution \citep{2010ARA&A..48...47A}. Another important process occurring in a clustered environment are stellar encounters. Most of the previous work done on this problem has concentrated on modeling a given existing stellar cluster. These studies \citep{2001MNRAS.325..449S,2006ApJ...642.1140O,2008A&A...487L..45P,2012ApJ...756..123O,2013ApJ...769..150C} involved either semi-analytic solutions or pure N--body simulations in which close stellar encounters are recorded and the effect of single encounters on a putative disc is inferred \emph{a posteriori} (using results from simulations with pure N-body techniques, or including also hydrodynamical effects).

\cite{2001MNRAS.325..449S} performed N--body simulations using 4 000 stars in virial equilibrium in an $r^{-2}$ density distribution with a half--mass radius of $\sim1$ pc to model the ONC. The ONC is a popular target, being the nearest massive star--forming region, where many protostellar discs are observed in silhouette against the bright nebula \citep{ricci2008,robberto13}. The ONC contains $\sim 4 000$ stars (i.e. $\sim2\times10^{3}$ M$_{\odot}$) in a $\sim$5 pc diameter volume, has a one--dimensional velocity dispersion of 2.5 km s$^{-1}$ and a core density of 4.7$\times$10$^{4}$ pc$^{-3}$. \cite{2001MNRAS.325..449S} found that $\sim8\%$ of all stars and $\sim30\%$ of core stars suffered a sub--100 $\mathrm{au}$ encounter after 12.5 Myr of integration and concluded that encounters were unlikely to significantly affect the disk population. However, they cautioned that the sharp outer cutoff in their stellar distribution caused their models to expand significantly. This naturally lowers the encounter rate.

\cite{2006ApJ...642.1140O} used very similar initial conditions to \cite{2001MNRAS.325..449S} in their N--body study of the ONC, except that they also modelled sub-virial clusters. Instead of recording the single closest encounter, as in \cite{2001MNRAS.325..449S}, \cite{2006ApJ...642.1140O} recorded the complete encounter history of objects on the following grounds: (a) the closest encounter may not be the most destructive, since a distant flyby of a massive perturber can do more damage than a near--miss with a low--mass object (\cite{2007ApJ...656..275M} found that unequal--mass encounters are more destructive); (b) some stars will experience several encounters whose effects may be cumulative. They also concluded that the fraction of stars experiencing sub-$100$ \au encounters in 12.5 Myr was small, at most $\sim12\%$. They estimated disk mass--losses explicitly using a fitting formula from an extension of the parameter--space study of \cite{2005A&A...437..967P} and found that serial encounters and flybys of massive perturbers were able to affect the disk population, at least in the dense core of the cluster. They concluded that, over 12.5 Myr, $\sim4\%$ of disks in the ONC and $\sim10\%$ of disks in the core would be destroyed outright, assuming initial disk radii of 100 \au. This fraction is increased to $\sim9\%$ and $\sim20\%$ respectively if initial outer disc radii of 200 $\mathrm{au} $ are assumed instead.

\cite{2008A&A...487L..45P}, again considering the ONC, pointed out that close encounters involving disc--bearing stars in clusters can also result in bursts of accretion due to spiral arms induced in the disks. They concluded that this is a common phenomenon in dense cluster cores, driving accretion rates up by orders of magnitude for short periods ($10^{2}$--$10^{4}$ yr), during which 5--10$\%$ of the disk may be accreted. \cite{2008A&A...492..735P} speculated that such events may be observed as FU Orionis outbursts.

\cite{2012ApJ...756..123O} studied star--disk interactions in the Arches cluster. The Arches is more massive ($\sim3\times10^{4}$M$_{\odot}$), more compact (with a half--mass radius of $\sim0.4$ pc) and therefore much denser ($\sim2\times10^{5}$ pc$^{-3}$) than the ONC. It also has a higher one--dimensional velocity dispersion (5.4 km s$^{-1}$). Encounter rates are therefore expected to be substantially higher in this system and, since its age is comparable to that of the ONC (a few Myr), the total number of encounters that have already occurred should also be much higher. Observations by \cite{2010ApJ...718..810S} revealed that the disc fraction in the Arches cluster is an increasing function of distance from the cluster centre, rising from a few percent in the core to around ten percent at a radius of 0.3 pc. \cite{2012ApJ...756..123O} again employed N--body modelling, and post--processing with techniques similar to \cite{2006ApJ...642.1140O} to infer disk mass--losses. They found disc destruction fractions of 10$\%$ in the entire cluster and 30$\%$ in the core over 2.5 Myr.

\citet{malmberg2007,malmberg2011} performed N-body simulations of clusters containing a number of stars ranging from 150 to 1000 and half-mass radii ranging from 0.38 to $7.66 \ \mathrm{pc}$. They quantified from the simulations the fraction of singletons, which they define as those stars that never had encounters closer than $1000 \ \mathrm{au}$. They found that in some cases almost $\sim 85 \%$ of stars are non-singletons, with potential impact on planet-forming protoplanetary discs and already existing planetary systems. They also found frequent exchange of stars in binaries. The effect of fly-bys on already formed planetary systems is to lead to planet ejection and eccentricity excitation in planets that are left in the system, as well as increasing the probability of planet-planet scattering after the fly-by. These authors note that due to binary heating, which will lead to a significant cluster expansion, most encounters happen when the cluster is very young, and therefore the impact on proto-planetary discs can be significant.

\cite{2013ApJ...769..150C} performed N--body simulations of a set of idealized, fractally--substructured clusters. Since the local stellar density in cluster subgroups can greatly exceed the average density, the encounter rates in a structured cluster should be considerably higher than in a smooth cluster with otherwise comparable properties. However, stellar subgroups are dynamically erased on a crossing time in bound systems, so it is not obvious that the total \emph{number} of encounters will be higher in a structured cluster. \cite{2013ApJ...769..150C} found that the overall enhancement in the number of encounters due to substructure is only a factor of a few, and that discs in such clusters are not likely to be significantly dynamically influenced in this way.

In this paper, we present results from hybrid N-body - smoothed particle hydrodynamics (SPH) simulations of coupled cluster and protoplanetary disc evolution. Therefore, we do not need to infer \emph{a posteriori} the effect of encounters on discs, but we compute it self-consistently together with the stellar dynamics. This allows us to include effects that were neglected in previous studies:
\begin{itemize}
\item disc spreading and truncation by encounters;
\item accretion onto the central star;
\item the finite time for a disc to regain equilibrium after an encounter;
\item the inclination of the rotation axis with respect to the inclination of the two stars' orbital plane, which has an important effect (it is well known, for example, that a retrograde passage is much less harmful for the disc than a prograde one);
\item disc-disc interactions, if both stars in an encounter have a disc;
\item the mass transfer between stars, possibly leading to the formation of a new disc.
\end{itemize}

Rather than trying to accurately reproduce one particular stellar cluster, we concentrate here on an idealized model. This allows us to work in a controlled environment, identifying the new phenomena that arise due to the new computational method adopted. At this stage, we are able to make some preliminary comparison with observations. The questions we want to answer are:
\begin{itemize}
\item How important are stellar encounters for disc dispersal? 
\item What are the conditions under which disc sizes are set by stellar encounters? 
\item Are there observables in protoplanetary discs that can tell us if a disc or a disc population experienced significant encounters?
\end{itemize}

Our paper is organized as follows. After describing the computational method in section \ref{sec_model}, we present our results in section \ref{sec_results}. We discuss them in section \ref{sec_discussion}, comparing with results from a simple semi-analytical method and with observations, and we draw our conclusions in section \ref{sec_conclusions}.

\section{Model}
\label{sec_model}

\subsection{Numerical method}

We use the SPH code \textsc{seren} \citep{SerenPaper}.  \textsc{seren} is capable of modelling both the hydrodynamics and stellar dynamics individually, or coupled together in the same simulation \citep{HybridNbodySPH}.
The equations of motion are derived via the Euler-Lagrange equations, similar to the derivations of \citet{2002MNRAS.333..649S} and \citet{2007MNRAS.374.1347P}, but including the coupled gas-star terms to maintain energy conservation.  The SPH particles are integrated using a 2nd order Leapfrog kick-drift-kick integration scheme and the star particles are integrated with a 4th-order Hermite integration scheme.  The equations of motion for both the SPH and star particles are integrated on hierarchical block timesteps. The smoothing lengths of SPH particles are calculated with the relation,
\begin{equation}
h_i = \eta\,\left( \frac{m_i}{\rho_i} \right)^{1/3}
\end{equation}
where $\eta = 1.2$ and $\rho_i$ is the SPH density.  We use the M4 kernel \citep{1985A&A...149..135M} for calculating all SPH sums.

We employ an ideal gas equation of state, assuming a mean molecular weight $\mu=2.35$. Due to the already high computational demands of running a cluster simulation with gas dynamics, we use a simplified approach to modelling the thermal and radiation physics: the temperature of each SPH particle depends only on its position relative to the central star (we explain in detail in section \ref{sec_ic_disc} how particles temperatures are assigned). Therefore, we do not need to solve the energy equation, nor alternative forms such as the entropy equation. 

In order to capture shocks and prevent interpenetration of particles, SPH needs to include an aritifical viscosity term. We use the term proposed by \citet{Monaghan1997} of the form,
\begin{equation} \label{EQN:ARTVISC}
\left. \frac{d{\bf v}_i}{dt} \right|_{_{\rm AV}} = \sum \limits_{j=1}^{N}\,{ \frac{m_j\,\alpha_{_{\rm AV}}\,v_{_{\rm SIG}}\,{\bf v}_{ij} \cdot \hat{\bf r}_{ij}}{\overline{\rho}_{ij}}\,\overline{\nabla_i\,W_{ij}}}\,,
\end{equation}
where $\alpha_{_{\rm AV}}$ is a dimensionless factor of order unity and $v_{_{\rm SIG}}= c_i + c_j - \beta_{_{\rm AV}} \,{\bf v}_{ij} \cdot \hat{\bf r}_{ij}$ is the signal speed between neighbouring SPH particles, with $\beta_{_{\rm AV}} = 2 \, \alpha_{_{\rm AV}}$.  In order to capture shocks, $\alpha_{_{\rm AV}} = 1$ usually suffices for adiabatic shocks for all Mach numbers (e.g. \citealt{2013MNRAS.432..711H}). It can be shown \citep{artymowiczlubow1994,murray96} that the artificial viscosity term can also be used to model the physical viscosity that is responsible in accretion discs for the redistribution of angular momentum. An unwanted effect is that in this case the artificial viscosity results in both bulk viscosity, which is required to capture shocks, and shear viscosity, which is the only one required in accretion discs. In practice, this is usually not a major problem in simulations of accretion discs, as bulk viscosity acts on strongly convergent flows, which are usually not present in accretion discs.  

The effective shear viscosity is resolution-dependent, and since our simulations have a relatively low resolution on the scale of individual discs, the effective shear viscosity is very high, leading to rapid viscous evolution of discs.  Although a variety of viscosity switches exist in SPH that attempt to address this problem (e.g. \citealt{balsara1995}, \citealt{morris&monaghan1997}), we simply adopt a lower value of $\alpha_{_{\rm AV}} = 0.1$ in all our simulations. Although this has the unwanted side effect of reducing the ability to capture strong shocks, this is a smaller problem than the high shear viscosity for our simulations. In the next section, we discuss the link between our numerical parameters and the physical values of the viscosity expected in proto-planetary discs.

To model the accretion of gas onto stars, we use sink particles similar to those described by \citet{1995MNRAS.277..362B} and \citet{SerenPaper}.  The smoothing length of sink particles is simply $R_\mathrm{in}/2$ for close encounters between sinks, where $R_\mathrm{in}$ is the sink accretion radius, whose value will be given in the next section. We do not allow the formation of new sinks, only the accretion of SPH particles onto existing sink/star particles.  Indeed, we do not expect new sinks to form both for numerical and physical reasons. Physically, the masses contained in the discs are too low to form new stars and the discs are not gravitationally unstable. In addition, other works \citep{2007MNRAS.374..590L,2009MNRAS.400.2022F} have shown that encounters between stars prohibit the fragmentation of discs and stabilise them. Also, the numerical resolution is too low to follow the formation of planet-sized objects by gravitational instability \citep{1997MNRAS.288.1060B}.

\subsection{Physical set-up}
\label{sec_ic}

\subsubsection{Cluster set-up}
Our model comprises two particle species, stars and gas. We choose to perform a controlled experiment and model the cluster dynamics as simply as possible by employing a Plummer sphere of 100 equal-mass stars. The Plummer sphere has a density profile given by:
\begin{equation}
\rho(r) = \frac{3M}{4\pi a^3} \left(  1+ \frac{r^2}{a^2}  \right) ^{-5/2}, \label{eq_plummer}
\end{equation}
where $M$ is the cluster mass and $a$ is a scale-radius, normally called the Plummer radius. The procedure used to generate the sphere is described in \citet{AarsethPlummer}, which describes also how to initialize the velocities under the assumption that the velocity distribution is isotropic. Because of their simplicity, Plummer spheres have been commonly used in previous works when dealing for the first time with a new numerical technique \citep{1984ApJ...285..141L,pelupessy12}. Since the Plummer sphere density profile theoretically extends to infinity, we truncate it at 20 times the Plummer radius. We use dimensionless units in which the radius and the mass of the Plummer sphere is 1, so that our results can be rescaled to different cluster sizes and masses. However, we note that for each given simulation the ratio between the star and disc mass, and correspondingly between the cluster and the disc size, stays constant when rescaling. In the rest of the paper, we assume that each star has a mass of $1 M_\odot$ and that the Plummer radius is $0.1 \ \mathrm{pc}$. We then scale all other quantities accordingly.

In the spirit of performing a controlled experiment, binaries are not included. As discussed by \citet{2012ApJ...756..123O}, this underestimates the effect of encounters by reducing the number of stars. In addition, binaries might have also an effect on the long term evolution of the cluster, since the binding energy of a single binary can easily be larger than the binding energy of the whole cluster. On the relaxation time scale, binary heating leads to expansion of the cluster, which would instead cause a decrease in the importance of encounters. As we discuss in section \ref{sec_results}, the relaxation time-scale is, however, longer than the time span we simulate. In addition, this level of sophistication would require a better dynamical model of the cluster than the Plummer sphere we employ here.

\subsubsection{Disc set-up}
\label{sec_ic_disc}

The only gas initially present is in the discs; we do not include any diffuse gas. Observations of star forming regions show that, even in very young regions, a fraction of young stars show no sign of infrared excess or accretion \citep{Fedele2010}. It could be that the discs of these stars have undergone a different, more rapid evolutionary path than the one of the discs we still observe. Recent work suggests that stars that do no longer have a disc are binary stars \citep{2012ApJ...745...19K}. Another possibility is that the age spread between stars must account for this difference (although this is probably small, see \citealt{longmore14}). Therefore, we add a randomly oriented gas disc around 50 per cent of the stars. 

The surface density profile of the gas is given by a power-law, with a slope of $p=3/2$ as estimated for the Minimum Mass Solar Nebula \citep{MMSN}:
\begin{equation}
\Sigma (R)= \Sigma _0 \left(\frac{R}{R_0}\right)^{-p}, \label{eq_sigma}
\end{equation}
where $R$ is the distance from the star in the disc plane, $R_0$ is a scale radius and $\Sigma_0$ is a surface density scale. The particles are distributed so as to attain a Gaussian density profile in the vertical direction, with thickness $H=c_\mathrm{s}/ \Omega$, where $c_\mathrm{s}$ is the gas sound speed and $\Omega$ the Keplerian orbital frequency around the star. We choose $\Sigma_0$ as to set the initial disc mass to be $5 \%$ of the stellar mass. The disc is truncated at the initial disc radius $R_\mathrm{out}$, and the inner disc radius is set to $R_\mathrm{in} = R_\mathrm{out}/5$. The true $R_\mathrm{in}$ is of the order of the star radius, but it is not possible to resolve it, because of the very short orbital time-scale there. Note that, as the disc expands, $R_\mathrm{in}$ stays constant, and the difference in the particle orbital time-scales increases. If particles move to within $R_\mathrm{in}$ of the star, they are accreted and removed from the simulation.


The temperature structure in the disc follows a power law distribution with radius: 
\begin{equation}
T(R) = T_0 (R/R_0)^{-q}. \label{eq_t_r}
\end{equation}
We choose the exponent $q$ to be $1.5$ for numerical convenience, although observations \citep[e.g.,][]{Andrews2005} find flatter distributions, with a median value of $0.58$. Our approach has the numerical advantage that it gives a constant vertical resolution and a constant $\alpha_\mathrm{_{SS}}$ parameter in the disc \citep{2007MNRAS.381.1287L,LodatoViscosity}, where $\alpha_\mathrm{_{SS}}$ is the standard viscosity parameter proposed by \citet{AlphaViscosity}. The $\alpha_\mathrm{_{SS}}$ parameter is related to the kinematical viscosity $\nu$ of the gas by the prescription $\nu=\alpha_\mathrm{_{SS}} c_\mathrm{s} H$. We fix the normalization $T_0$ in equation \ref{eq_t_r} so that the aspect ratio of the disc $H/R$ is 0.05 at the inner radius. At each timestep, we use the distance from the nearest star to set the temperature of each SPH particle. For simplicity, we use the spherical distance rather than the distance in the disc plane, as this would require to identify at each timestep the discs and find their axis, adding extra computational cost to the simulation. In practice, the difference introduced is marginal for particles in a thin disc, and it affects only particles that get ejected by the discs. For these particles, using the distance in the disc plane would be questionable anyway. Finally, to prevent unphysically low temperatures, we impose a lower threshold corresponding to the one at a distance of $7 R_\mathrm{in}$, which for run R10 corresponds to $~20 \ \mathrm{K}$.

We run simulations with $R_\mathrm{out}=$ 10, 30, 100 and 300 \au, which will be referred to in the text as R10, R30, R100 and R300, respectively. The parameter $R_\mathrm{out}$ can be interpreted not only as the initial disc size at the beginning of the class II phase, but also as the age of the disc at the beginning of the simulation, where bigger discs are ``old'' discs (due to viscous spreading) and the different simulations represent different evolutionary stages. Clearly, this is not fully self-consistent as we are ignoring the effect of the encounters in the expansion that brought the discs to reach these sizes. Nevertheless, since observations \citep{WilliamsCieza2011} show that protoplanetary discs can reach these sizes, it is interesting to know what is the effect of encounters on such discs.

We initially populate the discs with SPH particles by Monte-Carlo sampling the surface density distribution given in Equation \ref{eq_sigma}. This causes small random fluctuations which are erased on an orbital time-scale and do not affect our results. The particles are initialized in Keplerian orbit around their stars. Our resolution is $10^4$ particles per disc, resulting in a total number of $5 \times 10^5$ particles for each simulation. At this resolution, using the relations in \citet{2007MNRAS.381.1287L} shows that the discs are barely vertically resolved, that is, $h/H \simeq 1$, where $h$ is the SPH smoothing length. We also check in the simulation output files that this is the actual resolution in the vertical direction. We compute the scale-height of the disc through the standard deviation of the vertical coordinate of the particles comprising the disc and by fitting a gaussian profile to the density. We find that the two methods give the same answer within a factor of $\sim 1.5$. In all cases, the ratio $h/H$ is between 1 and 2.

Our choice of the resolution allows the possibility of simulating a long timespan (for run R10, we simulate $~170000$ orbits at the inner radius) and many discs at the same time, rather than being able to follow the detailed behaviour of the individual discs. A similar resolution has been employed in studies of star-star encounters \citep{1995ApJ...455..252H,2005ApJ...629..526P}. In particular, \citet{2005ApJ...629..526P} reported no significant difference between a simulation run with $10^4$ particles and one with $10^6$ particles. However, due to the accretion of particles on to the star, our spatial resolution degrades as the simulation progresses. For example, at the end of simulation R10, only $\sim 10$ per cent of the particles are left in the discs; when combined with the disc spreading, the analytical relations predict a degradation in the spatial resolution of a factor of $~3$. The actual values extracted from the simulation are however still in the same range as at the beginning of the simulation. We warn however that this does not mean that the actual resolution is higher than what we would expect; rather, it means that by definition it is not possible to resolve in the simulation features that are smaller, within a factor of order unity, than $h$. The interpretation of this result is thus that at the end of our simulations the discs are, for numerical reasons, thicker than what is expected from their temperatures.

It is possible to compute the resulting viscosity from the chosen density and temperature profiles. According to \citet{LodatoViscosity}, in our simulations the Shakura-Sunyaev parameter is $\alpha_\mathrm{_{SS}} \simeq 0.004$, which is in line with the observational results \citep{Armitage2011}. A possible concern is that the analytical relations do not hold at the resolutions employed in this paper. For this reason, in the next section we measure explicitly the value of the viscosity by measuring the rate of spreading of the disc. Indeed, we find the effective viscosity is higher than predicted by this estimate, yet still marginally compatible with the values found in observations.


\subsection{A semi-analytical model for the disc size}
\label{sec_semianal}

In this section we present a semi-analytical model that we will use in section \ref{sec_understanding} to understand the results of the simulation in terms of disc sizes. A class of widely used models for a disc in isolation are the family of self-similar solutions derived by \citet{lyndenbellpringle}. They describe the evolution of a disc whose viscosity profile is a power-law. The radius time evolution is given by:
\begin{equation}
R_\mathrm{disc} (t, R_0, t_{\nu,0}) =\left( 1+ \frac{t}{t_{\nu,0}} \right)^{1/(2-\gamma)} R_0,
\label{eq_r_t}
\end{equation}
where $R_0$ is the initial radius, $t_{\nu,0}$ the viscous time at the initial radius, and $\gamma$ is the exponent of the viscosity with respect to radius. The viscous time can be related to the $\alpha_\mathrm{_{SS}}$ (see Section \ref{sec_ic_disc}) parameter using the definition of viscous time (see equation 20 of \citealt{Hartmann98}) and standard relations:
\begin{equation}
\alpha_\mathrm{_{SS}} = \frac{1}{6 \pi} \left(\frac{t_{\nu,0}}{t_\mathrm{dyn}}\right)^{-1} (H/R)^{-2} \frac{1}{(2-\gamma)^2},
\label{eq_alpha}
\end{equation}
where $t_\mathrm{dyn}$ is the orbital time scale at $R_0$ and $H/R$ is the aspect ratio. For simplicity we will use $H/R=0.05$ when evaluating this relation numerically. We note that this is a worst-case scenario (that is, a slightly overestimate of $\alpha_\mathrm{_{SS}}$), as this is the value at the inner radius and the aspect ratio is a midly growing function of radius. In addition, if the disc is vertically under-resolved, the effective $H$ will be thicker than the one due to thermal pressure, therefore leading to a higher $H/R$ and therefore to a lower $\alpha_\mathrm{_{SS}}$ than the estimate we get.

The expansion law has the nice feature that, being self-similar, one is always free to ``reset'' what we call initial radius, and start the evolution again from there, without changing the results (provided we also update the viscous time). In mathematical terms,
\begin{equation}
R_\mathrm{disc} (t'', R_0, t_{\nu,0}) =R_\mathrm{disc} (t''-t', R_1, t_{\nu,1}),
\end{equation}
where $R_1=R_\mathrm{disc} (t', R_0, t_{\nu,0})$ and $t_{\nu,1}$ is the viscous time at $R_1$. 

We exploit this property in our simple semi-analytical model.  We let the disc increase in size at each timestep following Equation \ref{eq_r_t}. In order to derive the parameters, we use numerical fits to the results of the evolution of discs in isolation. At each timestep, we look in the results of the simulation for the closest star at that time to a given disc and record its distance $d$. We assume that the encounter would have truncated the disc at $d/3$ \citep{2010ARA&A..48...47A}. If the radius is larger than this value, we truncate the disc, otherwise we leave the disc unperturbed. We then let the viscous evolution start again. To summarise, we can compute the final disc radius assuming that:
\begin{enumerate}
\item the radius evolution is always given by Equation \ref{eq_r_t}, i.e. by the \citet{lyndenbellpringle} solution;
\item the encounter distances come from the results of the simulations;
\item the encounters simply truncate the discs at $d/3$.
\end{enumerate}

\begin{figure*}
\includegraphics[width = \textwidth]{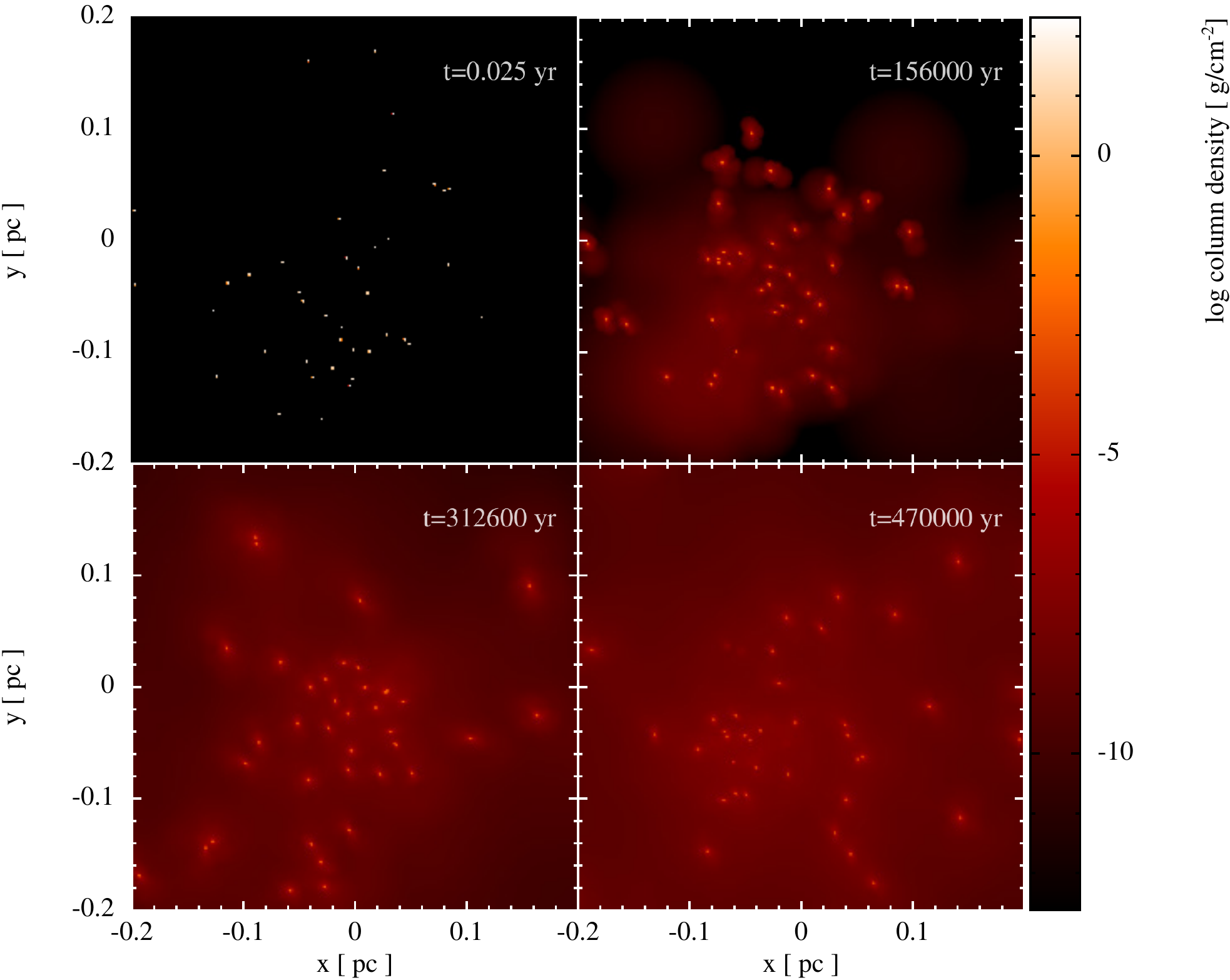}
\caption{Time evolution of the gas column density in run R10. The view is restricted to the central region of the cluster to help visualizing the gas distribution. Note that stars are not plotted.}
\label{fig_snapshot}
\end{figure*}

We do not expect such a simple model to be able to capture the full results of the 3D hydro simulation, however it is useful to assess if the encounters produce a simple truncation or have more complicated effects. As we will show, the fact that there are cases where the model is capable of correctly reproducing some of the results shows that it is a useful tool. In particular, it highlights that in these cases the assumptions that have been used to build it are valid. On the contrary, when the model breaks down it shows that these assumption must have broken down.

\section{Simulations}
\label{sec_results}

We evolve the clusters for 10 dynamical time-scales, where the dynamical time is defined as:
\begin{equation}
t_\mathrm{dyn}=\left(\frac{r_\mathrm{cluster}^3}{G M_\mathrm{cluster}}\right)^{1/2}.
\end{equation}
Here, $G$ is the universal gravitational constant, $M_\mathrm{cluster}$ is the total mass of the cluster and $r_\mathrm{cluster}$ is the scale length $a$ of the Plummer sphere (see Equation \ref{eq_plummer}). For the scale length ($0.1 \ \mathrm{pc}$) and mass values ($100 \ M_\ast$) introduced in section \ref{sec_ic}, the dynamical time-scale is $\simeq 47000 \ \mathrm{yr}$. Therefore, the simulations are evolved for $t_\mathrm{end} = 0.47 \ \mathrm{Myr}$. For example, after this time in simulation R10, nearly $90 \%$ of the initial mass has accreted onto the stars.

Our cluster will evolve on the relaxation time, which is given by:
\begin{equation}
t_\mathrm{relax}=\frac{N}{\ln (0.4 N)} t_\mathrm{dyn} \simeq 27 t_\mathrm{dyn} \simeq 1.27 \ \mathrm{Myr}.
\end{equation}
Given that this time is longer than the one we simulate, we do not expect a significant evolution of the cluster during the simulation due to pure N-body effects. The relaxation time is of the same order as the lifetime of protoplanetary discs ($~3 \ \mathrm{Myr}$, \citealt{Fedele2010}). This means that during the life of a protoplanetary disc there will be some evolution of the cluster.  This is important when interpreting discs with large outer radii as discs which have already evolved due to viscous spreading. In this case, the simulation is not fully self-consistent since we do not simulate the early dynamical evolution of the cluster but start from the same cluster initial conditions.

\begin{figure}
\includegraphics[width=\columnwidth]{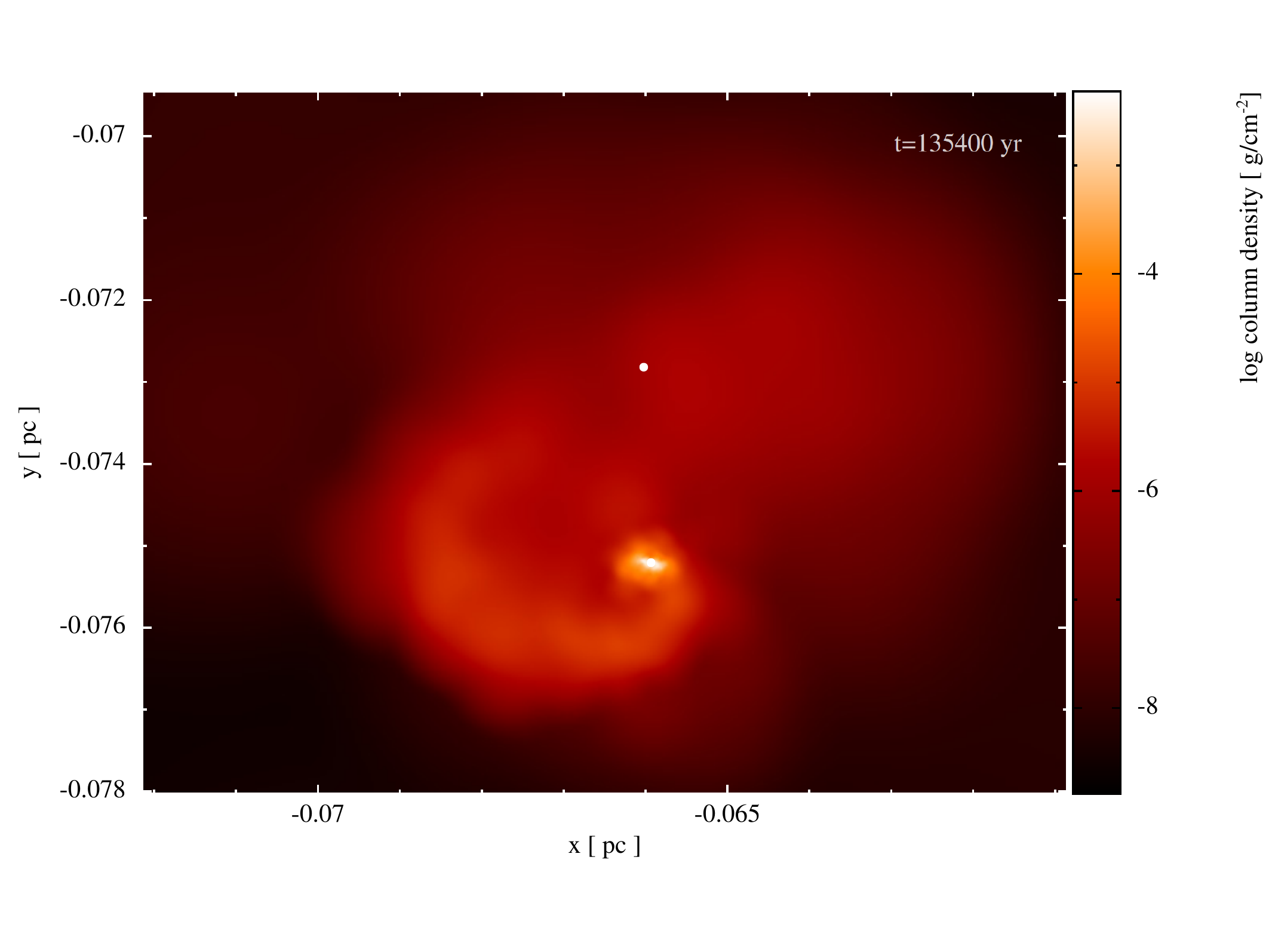}
\caption{Column density distribution at time $t=135400 \ \mathrm{yr}$, showing the stripping of a disc shortly after the interaction of a star with a disc with a second star. The two stars are represented by the white dots. The tidal tail created during the interaction is clearly visible.}
\label{fig_snap_spiral}
\end{figure}

\subsection{Extracting the discs from the simulation}
To analyze the results of the simulations, we apply a procedure to extract the discs. Each gas particle is assigned to the star that it is most bound to. We also apply a cut off in eccentricity of $0.9$, but in practice we find that very few bound particles have such high eccentricities. We define the \textit{ambient gas} as particles that are not bound to any star. Once we have identified a disc, in order to find its plane we apply the algorithm already used in \citet{2010MNRAS.402.2253W}. We compute the inertia tensor of the disc, defined as \citep{LandauLifshitz}:
\begin{equation}
I_{ij}=\sum_a m_a (x_a^2 \delta_{ij} - x_{i,a} x_{j,a}),
\end{equation}
where $\bmath{x}_a$ is the position vector of each particle in the disc with respect to the star, $\delta_{ik}$ the Kronecker delta, $m_a$ the mass of each particle and the summation over index $a$ is running over all the particles in the disc, while the indices $i,j=\{x,y,z\}$ are for coordinate axes. It can be shown that, in the limit of a razor thin axisymmetric disc, the eigenvalues of the tensor $I_{x'}, I_{y'}, I_{z'}$ are such that $I_{x'} = I_{y'} + I_{z'}$, where $I_{z'}$ is the eigenvalue corresponding to the eigenvector along the rotation axis $z'$ of the disc, while the other two eigenvectors lie in the plane of the disc. Therefore, to identify the plane we diagonalize the tensor and define the direction of the eigenvector with the largest eigenvalue as the direction of the disc axis. Since the disc is a continuous structure without an abrupt end, we compute the radius of the disc as the half-mass Lagrangian radius, that is, the radius that contains half the mass of the disc. If the evolution of the disc remains purely viscous, the surface density follows one of the self-similar solutions found by \citet{lyndenbellpringle} and the Lagrangian radius is proportional to the exponential tapering radius of the self-similar solution \citep[e.g.,][]{Hartmann98}. The self-similar solution is used in sub-millimeter observations to fit the surface density profile and derive the disc size \citep{WilliamsCieza2011}; therefore, it is important that our method is able to give consistent results in this case.

\subsection{Spreading in isolation}
\label{sec_isolation}

In order to compare the model presented in section \ref{sec_semianal} with the results of the simulation, we need to know the disc expansion law. In principle, if we were to know exactly our viscosity law, we could use it to derive our expansion law. In practice, since we cannot afford a very high resolution due to computational limitations, we have to instead rely on calibrations, that are derived by fitting the evolution of a disc in isolation.

\begin{table}
\begin{tabular}{ccccccc}
\toprule
Run & $R_\mathrm{out} [ \mathrm{au} ]  $ & $\gamma$ & $t_\nu \ [\mathrm{yr}]$ & $\alpha_\mathrm{_{SS}}$ & $t_\mathrm{spread} [ \mathrm{yr}]$ & $\boldsymbol{\alpha_\mathrm{_{SS,local}}}$\\
\midrule
R10 & 10 & 1.11 & 18891 & 0.045 & 16800 & 0.1\\
R30 & 30 & 0.44 & 23218 & 0.062 & 36220 & 0.45\\
R100 & 100 & -1.69 & 11762 & 0.133 & 43400 & 5.4\\
R300 & 300 & -3.19 & 25432 & 0.161 & 132000 & 13\\
\bottomrule
\end{tabular}
\caption{For each simulation run, we show the parameters of the fit to the radius-time relation with the analytical solution given by equation \ref{eq_r_t}.}
\label{tab_param_fit}
\end{table}

We show in table \ref{tab_param_fit} the parameters of the fit, together with the corresponding value of the effective viscosity $\alpha_\mathrm{_{SS}}$ computed using Equation \ref{eq_alpha}. We note that these values are higher than those predicted from the analytical relations (by a factor of $40$ in the worst case). Typical values for $\alpha_\mathrm{_{SS}}$ range from $10^{-2}$ to $10^{-4}$, so that our discs are quite viscous. From the analytical relations, we would expect a constant $\alpha_\mathrm{_{SS}}$, and therefore a constant viscosity, which should translate to $\gamma=0$. Since we get a different value, this means at these resolutions the analytical formulae for SPH viscosity are not valid. On the other hand, this is an effect that we can calibrate for. Although we can not decide which viscosity law to apply, we can still derive it \textit{a posteriori} by looking at the evolution of a disc in isolation. This also means however that care should be taken when interpreting the value of $\alpha_\mathrm{_{SS}}$ reported in the table. This value is to be interpreted as a \textit{global}, effective value that describes how fast overall the disc is expanding. However, the \textit{local}, that is, at $R_\mathrm{out}$, level of angular momentum transport is higher than this. This is important as it is this local level that determines the ability of the disc to wash out local perturbations. For reference we report in the table also the \textit{local} values of $\alpha_\mathrm{_{SS}}$, that we compute from the formula \citep[e.g., ][]{Armitage2011}:

\begin{equation}
\alpha_\mathrm{_{SS,local}} = \frac{1}{2 \pi} \left(\frac{t_{\nu,0}}{t_\mathrm{dyn}}\right)^{-1} (H/R)^{-2}.
\end{equation}

We note that the different runs have quite different expansion laws. In particular, it is the exponent in the relation that tends to vary the most. Some of the discs show values of $\gamma$ that are clearly unphysical: for example, the value $-3$ for R300 implies a very steep and increasing dependence of the temperature with radius, which is not present in our model. Therefore, one should regard the \citet{lyndenbellpringle} similarity solution as a fitting formula for the evolution of these discs, and not as a \textit{physical} description of their evolution. For this reason, the local estimate of $\alpha_\mathrm{_{SS}}$ is a more accurate description of what is going on in these simulations. Viscosity values are expected to be lower in real discs, which implies that those encounters' effects that in our simulations could not be washed out by viscosity would be even stronger in real discs. The variation of the exponent means that, despite the fact that the viscous times are similar, run R10 is the one varying the fastest (indeed, it even overtakes the other ones by the end of the simulation), while the other ones expand more slowly. For this reason, we stress that looking only at the viscous time might be misleading, since this value alone does not fully describe the evolution of the disc. As another reference, we list in the table also the value of $t_\mathrm{spread}$, which we define as:
\begin{equation}
t_\mathrm{spread} = \frac{R_\mathrm{disc}}{dR/dt},
\end{equation}
that is, the timescale for a significant change in the disc radius. We evaluate the denominator by computing analytically the derivative of equation \ref{eq_r_t}, and use the value of $R_\mathrm{disc}$ at $t=0$. It can be seen that the disc R10 is the one that is varying the fastest. It is reassuring that the disc in run R300, the one with the highest value of $\alpha_\mathrm{_{SS}}$, has a very long $t_\mathrm{spread}$, so that its spreading is quite limited during the course of the simulation (see also figure \ref{fig_medians}).

In order to quantify the effect of the limited resolution available on the viscosity, we run resolution tests of the discs in isolation, that we report in the appendix.

The viscosity is also important after an encounter, as it allows for the particle orbits to circularize. As \citet{1993MNRAS.261..190C} pointed out, the exact form of the viscosity law probably does not matter (in fact, some of the simulations in the literature have been done only with pseudo-viscosity) for what the final surface density distribution after an encounter is. Indeed, the final surface density is given by the particle specific angular momentum, which sets the final distance from the star once all the orbits have circularized. This is true if there are no further encounters, but the exact viscosity that one assumes decides the timescale over which the disc finds a new equilibrium after an encounter. This is important in our simulation, as the outcome of a second encounter will be different depending on whether the disc had time to gain a new equilibrium or not. However, since it is not clear from the physical point of view what the viscosity after an encounter is going to be, we need to just rely on the value that comes out of the SPH algorithm.

\subsection{Simulation R10}

\begin{figure*}
\includegraphics[width=\columnwidth]{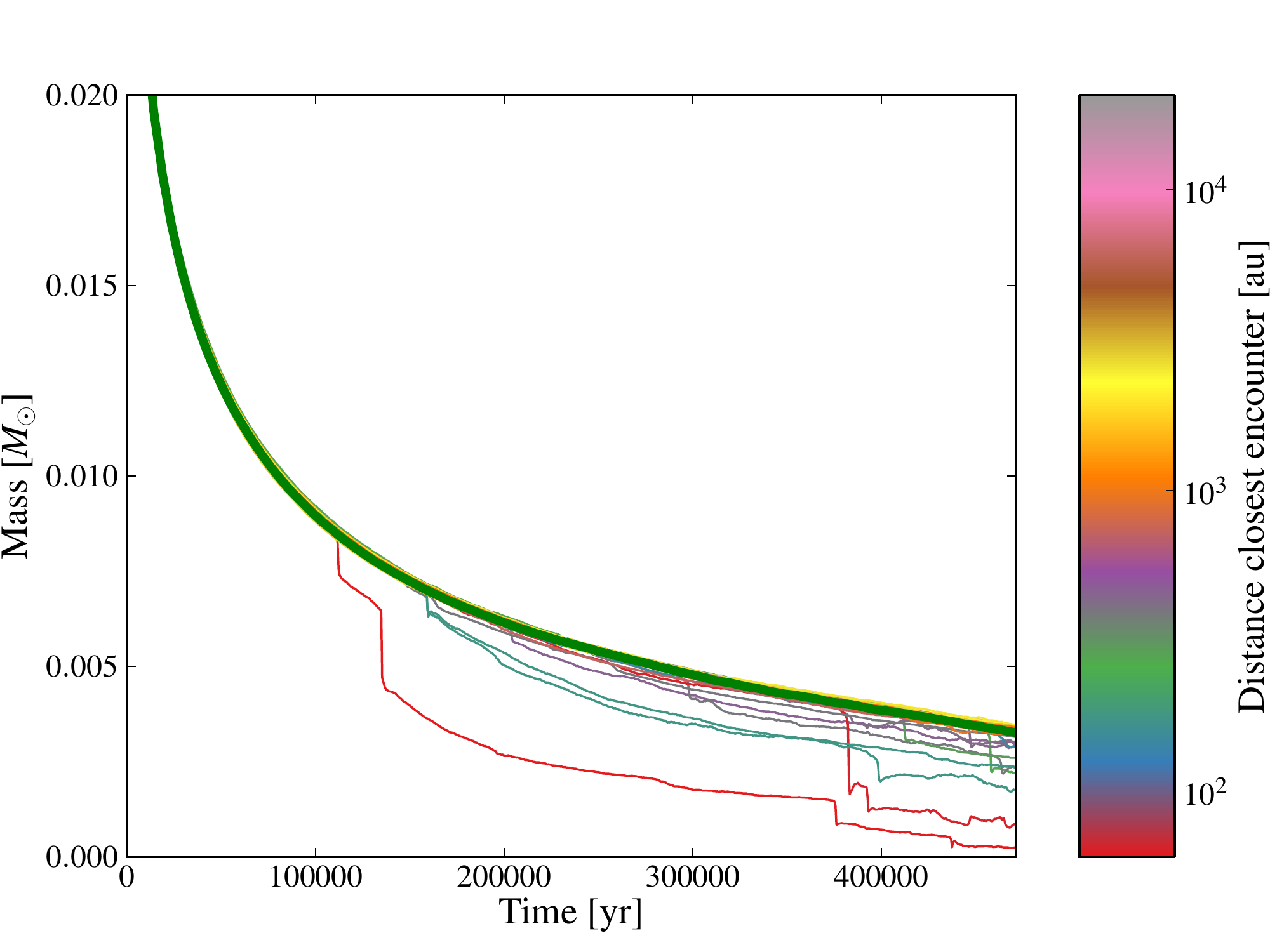}
\includegraphics[width=\columnwidth]{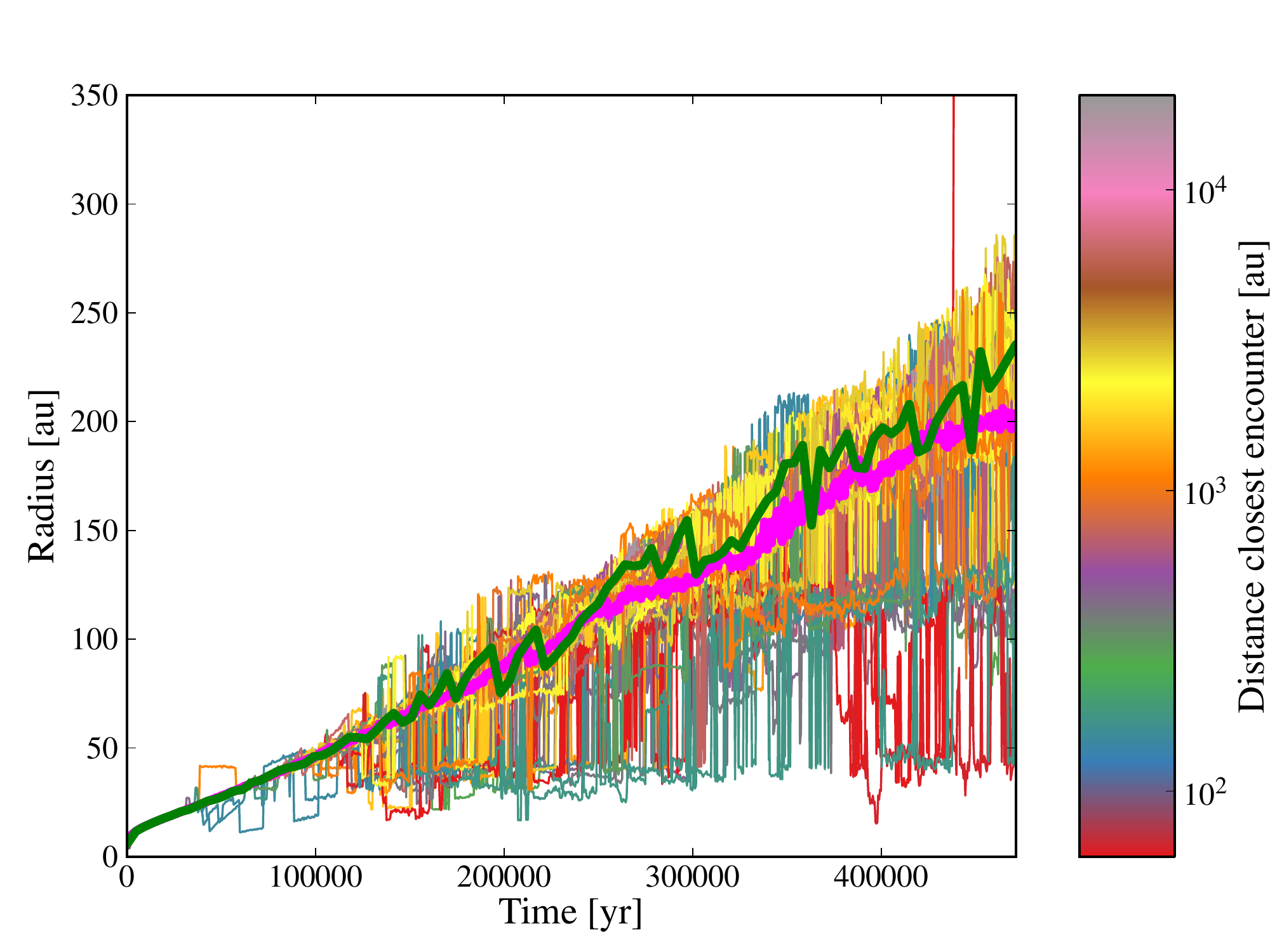}
\caption{Left panel: disc masses as a function of time. Right panel: disc radii as a function of time. In both panels, lines are colored by the distance of closest encounter that the star experienced, and the color gradient shows how the closer the encounter, the more destructive the effect. The thick green line is the disc run in isolation. The thick magenta line in the right panel is the median among all the discs, and shows the truncation effect of the encounters.}
\label{fig_mass_radius_vs_time}
\end{figure*}

We first comment in depth on the results from simulation R10, and use this as a reference to compare to the other simulations. Figure \ref{fig_snapshot} shows four column density snapshots from simulation R10. While at the beginning of the simulation the discs are so small that they are barely visible on the scale of the cluster, they expand significantly due to viscous spreading. Due to this expansion, the discs become large enough to be influenced by encounters. The interactions between stars produce some unbound gas, which is visible as a non-zero background density. The amount of gas that becomes unbound is small, and at the end of the simulation, the mass of the unbound gas is one order of magnitude less than the mass in all discs at $t_\mathrm{end}$. Figure \ref{fig_snap_spiral} shows the detail of a disc during an interaction. The two stars are represented by the white dots (note that only one of them has a disc in this particular case). The tidal tail of gas that has been ejected from the disc \citep{1972ApJ...178..623T,1993MNRAS.261..190C} is clearly visible. We concentrate now on how the encounters affect the disc properties.

Figure \ref{fig_mass_radius_vs_time} shows the evolution of the discs in the cluster as a function of time for simulation R10. The left panel shows the mass of the discs as a function of time, while the right panel shows the radius of the discs as a function of time. The thick green line in the plot is a control run with a single star-disc system run in isolation. It is shown here as a reference, as it allows us to distinguish the effect due to the encounters. In isolation, the radius increases due to the redistribution of angular momentum due to viscosity and the mass decreases due to accretion onto the star. 

\begin{figure}
\includegraphics[width=\columnwidth]{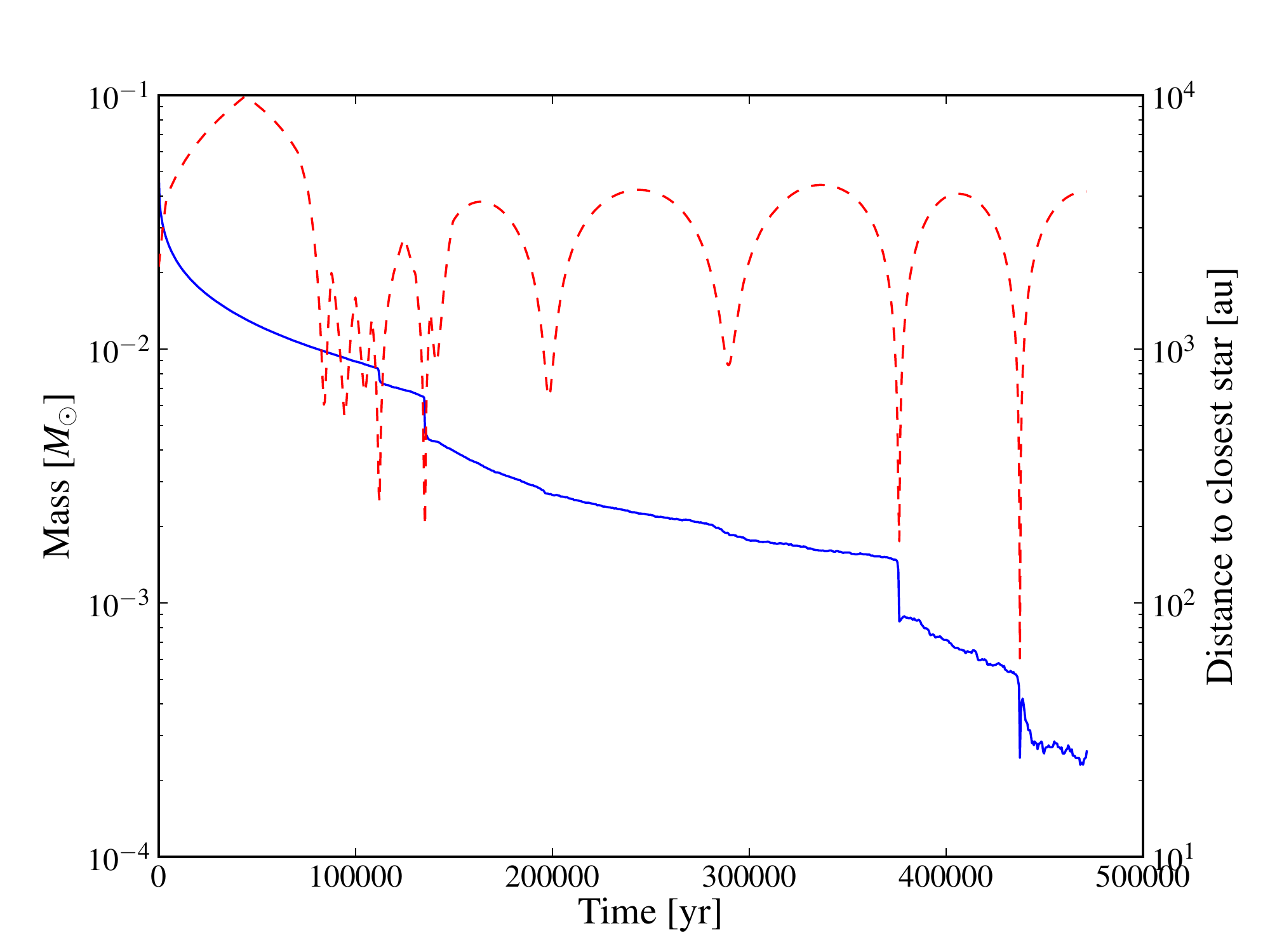}
\caption{Mass versus time (blue solid line) and distance to closest star (red dashed line) for one disc from simulation R10. Sudden drops in mass are clearly caused by close encounters.}
\label{fig_mass&distance_time}
\end{figure}

\begin{figure*}
\includegraphics[width=\columnwidth]{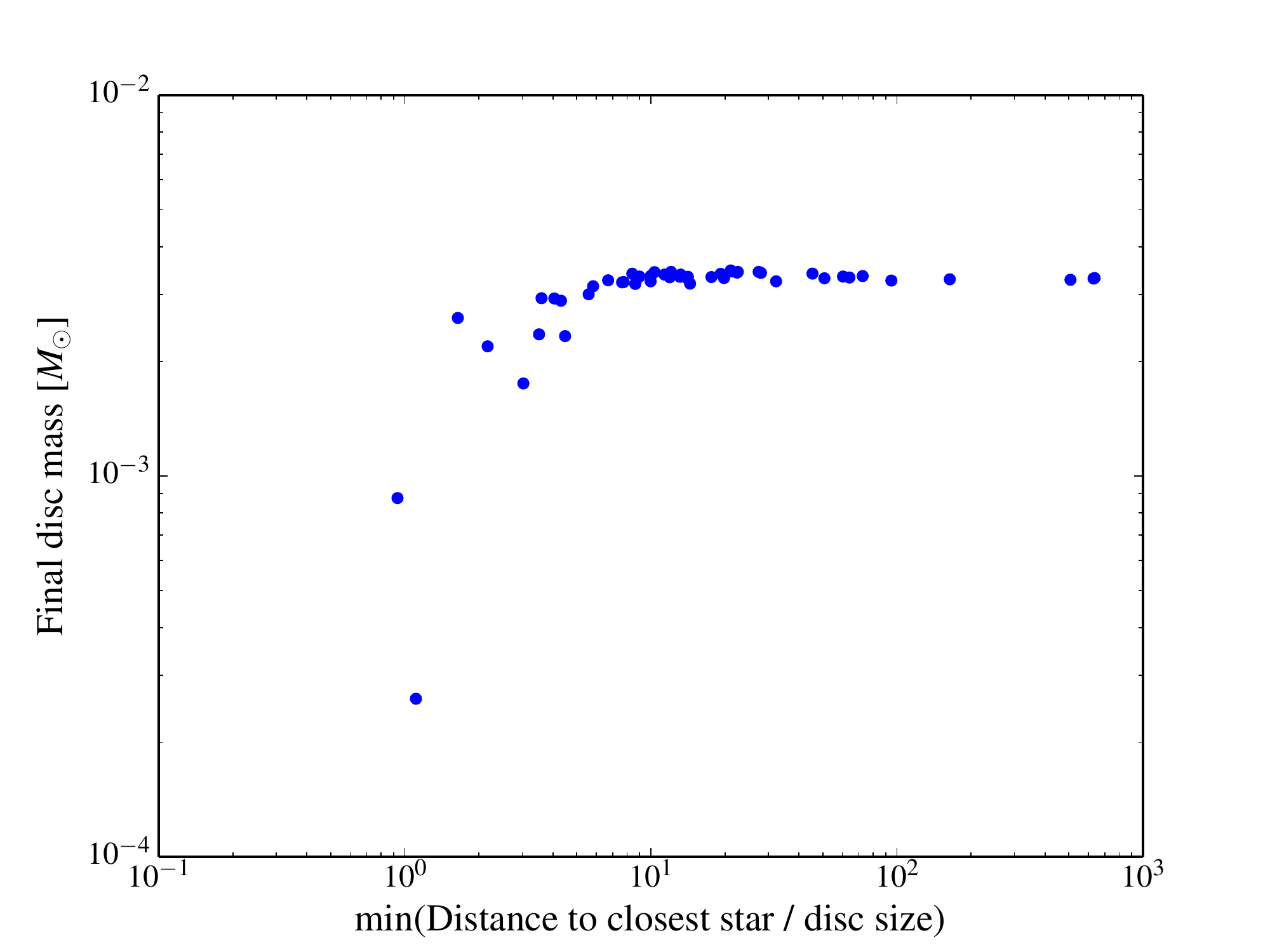}
\includegraphics[width=\columnwidth]{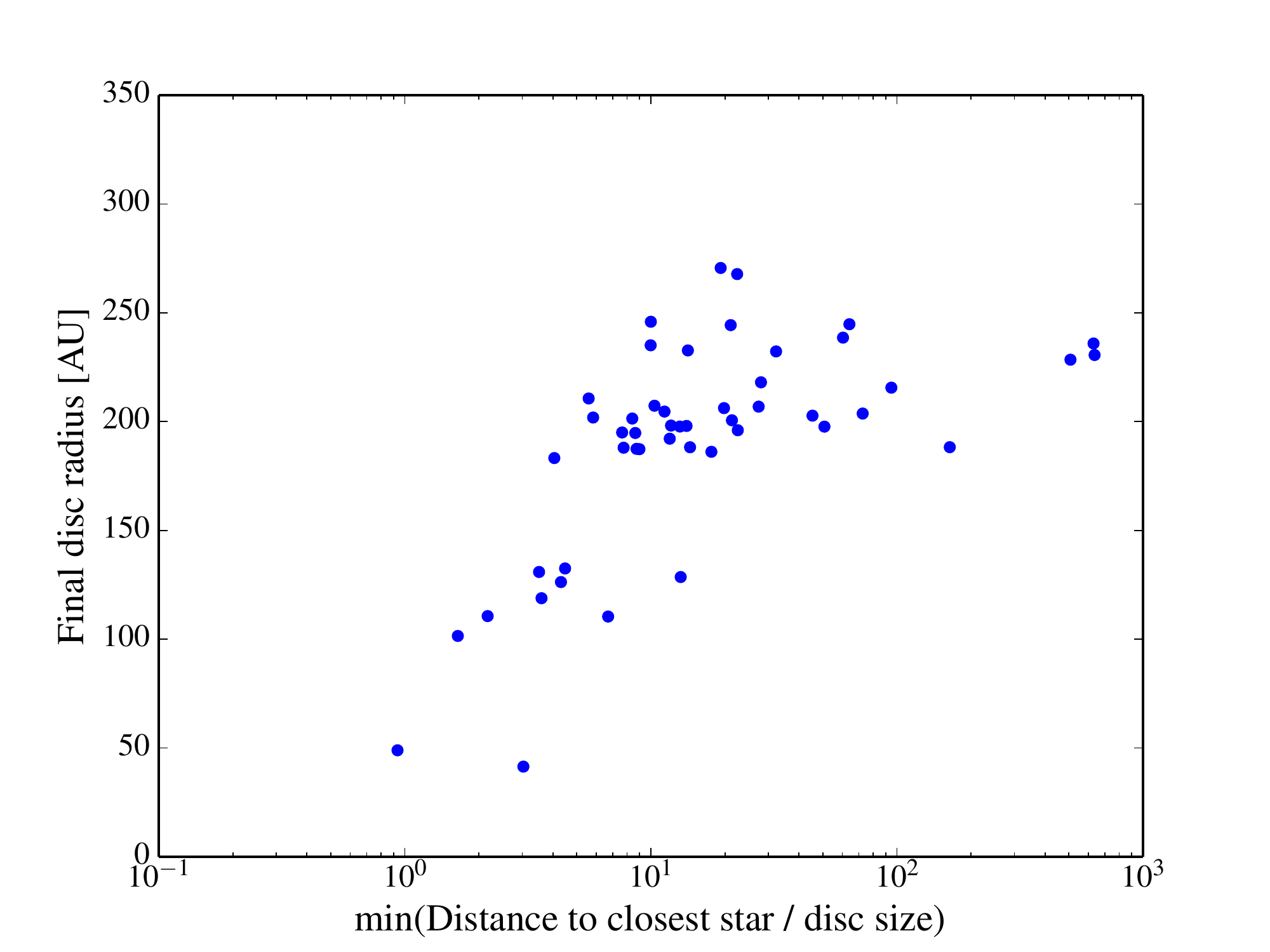}
\caption{Left panel: final disc mass as a function of the distance of closest encounter, in units of the disc size at the moment of the encounter. Right panel: final disc radius a function of the distance of closest encounter, in units of the disc size at the moment of the encounter. The correlation in the disc mass is quite remarkable, and shows how the distance of the encounter is a good quantity to use to derive the mass lost in the encounter. The correlation in the disc radius has more scatter (see the text for an explanation), which shows the importance of numerical simulations to quantify the influence of encounters on the disc sizes.}
\label{fig_mass_radius_distance_encounter}
\end{figure*}

The interactions between stars are stripping mass from the disc. The lines in the plot are colored according to the distance of the closest encounter that each star had during the course of the simulation. The color gradient shows that the closer the encounter, the stronger the effect in ejecting mass from the disc. In Figure \ref{fig_mass&distance_time} we pick one of the discs that had a very close encounter (minimum distance smaller than $100 \ \mathrm{au}$), and plot against time both the disc mass (blue line) and the distance to the closest star at the given time (red dashed line). This clearly shows that the drops in mass are caused by close encounters.

In the right panel of Figure \ref{fig_mass_radius_vs_time}, the radius evolution is not completely smooth even for the disc in isolation. Here the finite resolution certainly plays a more important role than for the mass, being an integrated quantity. Nevertheless, the effects of the encounters are much bigger than the noise for the disc in isolation. Some discs show big variations in the radius after an encounter, which is due to the fact that they are seeking a new equilibrium after they have been perturbed. It can be seen how nearly all the discs have smaller radii than the one run in isolation. To highlight this effect, we plotted also the median among all the discs, which shows the truncation effect of the encounters. The color gradient is not so clearly visible here as in the left panel. It can still be seen however that the discs that are significantly smaller than the disc in isolation experienced close encounters.

To further explore the dependency of disc parameters at $t_\mathrm{end}$ on the distance of closest encounter, we show in Figure \ref{fig_mass_radius_distance_encounter} the final disc radii and masses as a function of the distances of the closest encounter. Since the disc size varies in time, we normalize the distance of the encounter to the disc size at the moment of the encounter. The correlation in mass (left panel) is quite strong, and confirms that the distance of the closest encounter is a good quantity to derive the mass lost in an encounter, even when disc spreading and mass accretion onto the central star are taken into account. This confirms qualitatively the validity of previous studies \citep[e.g., ][]{2001MNRAS.325..449S,2006ApJ...642.1140O} that used this parameter, either recording the single closest encounter, either the history of the most destructive ones, to quantify the importance of encounters in mass removal from the disc. However, due to the presence of accretion and spreading in our work, a detailed comparison with previous work is not possible. Note that, since the discs all start from the same initial conditions and follow the same evolution in absence of external perturbations, discs that experienced only distant encounters end up with the same value for the mass. We do not expect a real disc population to exhibit such behaviour, due to a spread in the initial conditions and in the evolution.

The right panel shows that also the disc size at $t_\mathrm{end}$ correlates with the distance of the closest encounter. This shows the destructive effect of encounters, which are able to truncate the discs. It is instructive to compare this correlation with the mass one. The disc size is more sensitive to distant encounters than the disc mass (Figure \ref{fig_mass_radius_distance_encounter}). In particular, distant encounters (e.g., $10-20$ disc radii) do not affect the disc mass, but are able to modify the disc radius. Since a star also experiences other encounters than the closest one, they can also contribute to determine the final disc size. This is one of the reasons why the radius correlation has more scatter than the mass one.

In addition, after an encounter the disc spreads again, so that, given two discs that experienced a close encounter at the same distance but at different times, we do not expect the final sizes to be the same. This also means that, while we commented before that our simulations confirm the validity of semi-analytical approximations for inferring the disc mass, the same cannot be said for the disc sizes. Numerical simulations are of primary importance here to get accurate determinations of the importance of dynamical interactions in shaping the disc size. 

\begin{figure*}
\includegraphics[width=\columnwidth]{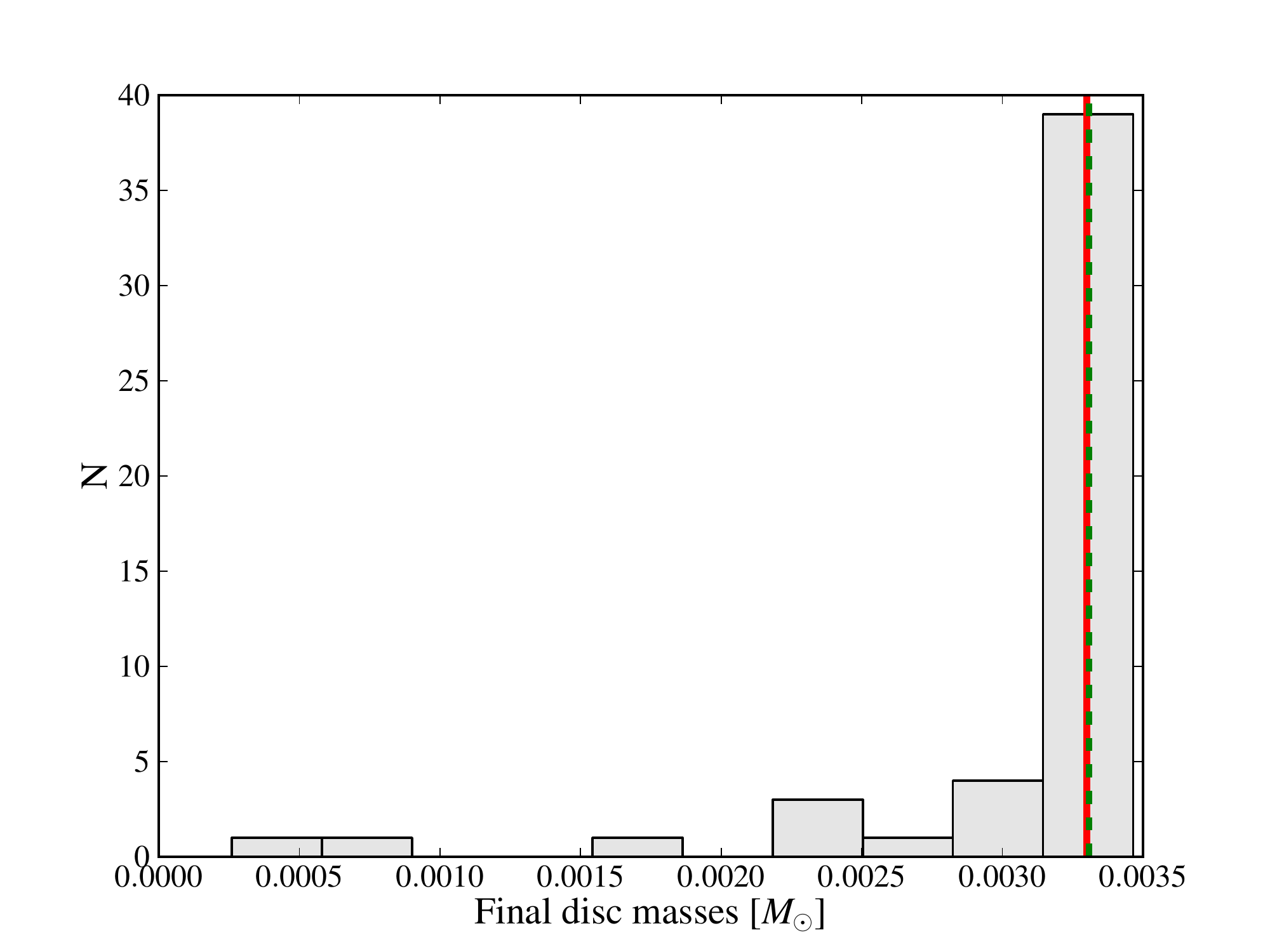}
\includegraphics[width=\columnwidth]{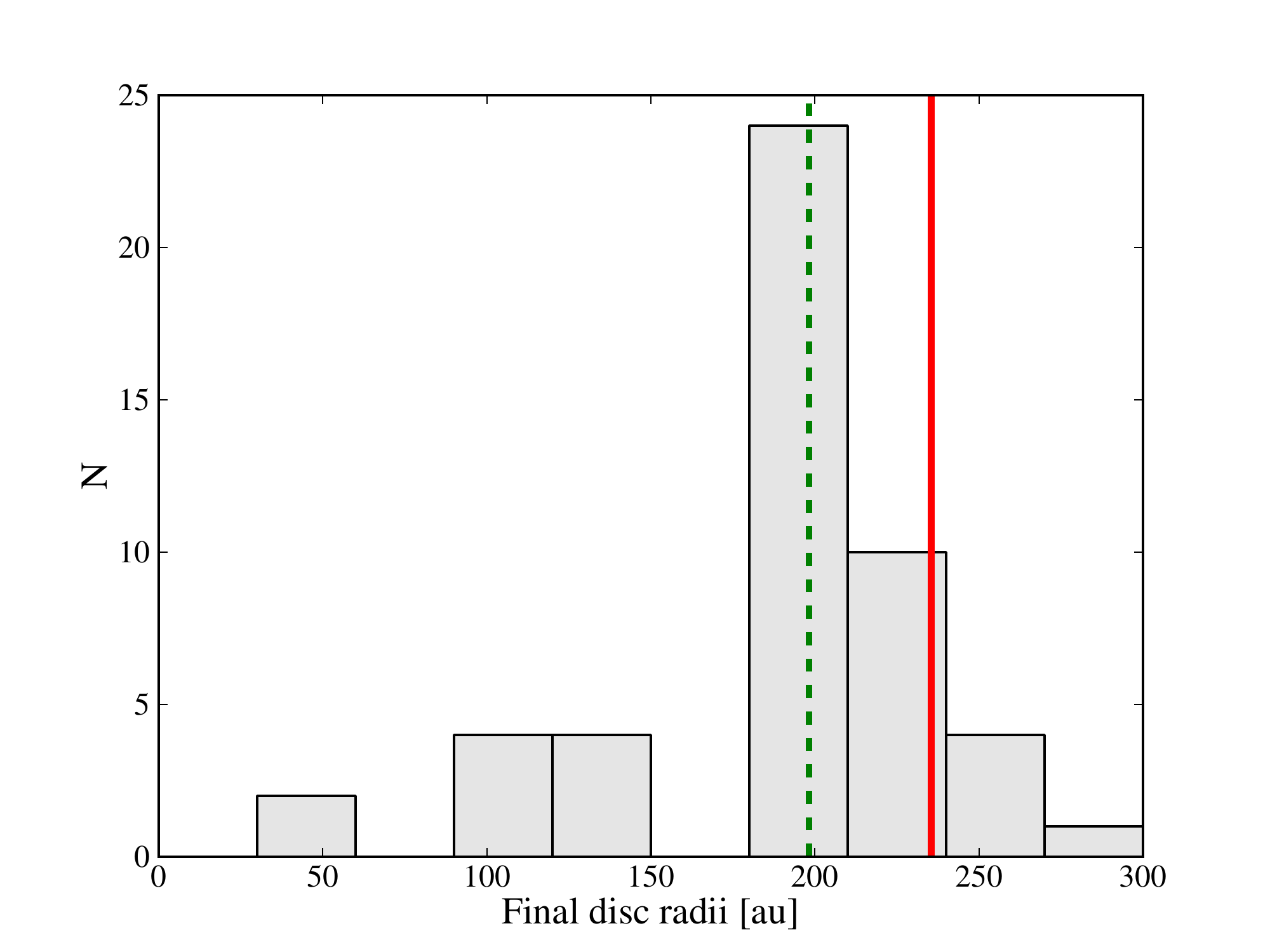}
\caption{Left panel: histogram of the final disc masses. Right panel: histogram of the final disc radii. The red solid vertical line in both plots shows the value for the disc run in isolation, while the green dashed vertical line shows the median of the distribution. Although a close encounter can have a dramatic effect on the mass of the disc, few discs had such close encounters, so that the median of the mass distribution is not significantly affected. However, even distant encounters can change the disc radius, so that we see a change in the median disc radius when comparing with the disc run in isolation.}
\label{fig_hists}
\end{figure*}

We show in Figure \ref{fig_hists} histograms for the final disc masses and radii. The red solid vertical line in both plots shows the value for the disc run in isolation, while the green dashed vertical line shows the median of the distribution. Few discs had close encounters that modified their masses significantly (Figure \ref{fig_hists}, left panel), as shown in Figure \ref{fig_mass_radius_distance_encounter}, so that the median of the distribution is only marginally smaller than the value for the disc run in isolation. Therefore, although a close encounter can have a strong effect on the specific disc, close encounters are not frequent enough to significantly alter the disc masses on average. This is highly dependent on the initial conditions for the cluster, and the absence of high-mass stars certainly plays an important role, since it removes the source of the most destructive encounters. This will be discussed in a subsequent paper.

The effect of encounters on the distribution of final disc radii (Figure \ref{fig_hists}, right panel) is more evident, as most of the discs experience a reduction in size due to the encounters. A number of discs are truncated at very small radii. We will show in section \ref{sec_discussion} that their final radii are compatible with having been truncated by the close encounters. There is also a number of discs which are not dramatically truncated, yet which are affected by more distant encounters. With respect to the disc in isolation, the final radius of these discs is reduced by $\sim 10-20 \ \%$.

\subsection{Simulations with larger initial radius}

\begin{figure}
\includegraphics[width=\columnwidth]{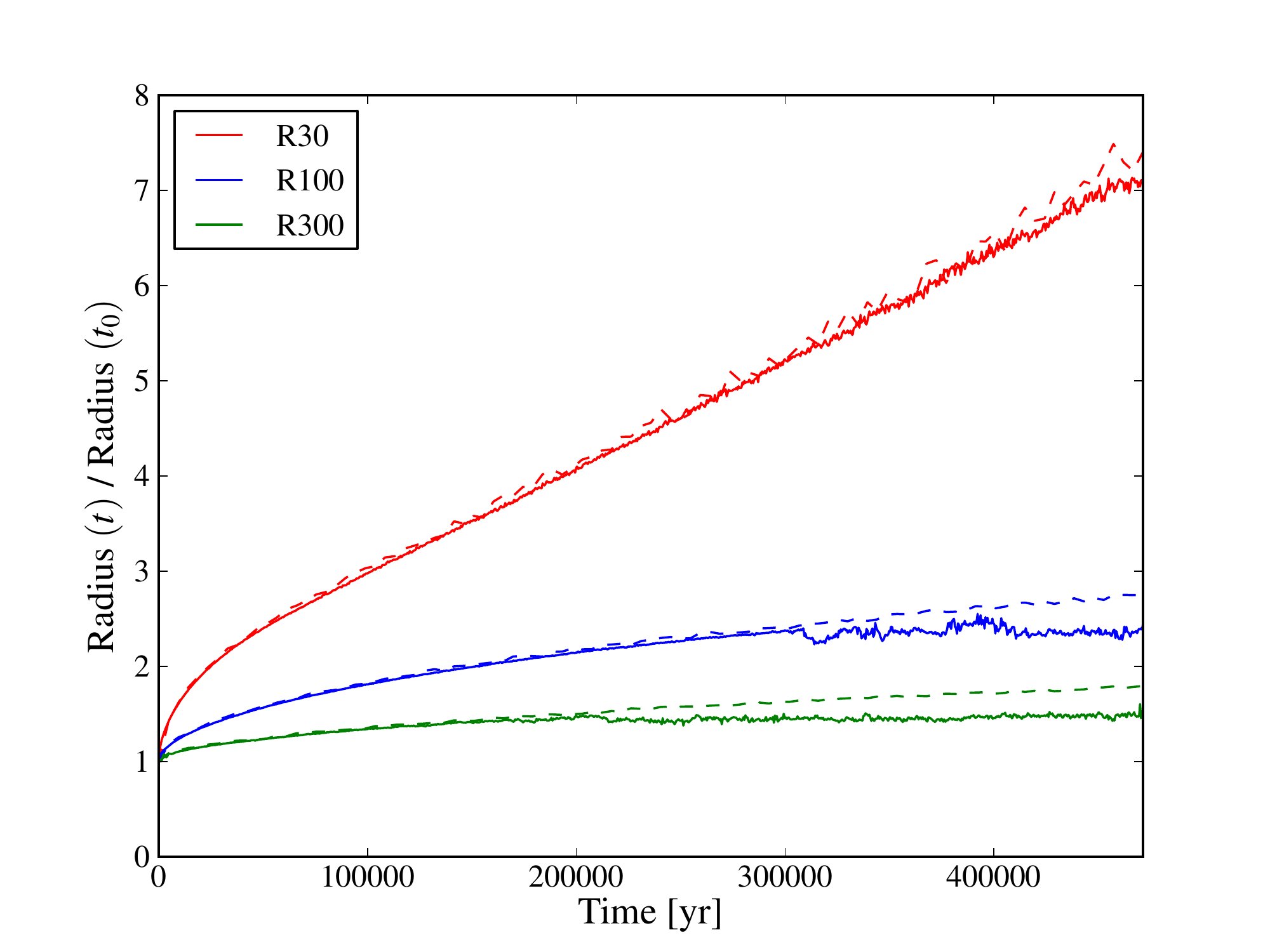}
\caption{Evolution of the median disc radius for runs R30, R100 and R300 (see legend), in units of the initial disc size. The dashed line is the disc in isolation, while the solid line is the median of all the discs in the simulation. The median disc radius in run R300 no longer increases after $\sim 2 \times 10^5 \ \mathrm{yr}$, showing that we have reached a regime where encounters are limiting the disc size.}
\label{fig_medians}
\end{figure}

\begin{figure*}
\subfloat[Run R30]{
\includegraphics[width=\columnwidth]{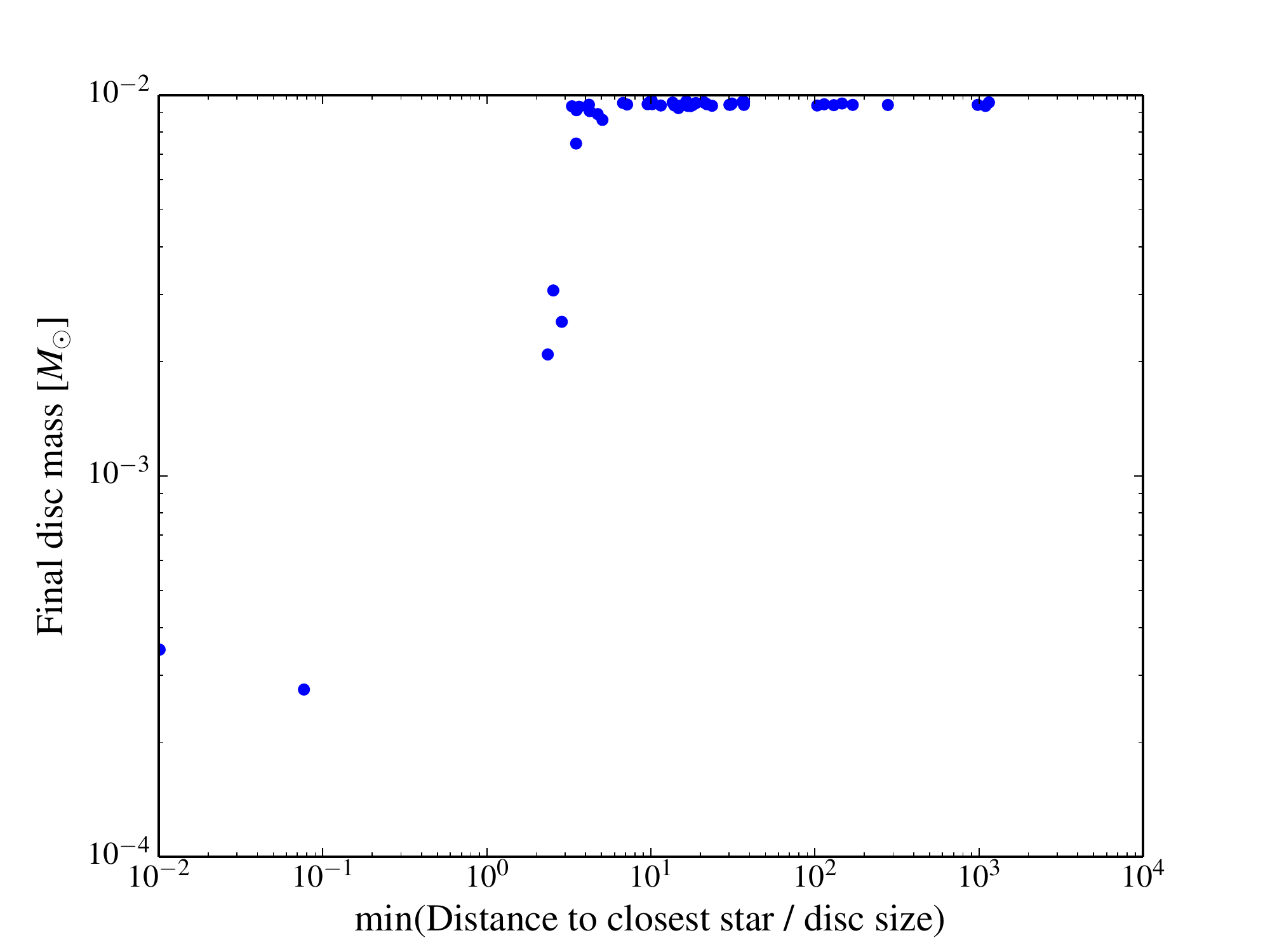}
\includegraphics[width=\columnwidth]{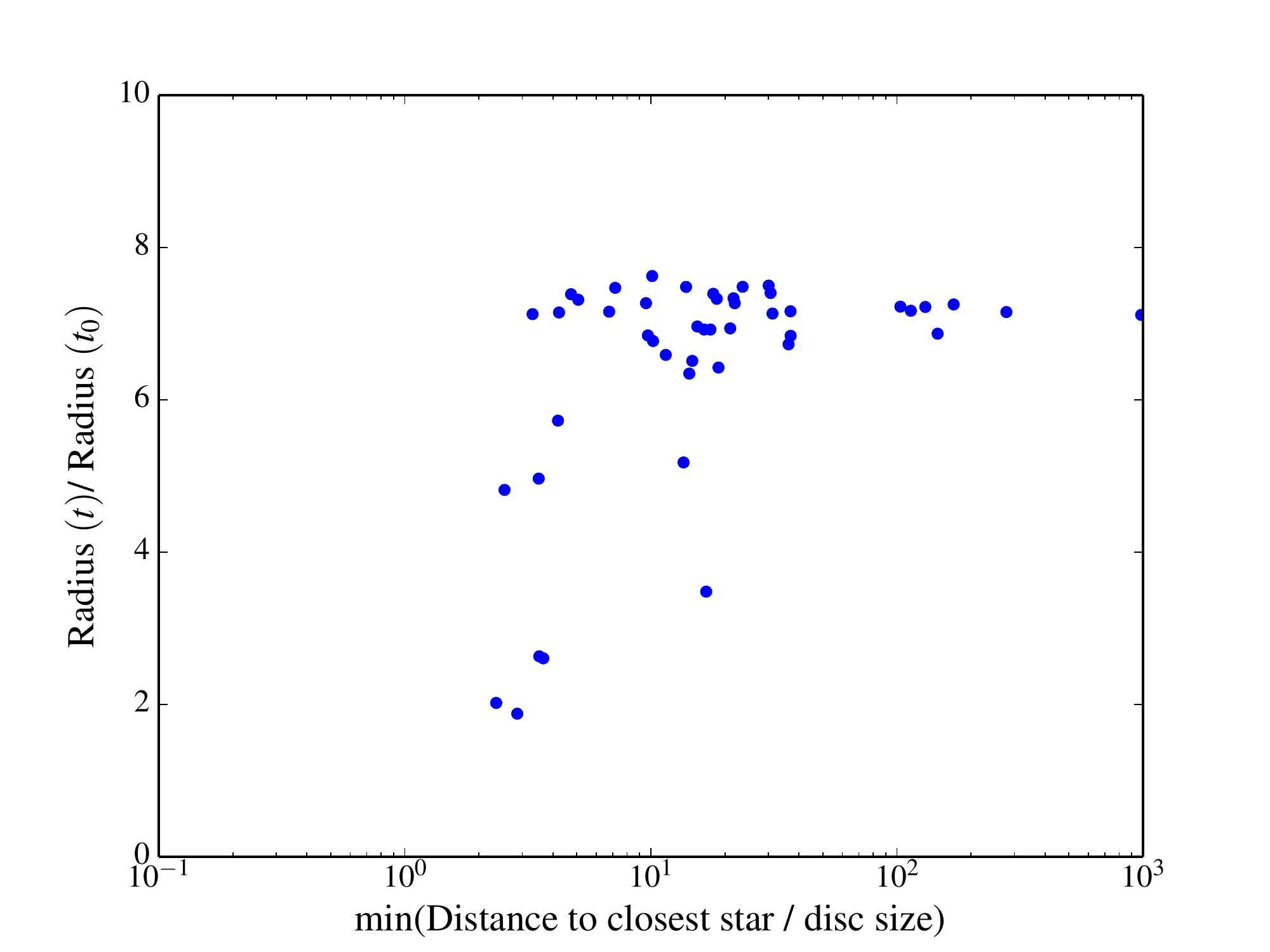}
}

\subfloat[Run R100]{
\includegraphics[width=\columnwidth]{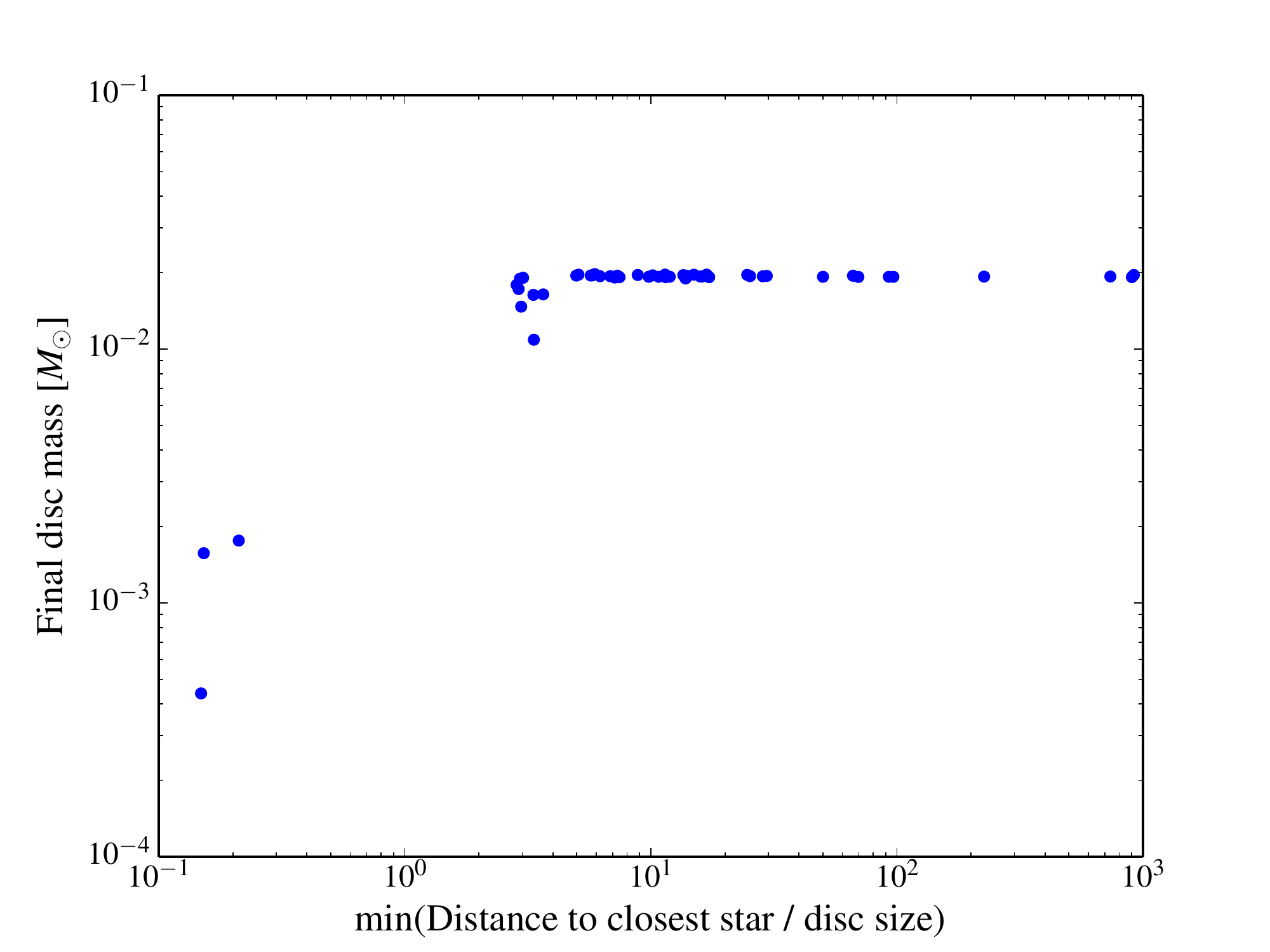}
\includegraphics[width=\columnwidth]{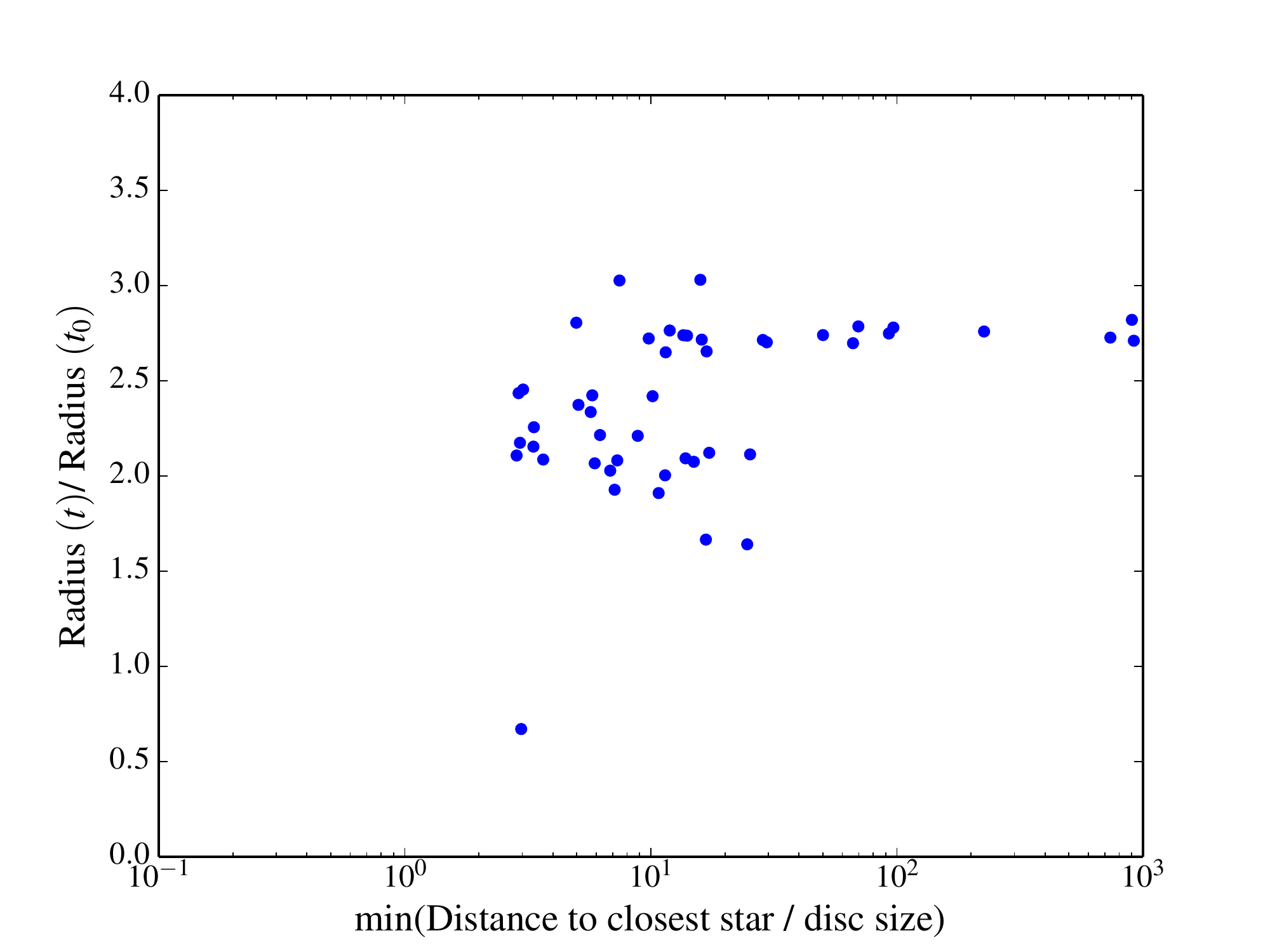}
}

\subfloat[Run R300]{
\includegraphics[width=\columnwidth]{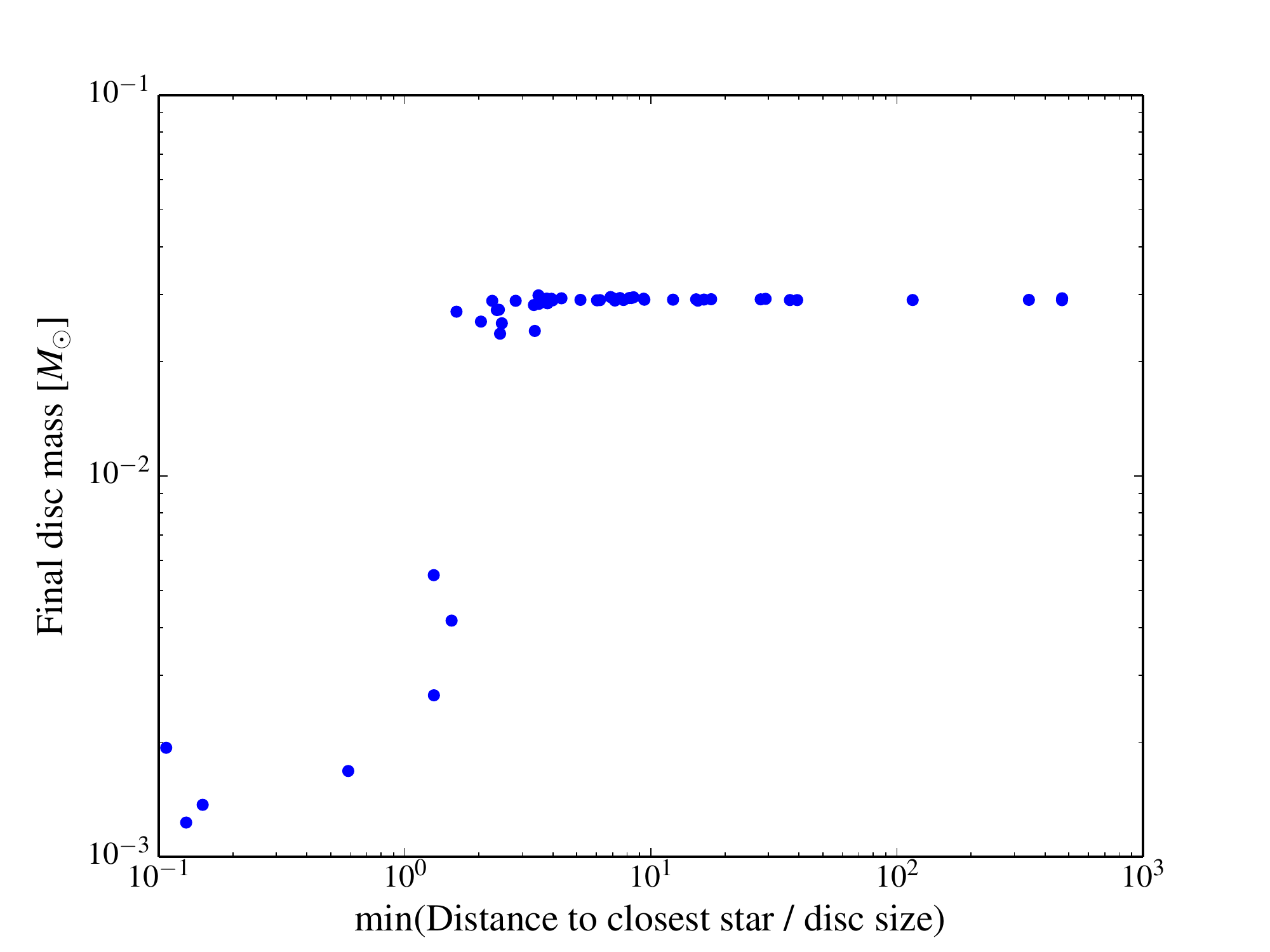}
\includegraphics[width=\columnwidth]{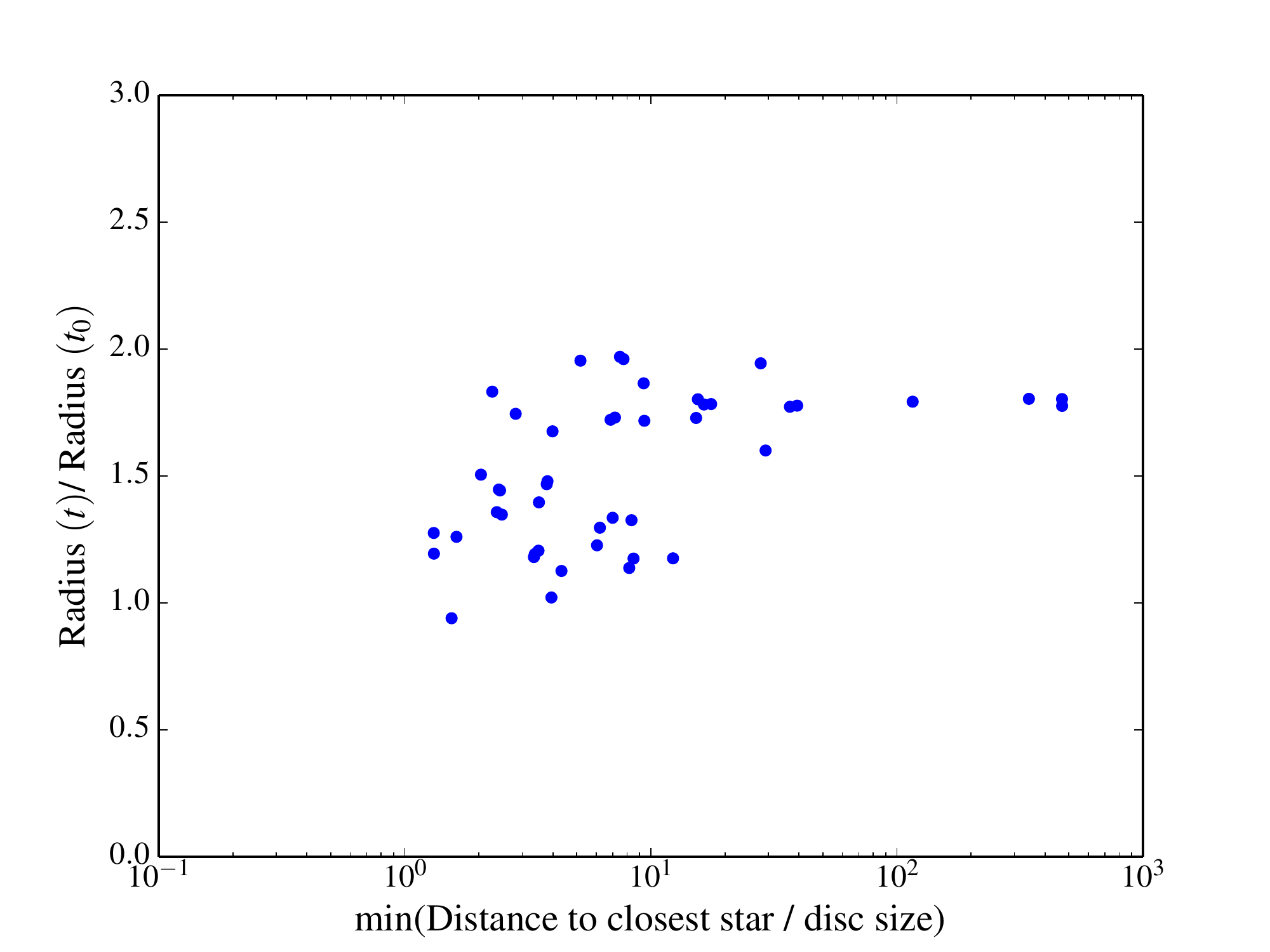}
}

\caption{Left panels: final disc masses as a function of the distance of the closest encounter, measured in units of the disc size at that moment. Right panels: final disc radii in units of the initial size as a function of the minimum value of the ratio of the encounter distance to the instantaneous disc size.}
\label{fig_dots_others}
\end{figure*}

\begin{figure*}

\subfloat[Run R30]{
\includegraphics[width=\columnwidth]{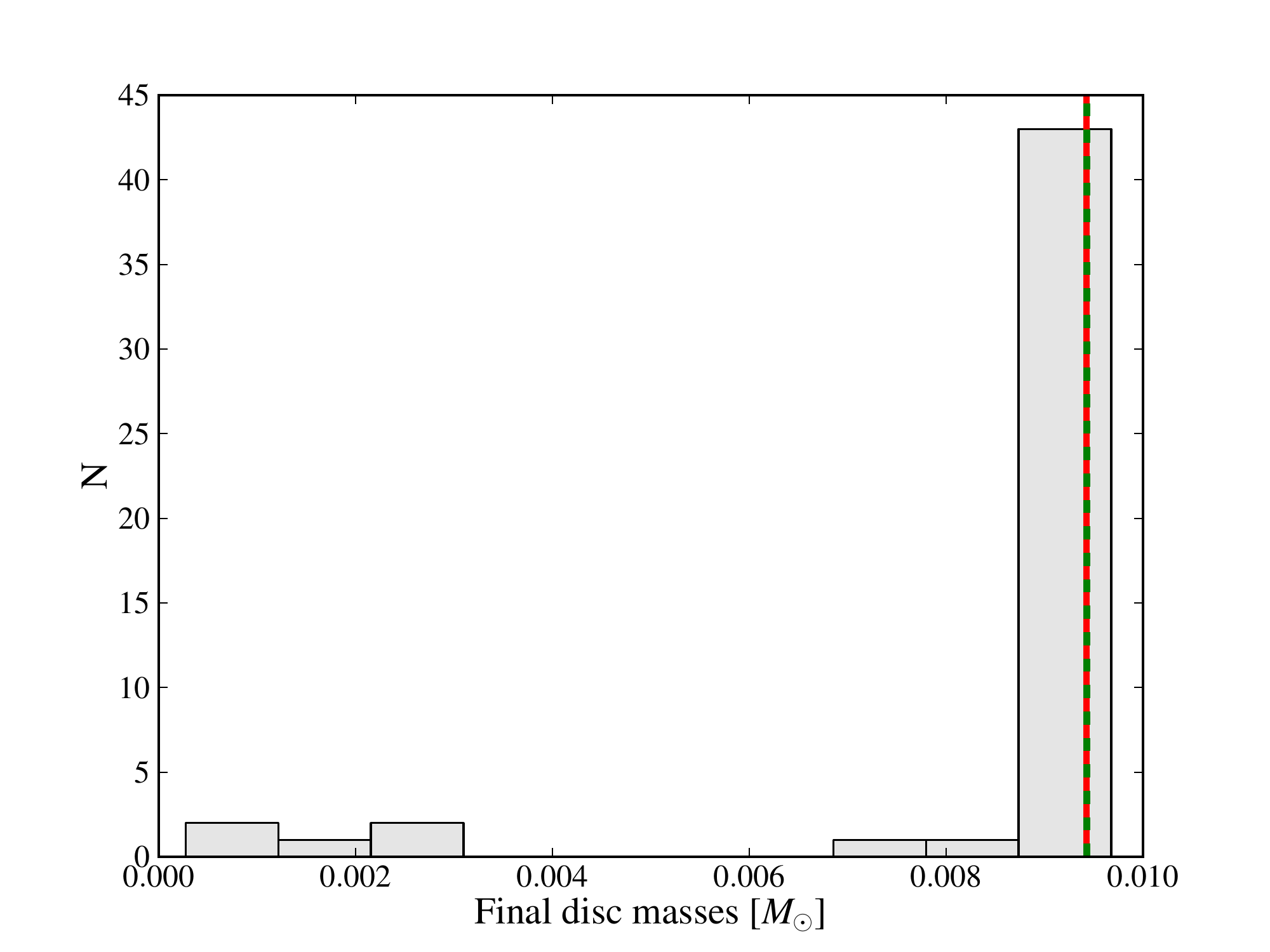}
\includegraphics[width=\columnwidth]{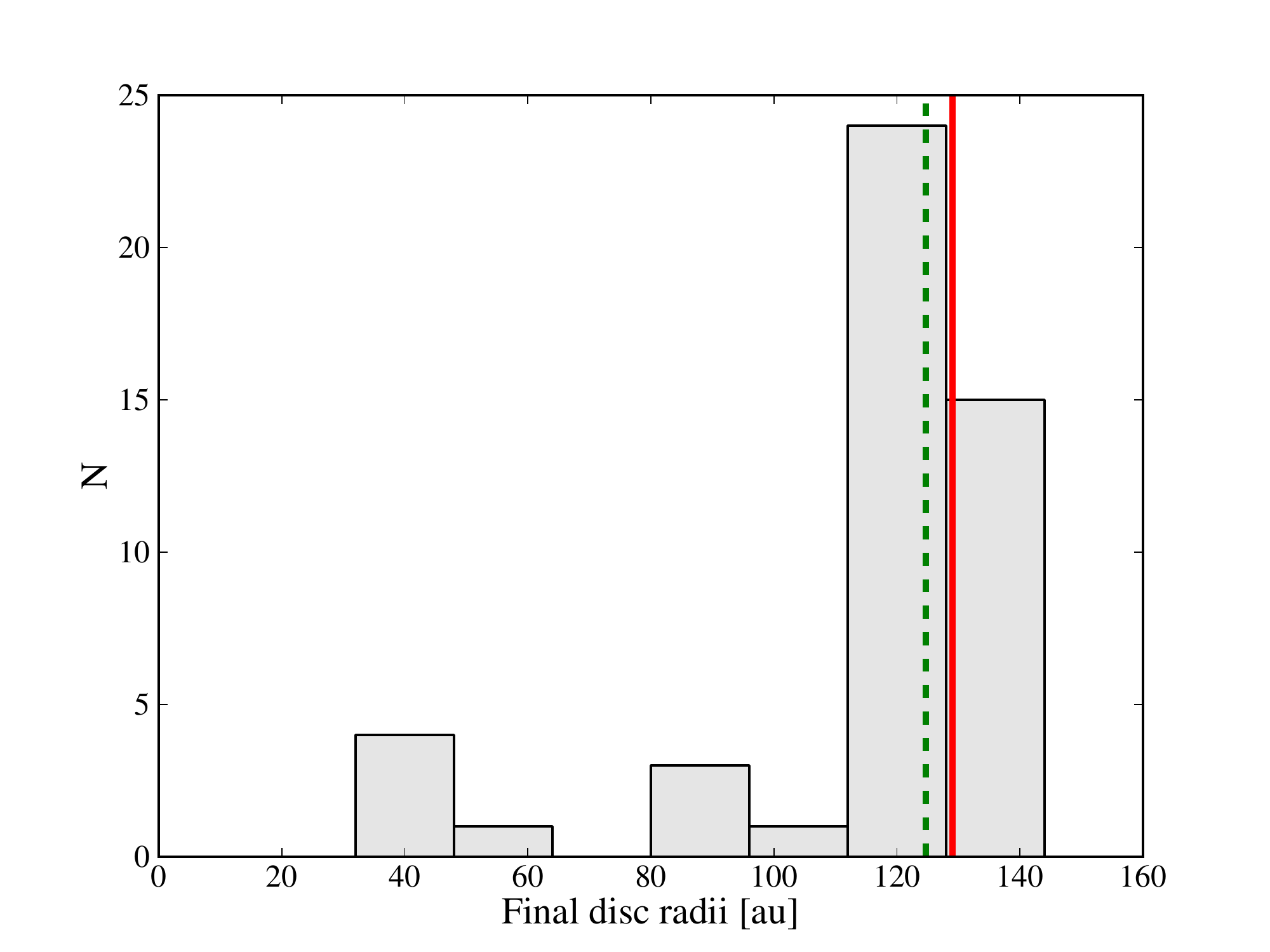}
}

\subfloat[Run R100]{
\includegraphics[width=\columnwidth]{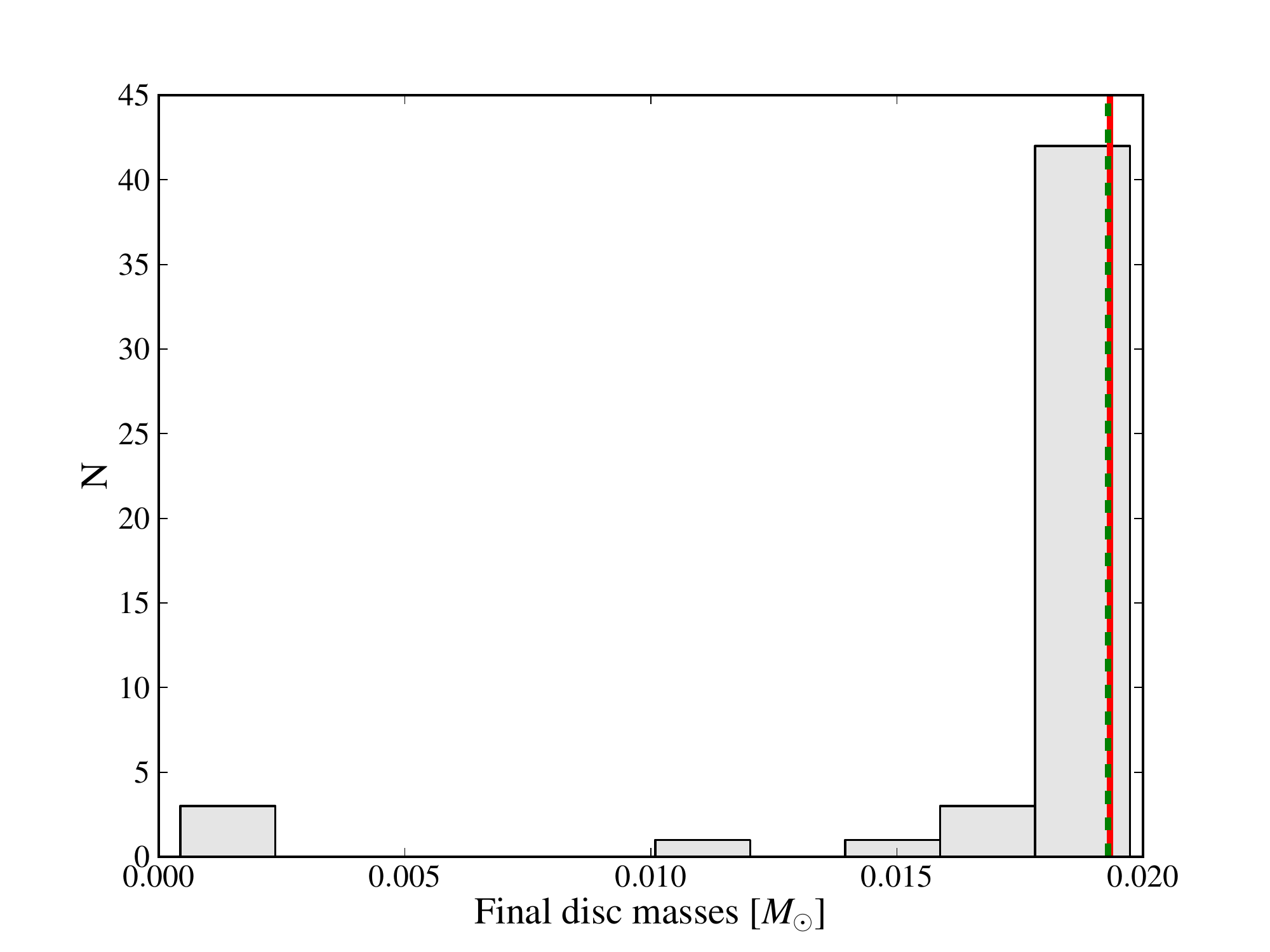}
\includegraphics[width=\columnwidth]{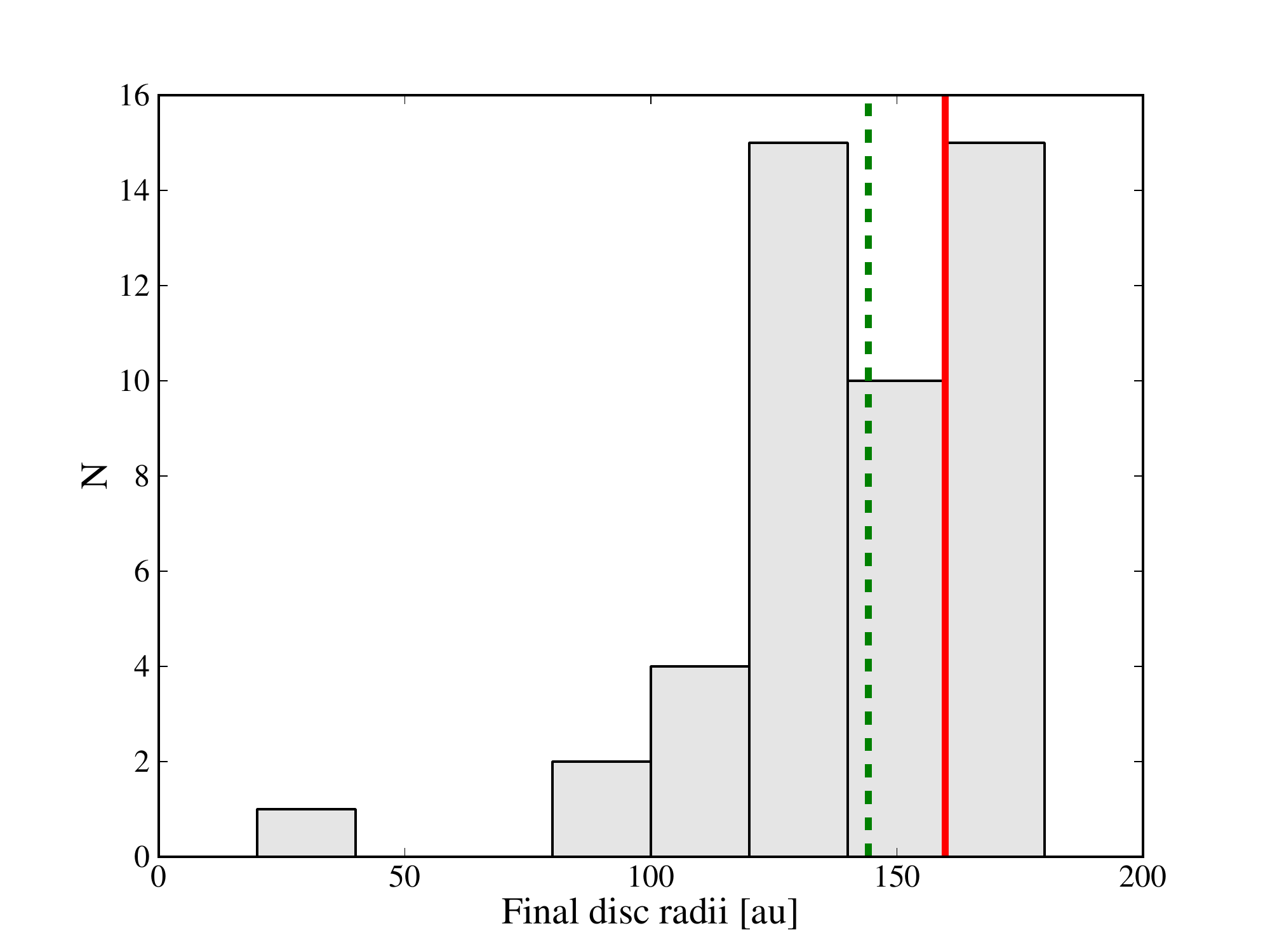}
}

\subfloat[Run R300]{
\includegraphics[width=\columnwidth]{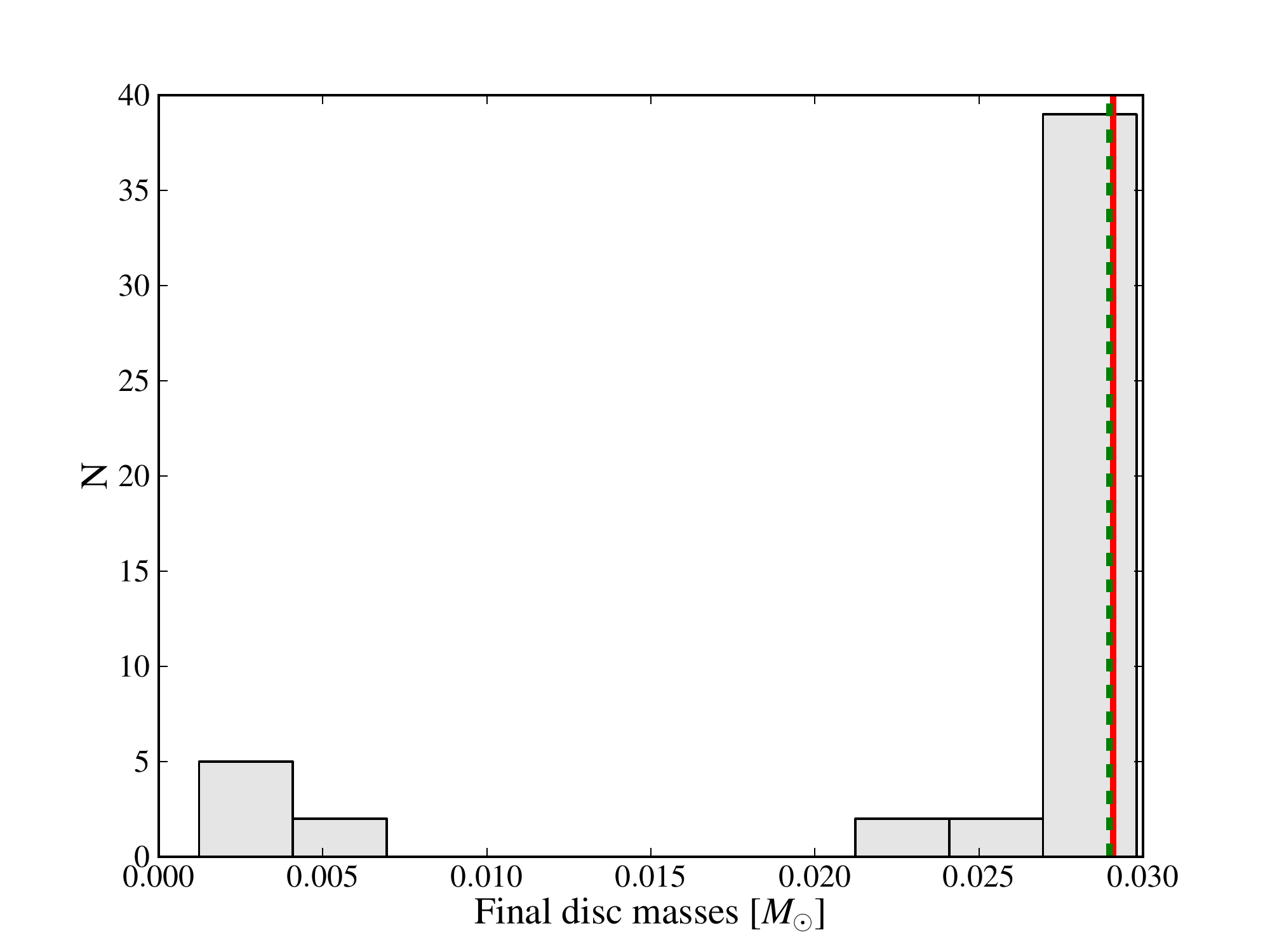}
\includegraphics[width=\columnwidth]{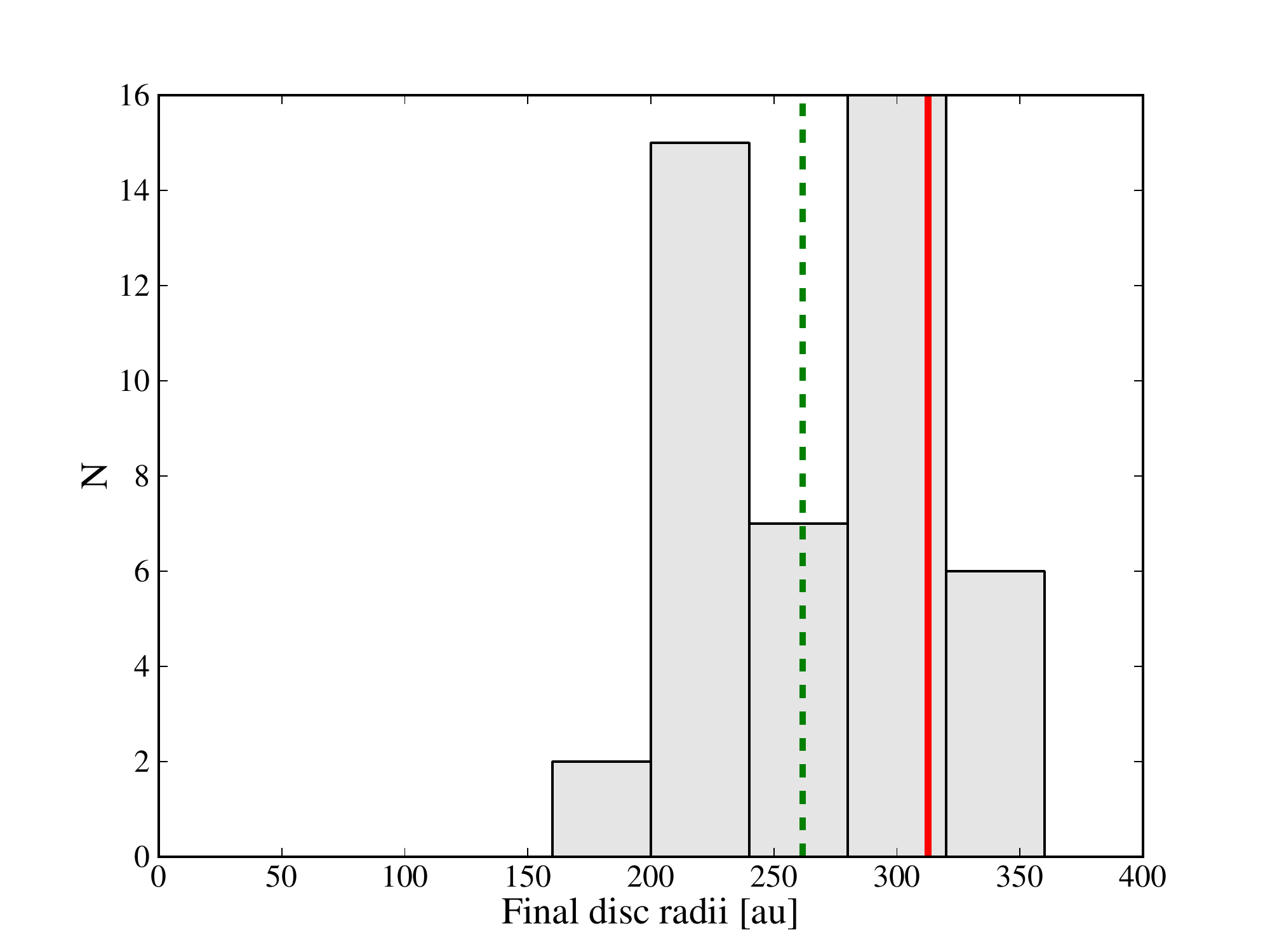}
}

\caption{Left panels: histogram of the final disc masses. The solid red line is the value for the disc in isolation, while the green dashed line is the median of the distribution. Right panels: histogram of the final disc radii. The solid red line and the green dashed line have the same meanings as in the radii plot.}
\label{fig_hists_others}
\end{figure*}

Figure \ref{fig_medians} shows the evolution of the radius for runs R30, R100 and R300. For simplicity, we show only the evolution of the median radius, compared with the disc in isolation. To compute the median, we excluded discs that experienced a significant mass loss (more than $99.5$ \%). As they are now only represented by a handful of SPH particles, the definition of radius ceases to be meaningful for them. Note that, for different disc sizes, the viscous time and the exponent in the expansion law change (see section \ref{sec_isolation}). It is clear that for run R300 we have entered a regime where the disc size is set by the encounters rather than from the size of the disc itself or from viscous spreading: the median no longer increases after $\sim 2 \times 10^5 \ \mathrm{yr}$. A similar behaviour, altough only towards the end of the simulation, can be seen for run R100. The spreading of these discs happens slower, while the encounter importance increases due to the smaller ratio between the encounter distance and the disc size. This behavior is not linear, however, as can be seen from the R30 and R100 runs: the radius of discs in run R30 is almost unaffected by encounters, while the effect on run R100 is visible, but smaller than in run R10. As discussed in section \ref{sec_isolation}, the discs in the different runs have different spreading rates. This brings the discs in run R10 to become almost as big as the discs in run R300 within the timescale of the simulation, while the discs in run R30 and R100 are eventually overtaken by the ones in R10. This effect comes from the different viscosities that we have in the different runs, which in the SPH method we cannot control. However, it is an effect that we can calibrate for, and which shows the interplay between disc truncation and spreading.

The mass histograms (left panel of Figure \ref{fig_hists_others}) shows that the runs with bigger discs had a handful of very destructive encounters, which produced discs with very small masses compared with the unaffected ones (note that in run R10, only a couple of discs had their mass reduced by an order of magnitude compared to 8 discs in R300). This is expected, since for bigger discs the ratio between the distance of the encounter and the disc size decreases and the encounters become more penetrating. At the same time, however, the dependence of the final disc mass on the distance of the closest encounter (left panels of Figure \ref{fig_dots_others}), measured in units of the disc size, changes. While for run R10 the dependence is quite shallow, it is much steeper, almost a step function, for the other three runs. Stated in another way, the bigger discs are also more resistant to encounters, as only very close encounters are able to affect them. Therefore, while the close encounters produced a lot of damage in the discs, overall the encounters were not able to modify significantly the mass of the discs, because only a few of them were able to probe the left part of this step function. This can be seen also comparing the difference between the median mass and the disc in isolation, which is very small for all the runs. In short, the masses of small discs are not significantly affected, because the encounters are not penetrating enough; big discs are also not significantly affected, because they are more resistant to encounters. 

One possible explanation for this different behaviour is that distant encounters cause a mass redistribution in the disc, ``hardening'' the surface density (see \citealt{1997MNRAS.287..148H}). While this is washed out easily in run R10 by the higher viscosity, through disc spreading, this effect is not strong enough in the other runs. This accounts for the steeper relation observed in the correlation between the final mass and the distance of the closest encounter, as a small difference in the encounter distance can make a big difference in the mass involved in the encounter if the surface density is steep. We discuss this idea further in section \ref{sec_understanding}.

\begin{figure*}
\begin{tabular}{cc}
\includegraphics[width=\columnwidth]{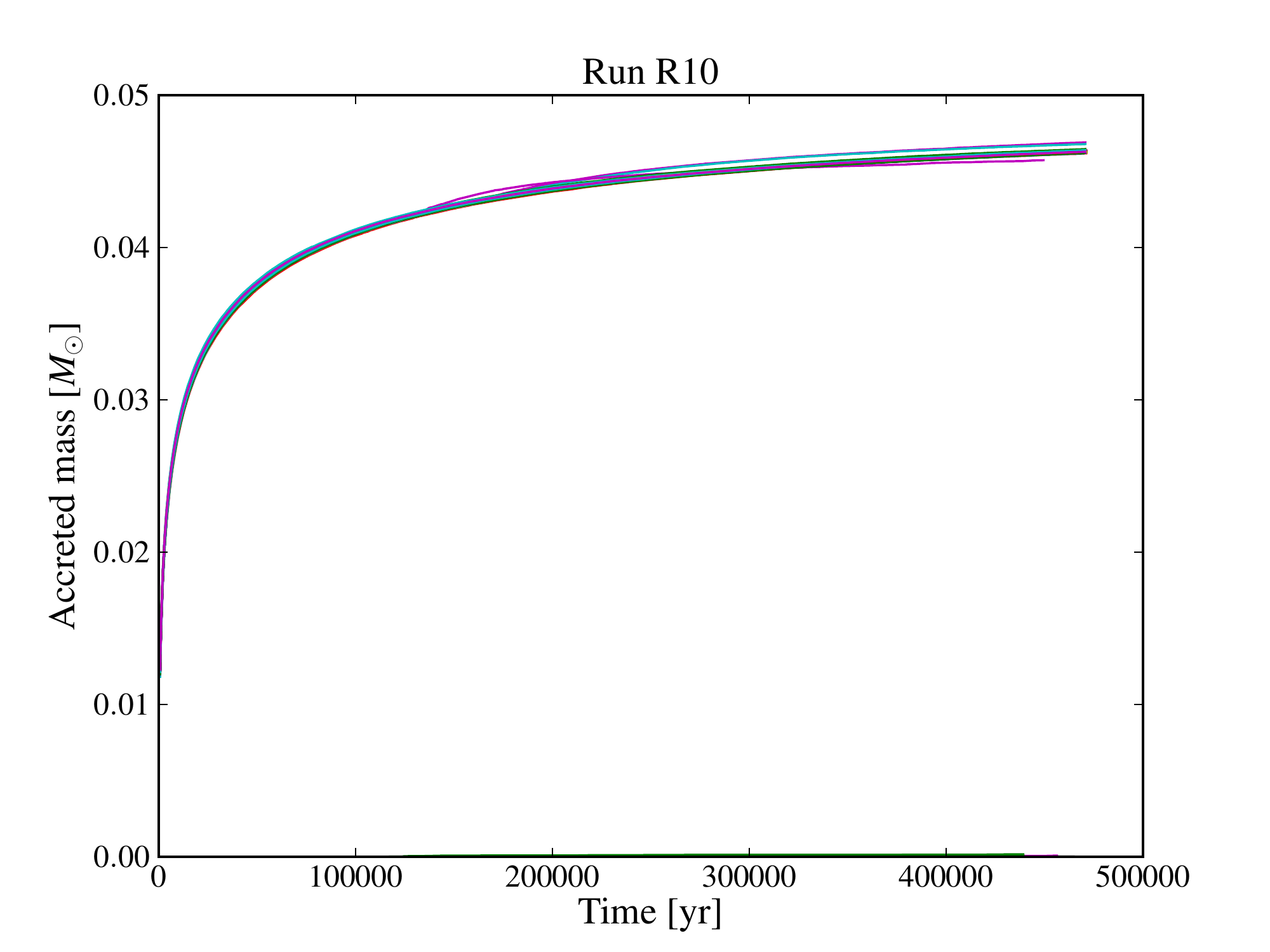}&
\includegraphics[width=\columnwidth]{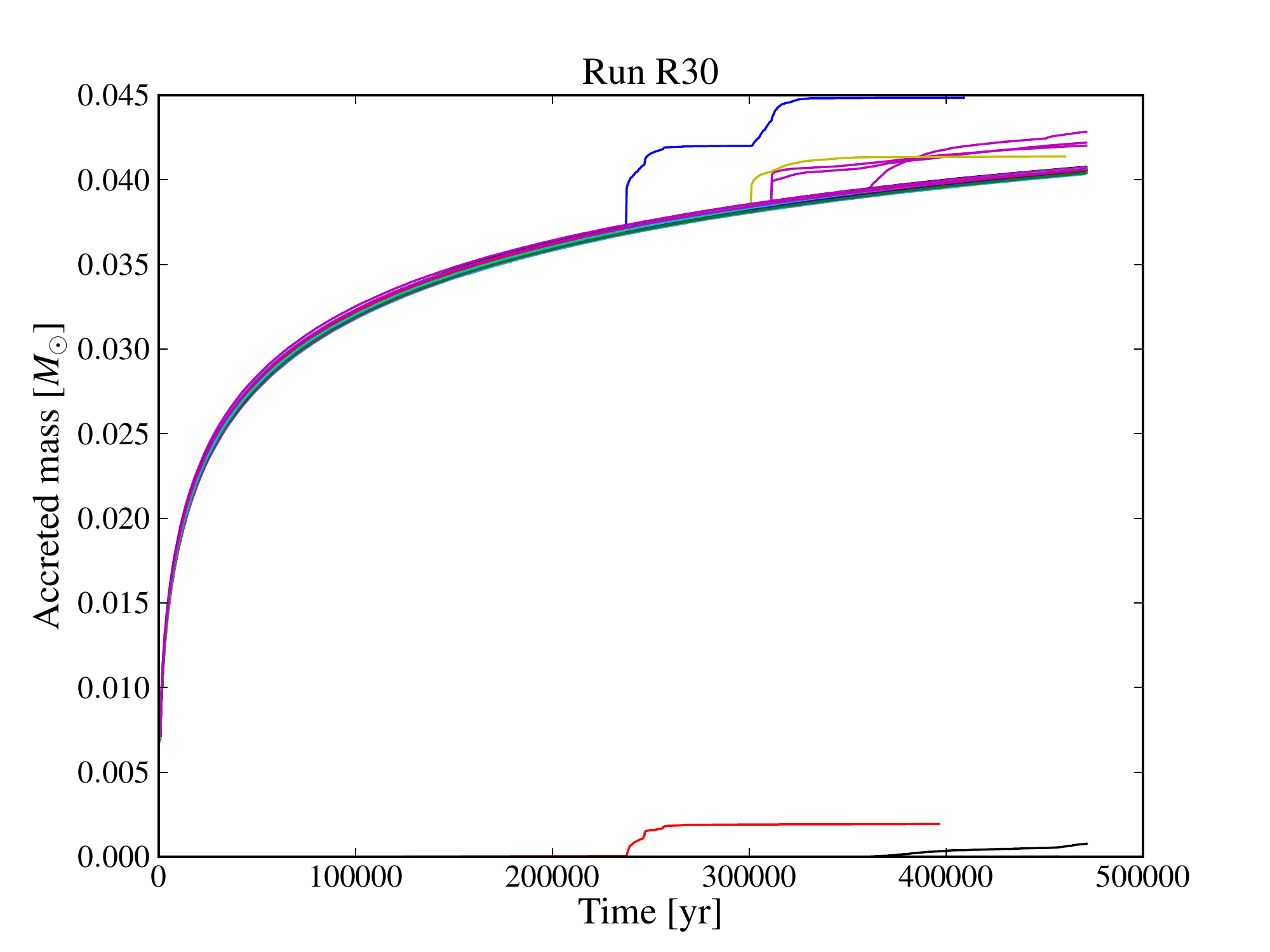}\\
\includegraphics[width=\columnwidth]{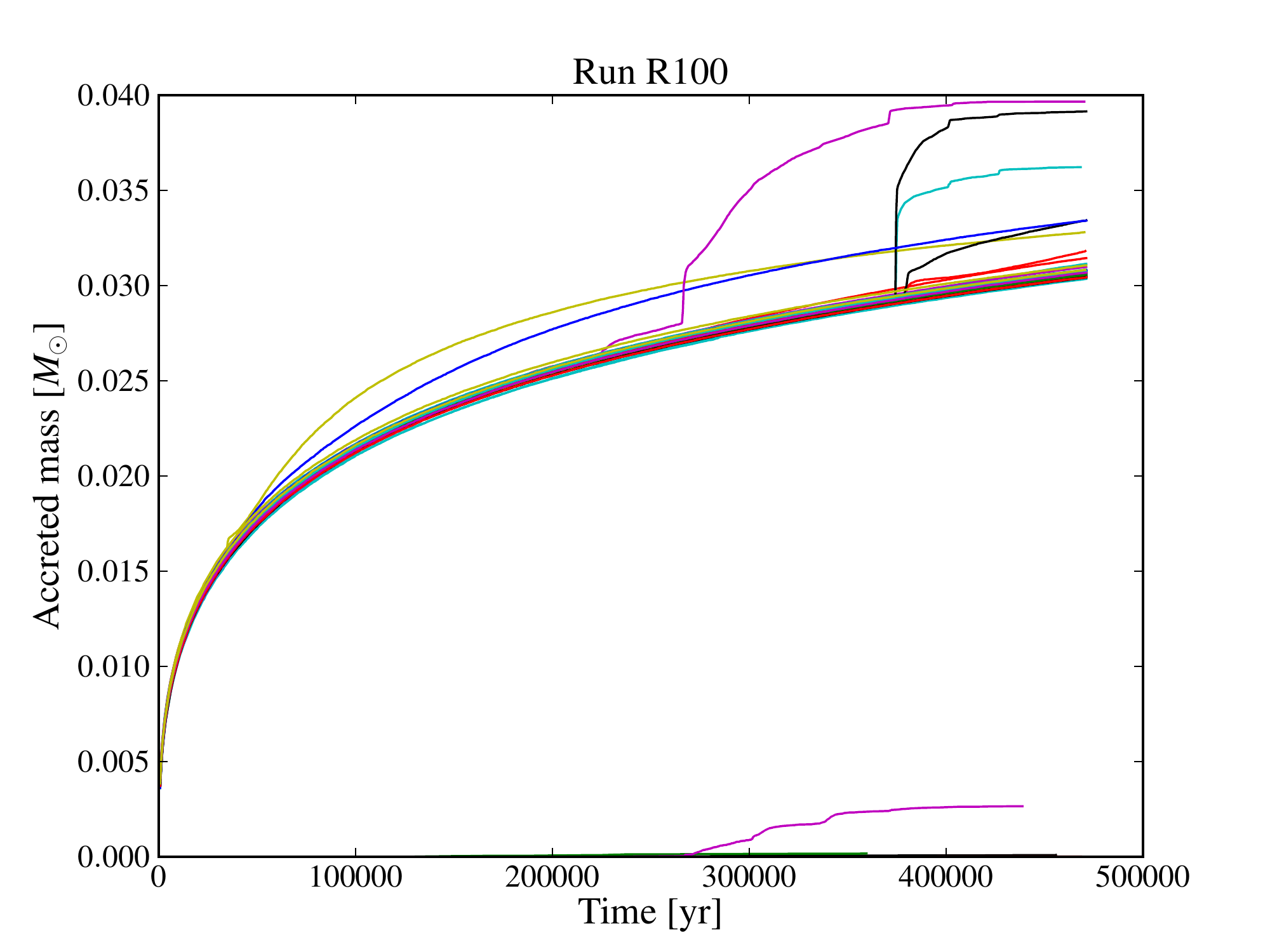}&
\includegraphics[width=\columnwidth]{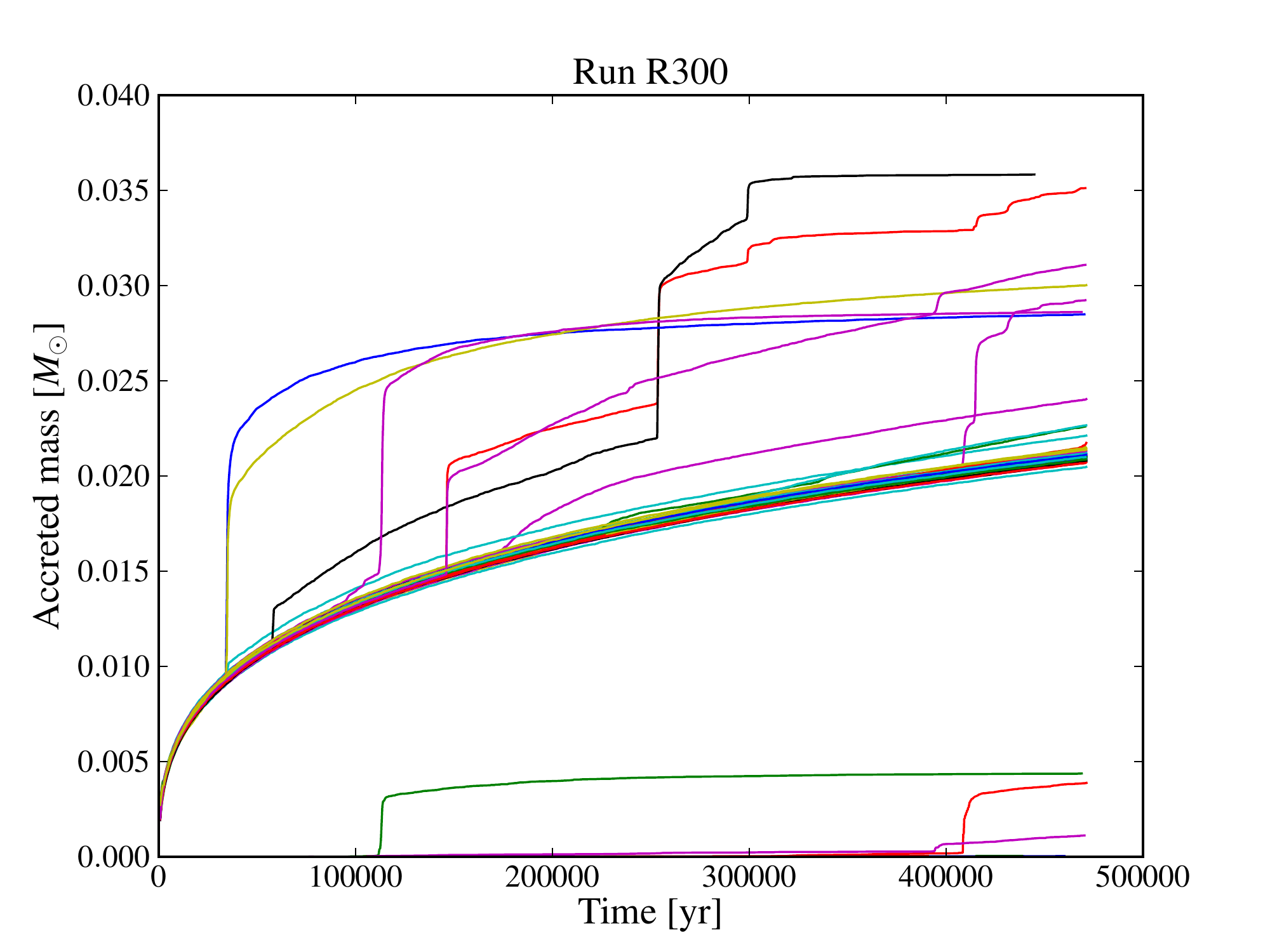}\\
\end{tabular}
\caption{Accreted mass on the central star as a function of time for the four simulations run. Each line has a different colour to help distinguish them.}
\label{fig_accretion_history}
\end{figure*}

Another quantity of interest is the mass accreted onto the star. We show in Figure \ref{fig_accretion_history} the accreted mass as a function of time for the discs in the simulation, with the four panels corresponding to the four simulations. While for the small discs there is little to no effect on the accreted mass, the encounters produce strong bursts in accretion for the bigger discs. Such bursts were already found in simulations by \citet{2008A&A...487L..45P}, and they have been proposed as an explanation for FU Orionis objects. However, we caution that numerical effects may also partially contribute to this result, since bigger discs also have bigger accretion radii. Therefore, while the result is interesting, further work is needed to assess its physical relevance. Interestingly, we note also that some stars that did not possess a disc at the beginning of the simulation may accrete some mass, which they have stripped in an encounter from a disc around another star. This could open the exciting possibility of reactivating accretion on a star that is already in the class III phase (i.e., has already dissipated its disc). Unfortunately it is difficult to quantify precisely how long such a burst would last and which accretion rates it could reach, due to the limited resolution available in terms of mass. We leave also this study for future work.


\section{Discussion}
\label{sec_discussion}
\subsection{Understanding disc sizes}
\label{sec_understanding}

\begin{figure*}

\begin{tabular}{cc}
\includegraphics[width=\columnwidth]{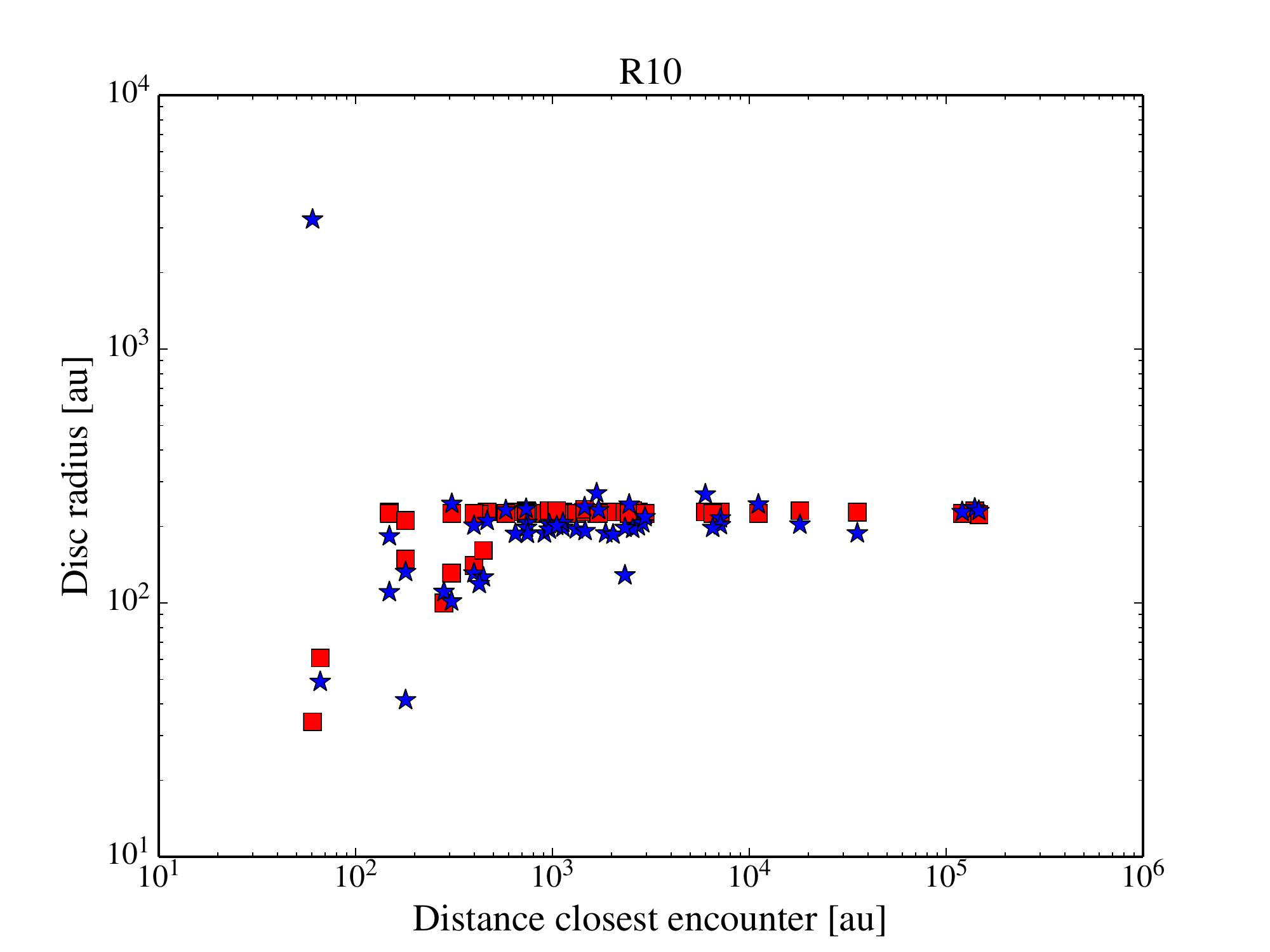}&
\includegraphics[width=\columnwidth]{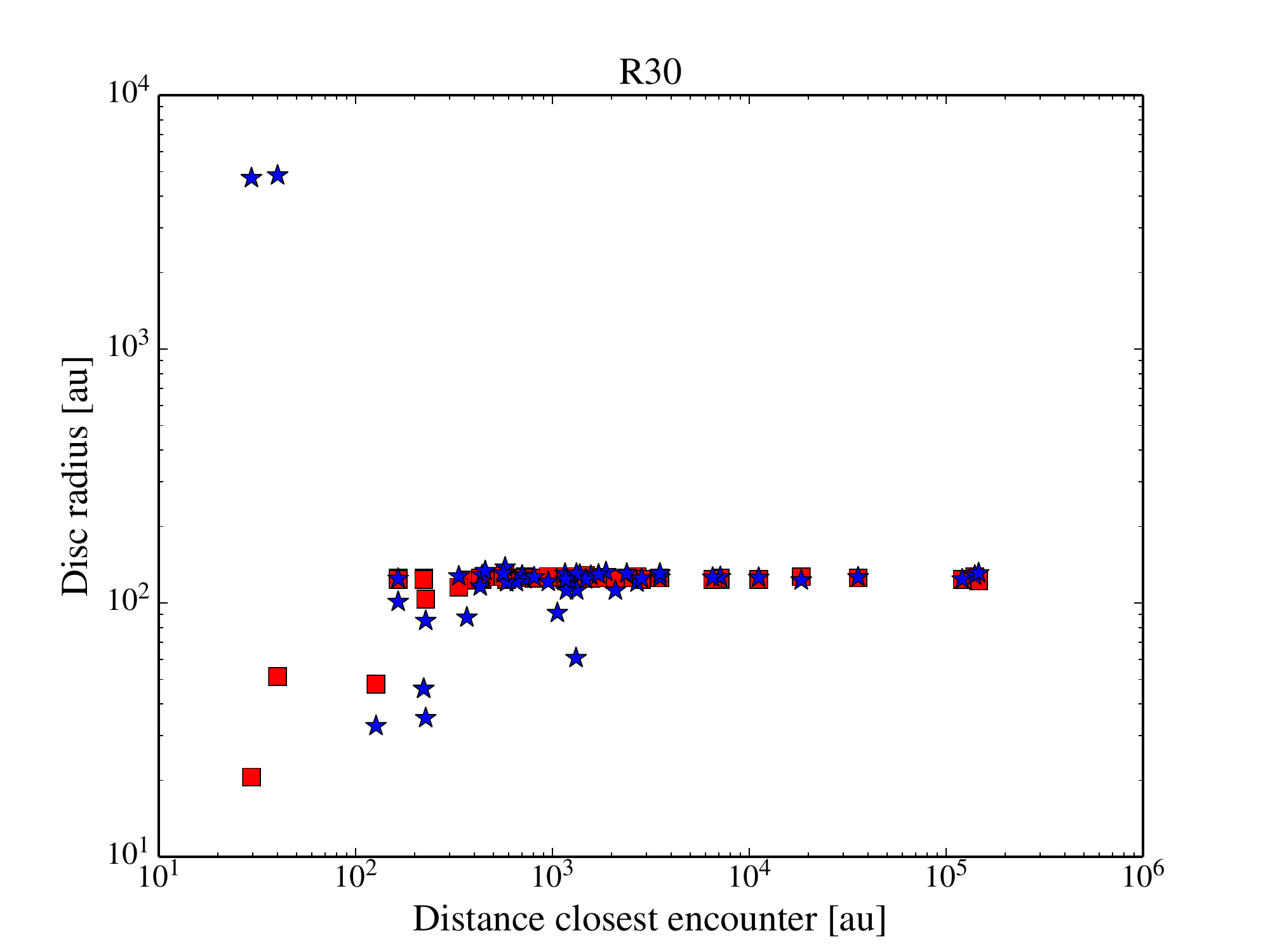}\\
\includegraphics[width=\columnwidth]{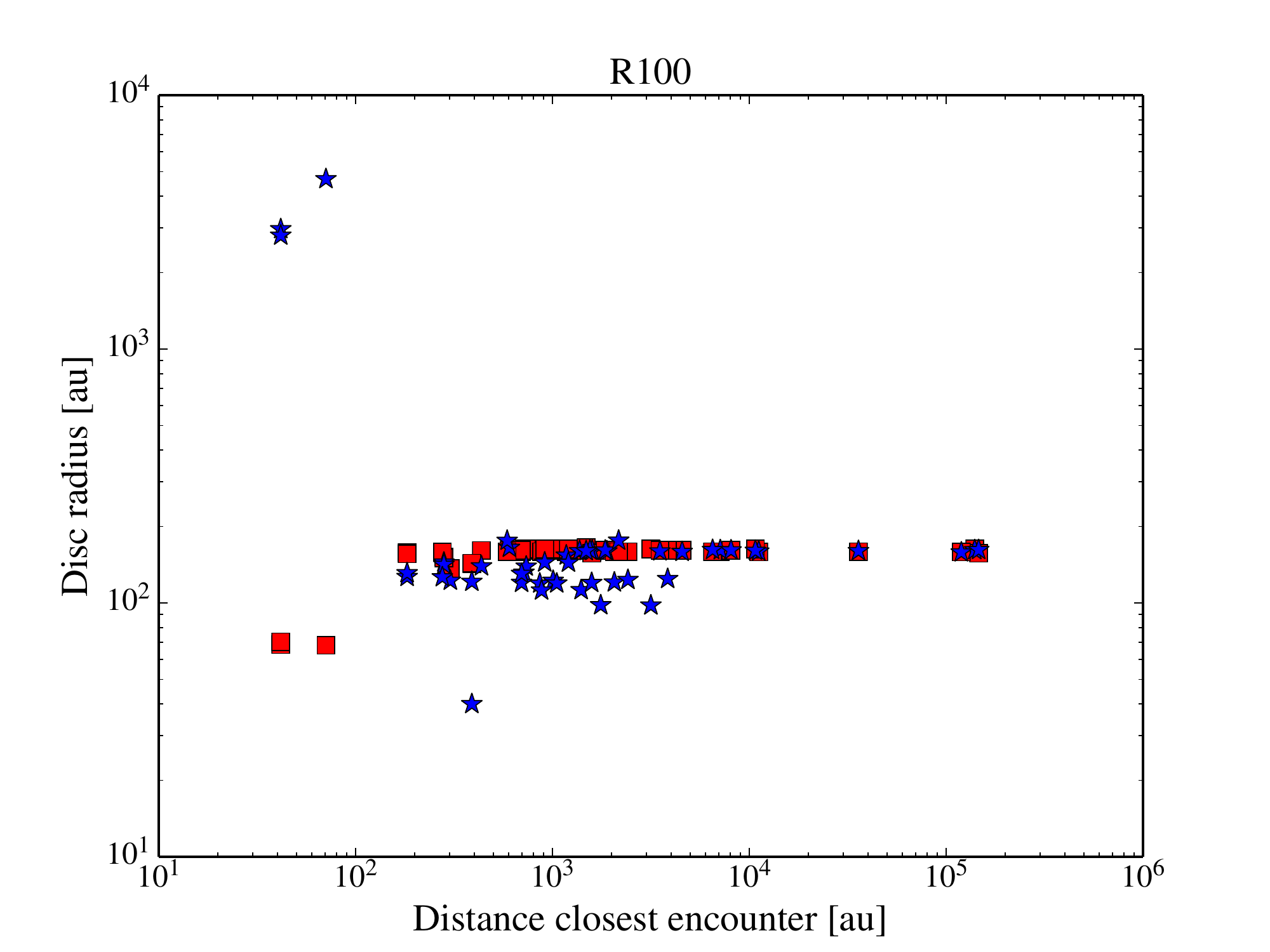}&
\includegraphics[width=\columnwidth]{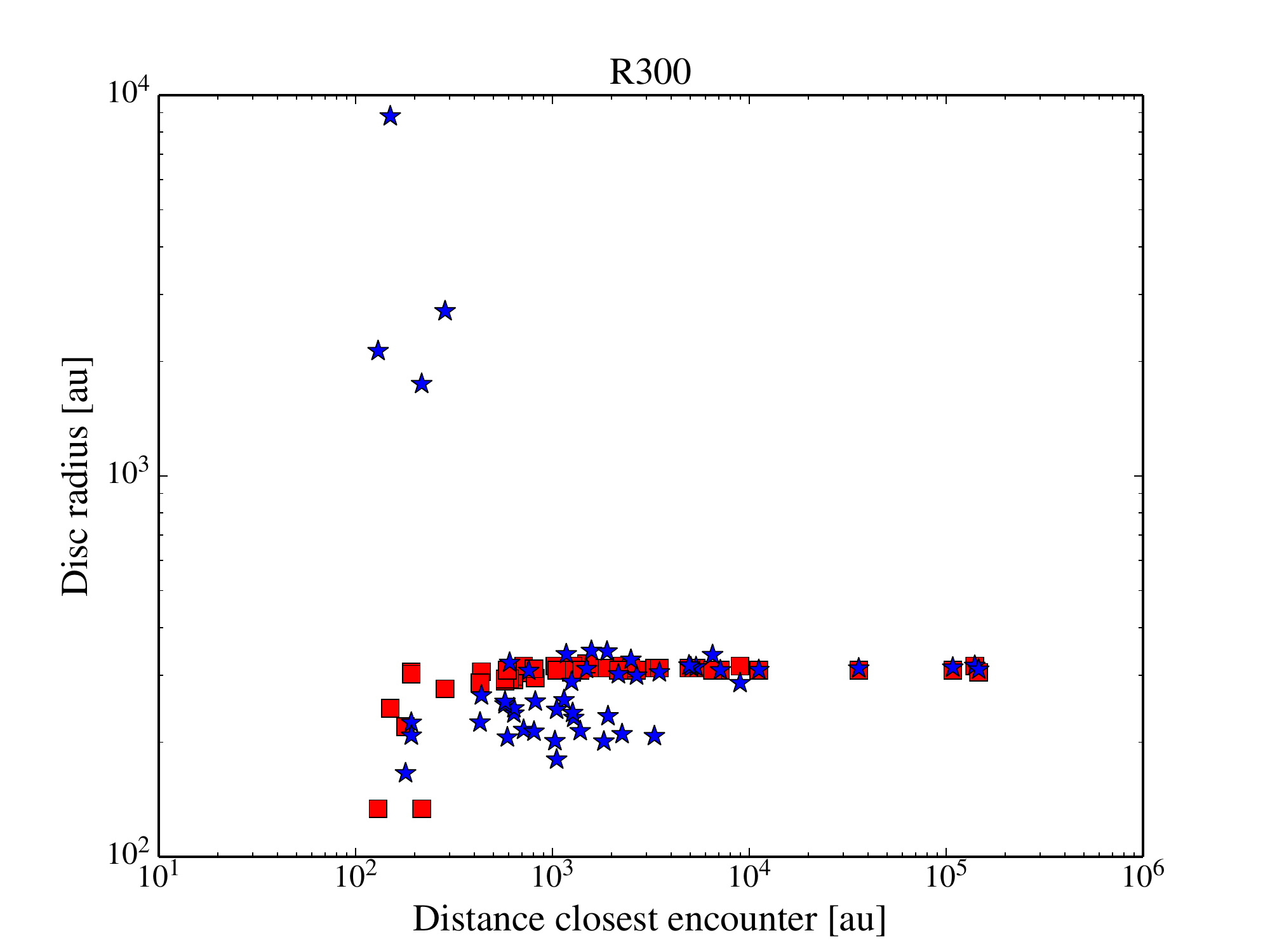}\\
\end{tabular}

\caption{Comparison between the predicted disc sizes by the model (red squares) and the results from the simulation (blue stars). The different simulations are shown in order, from left to right and from top to bottom.}
\label{fig_rdisc_rclose_theory}
\end{figure*}

We show in Figure \ref{fig_rdisc_rclose_theory} the results of the model presented in section \ref{sec_semianal}, compared with those from the simulation. For comparison, here we plot also the discs that experienced very close encounters. It can be noticed that sometimes the sizes derived from the simulation are much bigger than the disc in isolation; this is because the very close encounter destroyed the disc, and therefore the notion of disc radius is no longer meaningful.

We note the very good agreement for run R10, which is remarkable for such a simple model. While it does not correctly predict the sizes of all the discs, it is still quite effective for most of them, and it correctly reproduces the correlation between the two quantities. While the agreement is not as good for the other runs, the model still correctly identifies which discs have been severely affected by the encounters and which ones are not. In particular, by combining information from these plots and the ones shown previously, we can observe three different regimes:
\begin{enumerate}
\item Discs that had very penetrating encounters were significantly affected in their sizes. The assumption that discs are truncated at $d/3$ made in the model correctly captures which discs are in this regime, in line with previous findings \citep{1993MNRAS.261..190C}. Notice that in run R10 discs sizes are sometimes reproduced even in this regime, while the model clearly fails for the other runs. To interpret this result, we note that discs in run R10 are undergoing much faster spreading, and so they are in a ``spreading dominated regime''. Viscosity is for them the main driver of evolution, and encounters act simply to truncate the disc. In the other runs, instead, a more complicated interplay emerges;
\item Discs that had only distant encounters, but which nevertheless are smaller than the disc run in isolation. This population is present especially in runs R100 and R300. This is a feature that the simple model does not catch. We suggest that this is due to the cumulative effect of many distant encounters, which modify the mass distribution of the discs, ``hardening'' them \citep{1997MNRAS.287..148H} and violating the assumptions of our semi-analytical model. In run R30 the discs did not grow enough to be significantly influenced by this effect; in run R10, as previously mentioned the spreading is much faster, so that the steepening is been washed out by viscous evolution. Still, although this effect is smaller than in runs R100 and R300, also in run R10 there are discs that are a bit smaller than the prediction of the model, while still having had only distant encounters.
\item Discs that were largely unaffected by the encounters. The critical closest encounter to access this regime is of order $50$ disc radii, but notice that it is different in the different runs. In particular, the number seems to decrease with the initial disc size, which would point again in the direction of the big discs been ``hardened''. 
\end{enumerate}

\begin{table}
\begin{tabular}{ccc}
\toprule
Run & $p$ isolation & $p$ hybrid \\
\midrule
R10 & 0.56 & 0.55\\
R30 & 0.85 & 0.87\\
R100 & 2.34 & 2.47\\
R300 & 1.09 & 2.27\\
\bottomrule
\end{tabular}
\caption{We report for each run the power-law index $p$ (Equation \ref{eq_sigma}) when fitting the surface density distribution. We list the values for the disc in isolation and for the median of the indexes obtained when fitting the discs in the hybrid simulation. The effect of steepening is evident in run R300 that has a higher value of the median with respect to the disc in isolation.}
\label{table_p}
\end{table}

To check the hypothesis of modifications in the surface density distribution by the encounters, we fit the surface density at the end of the simulation with a power-law (Equation \ref{eq_sigma}). We report in table \ref{table_p} the results. We list the values for the disc in isolation and for the median of the discs in the hybrid simulation. In the latter case, we fit only the discs that in the course of the simulation went above the threshold value of $10^{-2} \ M_\odot$, to avoid artifacts. As expected, simulation R10 shows very little difference between the isolation run and the median value, as it is the case for run R30. The steepening is instead clear in simulation R300, where the value of the median is well above the value of the disc in isolation. A hint of steepening may already be seen for run R100, with some caveats however: the surface density of the discs in this run tends to be quite steep even in isolation, another feature that is due to the behaviour of SPH viscosity at these low resolutions. Therefore, while the analysis shows that some hardening is taking place in simulation R300, we caution that the low resolution does not allow us to measure quantitatively this effect. Future work will allow hardening to be studied more in detail. We note also that, if this effect is confirmed to be as big as measured here, it could be observationally probed by resolved observations of proto-planetary discs \citep{WilliamsCieza2011}, which which would allow one to verify if discs in dense environments have steeper surface density profiles than ones in sparse environments.  

\subsection{Comparison with observations}

\begin{figure}
\includegraphics[width=\columnwidth]{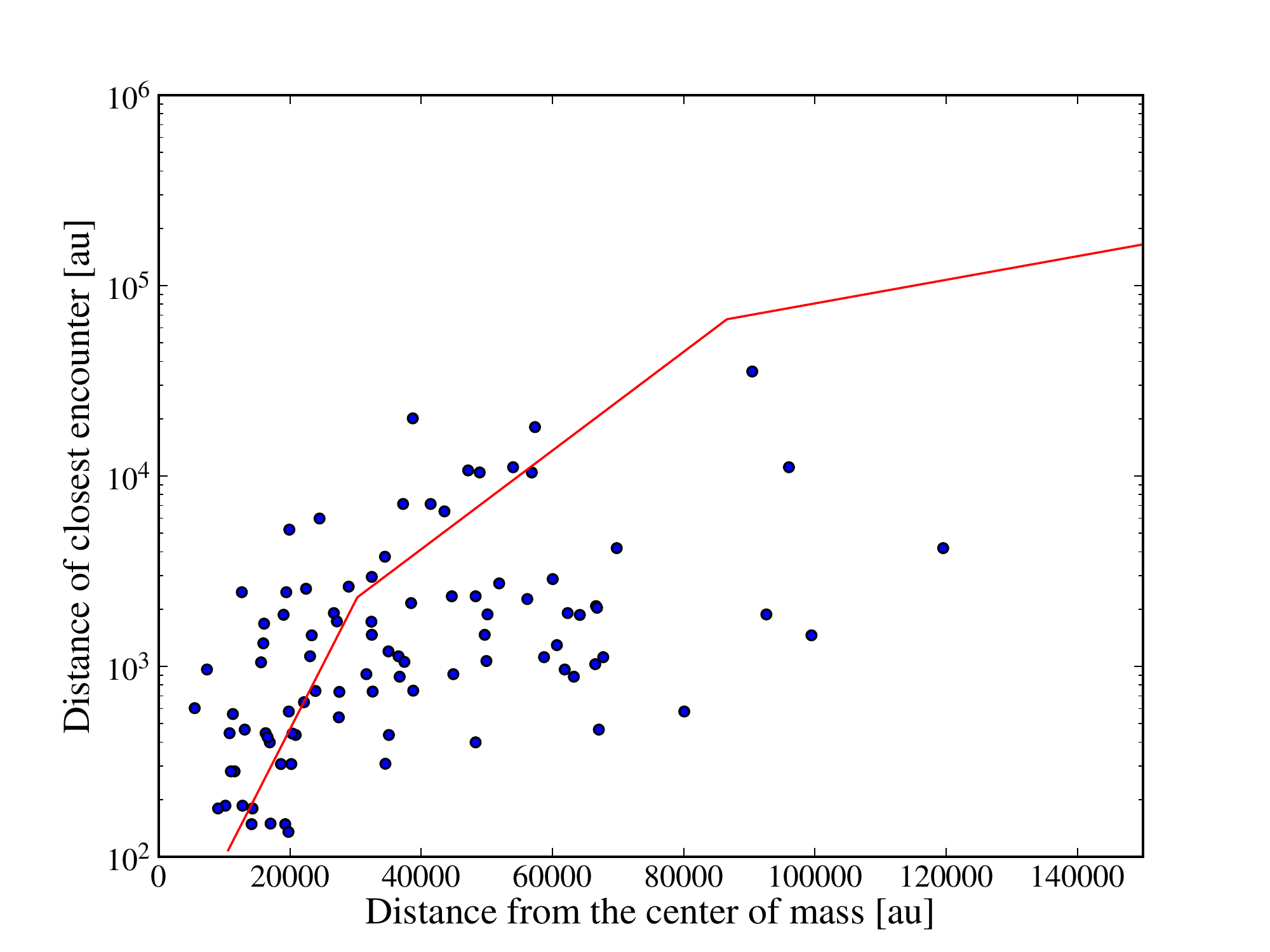}
\caption{Distance of the closest encounter versus the distance from the center of mass for each star. The dots are the results of the simulation, while the solid line is the result of the Monte-Carlo experiment described in the text, which yields a prediction for the median of the distance of the closest encounter at each location.}
\label{fig_r_close_theory}
\end{figure}
%

\begin{figure*}
\includegraphics[width=\textwidth]{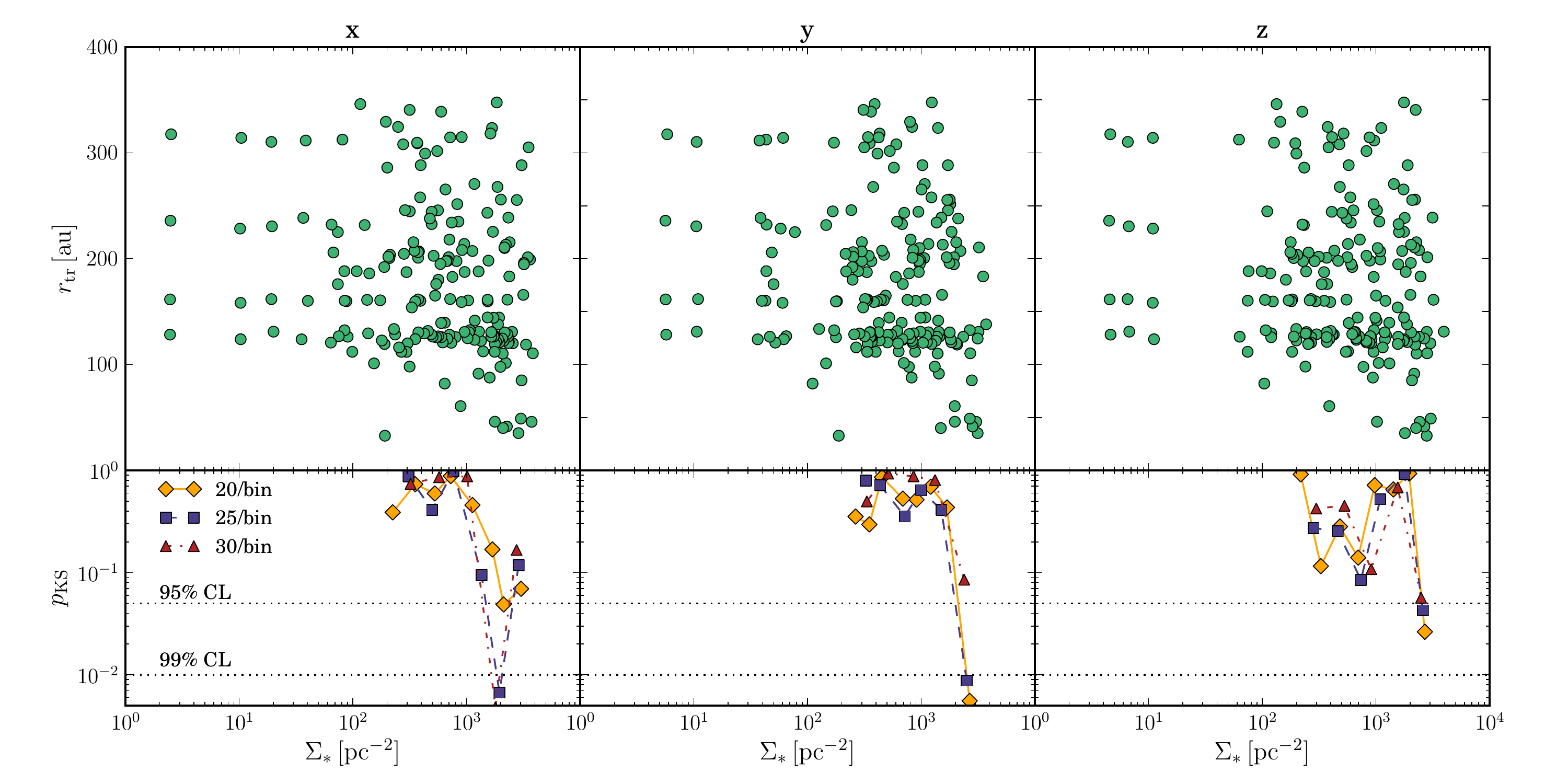}
\caption{Projections along the three coordinate axes of the radii vs. environmental surface stellar density ($\Sigma$) distribution of the population of discs resulting from the combination of all discs in each of our four runs. The lower panels show the results of a Kolmogorov-Smirnov test where we bin the data and compare the distribution in each bin with the one of the discs at a lower stellar densities. The different symbols are for different number of elements in each bin. Although the stars move in the cluster, they still retain information about their original position, so that there is statistical evidence of a cut in disc sizes at high stellar densities. In particular, the probability of the last bin to be compatible with the rest is always low. Notice however that this simulation did not explore the high stellar densities that are present in some of the real clusters (compare with Figure 1 of \citealp{2012A&A...546L...1D}).}
\label{fig_r_n}
\end{figure*}

\citet{2012A&A...546L...1D} pointed out that observational data suggests a reduced disc size in environments with high stellar surface densities ($\Sigma_\ast$). In particular, they looked at the measured disc size as a function of $\Sigma_\ast$. The population they consider is composed of $67$ Class II objects in nearby star forming regions (SFR) with radii measured through resolved imaging. The stellar density is computed with the \citet{1985ApJ...298...80C} method. Namely, one finds the $N$ closest stars, defines $d_N$ as the (projected) distance of the N-th nearest neighbour and computes the surface density as $(N-1)/\pi d_N^2$. The data shows a cut-off in disc sizes at stellar surface densities higher than $\Sigma_\ast>10^{3.5}\,\mathrm{pc}^2$. To highlight this cut-off, they perform a statistical test. They bin the data from higher to lower densities and for each bin they test with the Kolmogorov-Smirnov method if the distribution of the points in the bin is compatible with the distribution at lower surface densities.


We showed in the previous sections that the disc size is correlated with the closest encounter. In turn, the distance of the closest encounter depends on the environmental stellar density, but it is also a stochastic process; after experiencing a close encounter the star might move to a lower density region inside the cluster. Therefore, numerical simulations are of primary importance to assess what kind of correlation we expect theoretically in this parameter space, which can be probed by observations. It is important to have such predictions to distinguish from other candidates for the truncation of discs in clustered environments, such as external photoevaporation by massive stars.

First of all, we want to inquire how important is the movement of the stars in the cluster. The expectation to find smaller discs at high stellar densities relies on the fact that we expect the distance of encounters to depend on the local stellar density. However, this dependence could be washed out if a star experience a close encounter in a high density region and then moves to a region with lower density. It is then important to check how important is this effect. We follow \citet{2001MNRAS.325..449S} (see their Appendix A) to build a semi-analytical model of the N-body dynamics in the cluster. We make the assumption that the stars keep their distance from the cluster centre of mass fixed, and so they experience a constant stellar density throughout the simulation. We then compute with a Monte-Carlo experiment the distribution of the minimum encounter distance, and we compare it against the results of the simulation. We show in Figure \ref{fig_r_close_theory} the results. We plot the distance of the closest encounter (the blue dots are the results of the simulation) versus the distance from the centre of mass of the cluster. The red solid line is the median of the distribution drawn from the Monte-Carlo experiment, which agrees with the results of the simulation. Therefore it is indeed a good assumption to assume that the stars do not move systematically in the cluster over the simulation time-scale. This means that we expect the disc sizes to retain some information about the local stellar density. In addition, while we focused here mostly on the closest encounter, we note that a higher stellar density also enhances the number of encounters closer than a given distance, which also contributes to strengthen the correlation.


To check if dynamical encounters in our simulations can produce a feature like the one observed in the  \citet{2012A&A...546L...1D} study, we go through the same exercise they carry out. We chose $N=20$ to compute the stellar surface density. Since our data is three dimensional, we show the results of projecting along three different axes (we chose the coordinate axes for simplicity). However, our simulations are carried out for different values of initial disc radius separately. Our radii vs. ambient stellar density distribution is strongly influenced by this initial condition. To generate a more realistic population, and therefore more similar to that of the \citet{2012A&A...546L...1D} study, we combine the discs in all four simulations into a single population conserving the separated projections onto the $x$, $y$ and $z$ axes. We then perform the KS-test over this composite distribution for each axis separately. Figure\,\ref{fig_r_n} shows the radii vs. ambient surface stellar density distribution and the results of the test for each axis in the same format as in Figure\,1 of \citet{2012A&A...546L...1D}. To show the effect of the bin size, we compute the KS-test for $[20,25,30]$ elements per bin and our composite population has $192$ discs. To avoid edge effects we perform the test only over bins where the population at lower densities is larger than a sixth of the complete population.
 
The result is that, despite the fact that stars move in the cluster, there is still statistical evidence of a reduction in disc sizes at high stellar densities, namely above $\sim 2-3 \times 10^3 \ \mathrm{stars}/\mathrm{pc}^2$. In particular, the data in the last bins systematically show a low probability of being compatible with the rest of the distribution. Notice that these simulations did not explore the high stellar densities that are present in some of the real clusters (compare with Figure 1 of \citealp{2012A&A...546L...1D}), which can go up to the $10^4 \ \mathrm{stars}/\mathrm{pc}^2$ in the ONC, and we are thus just beginning to sample the cut-off. It is however very promising that the density at which this cut-off happens is consistent with the one found by \citet{2012A&A...546L...1D}. We leave future work to assess this regime, where the inclusion of massive stars is also important. Future work will also explore higher resolutions than what is currently possible, and measure with more accuracy the exact threshold at which the cut-off takes place.

We also note that in the literature there are indications of other influences of the environment on protoplanetary discs. For example, \citet{SiciliaAguilar11,SiciliaAguilar13} consider the Coronet cluster, which having $\sim 50$ stars inside $0.15 \ \mathrm{pc}$ is not very dissimilar from our simulations.  These authors find that discs in the Coronet cluster are more evolved than in the Tr 37 cluster. This is quite surprising, given that the Coronet cluster is 1-2 Myr old, while Tr 37 has an estimated age of 4 Myr. Their interpretation is that the difference is due to the much higher stellar density in the Coronet cluster. While our simulations do not yet allow for a detailed comparison with their results, this is an interesting path to be explored in future works.

Finally, in the present work we have ignored the effect of external photo-evaporation, with the goal to isolate the effect of the encounters. External photo-evaporation is also a process that limits the disc size. Using a time-scale argument, \citet{adams04} estimated that a proto-planetary disc around a solar mass star would have its size reduced by external photo-evaporation down to $30-60 \ \mathrm{au}$ (depending on the assumptions on the disc viscosity) in a time scale of several $\mathrm{Myr}$. \citet{clarke07} confirmed these results through the modelling of the viscous evolution of a protoplanetary disc undergoing external photo-evaporation, and found a significant shrinking of discs around solar-mass stars down to $\sim 100 \ \mathrm{au}$ after approximately $1 \ \mathrm{Myr}$.

These radii are much smaller than the final radii of the discs in our simulation. However, we remark that these authors simulate conditions ($G_0=3000$, where $G_0$ is the value of the far ultra-violet field in the inter-stellar medium) which are more relevant for a massive cluster such as the ONC than for the cluster we simulate here. Indeed, the ONC is a spectacular example of the potential impact of external photo-evaporation on proto-planetary discs. However, a cluster with only 100 stars is unlikely to have massive stars due to the limited sampling of the IMF, and therefore we do not expect any significant external photo-evaporation to happen in it.

More massive clusters have instead a higher probability of hosting massive stars, increasing the importance of external photo-evaporation. In addition, the importance of both external photoevaporation and encounters depend on the number density of stars, so that it is not trivial to understand which process would dominate. The picture is complicated even more by mass segregation that acts on different time scales, making massive stars sink to the central dense regions more rapidly in low mass clusters. Nevertheless, if it is confirmed that external photo-evaporation is more important than encounters in limiting disc sizes in massive clusters, then there must exist a threshold mass of the cluster where one switches from an encounter dominated regime to an external photo-evaporation dominated regime. Further work is needed to include the effects of external photo-evaporation in simulations like the one we conducted here and investigate these effects. While only a minority of all stars form in bound clusters \citep{lada03,kruijssen12d}, up to 50\% of all stars forming in bound clusters do so in clusters of $M<10^3~{\rm M}_\odot$. Hence, both high and low-mass clusters are worth exploring, which we plan to do in future work.

\section{Conclusions}

\label{sec_conclusions}

We have presented results from the first hybrid N-body - SPH simulations of coupled cluster and protoplanetary disc evolution. The discs in our simulation are expanding and accreting material onto the star due to viscous evolution, but they are also affected by close encounters between stars. Our simulations allow us to study whether a clustered environment, through the effect of encounters, modifies the protoplanetary disc evolution. We find that encounters can be very destructive for some of the discs, leading to almost complete dispersal for some of them. However, overall the median mass of the discs is not severely affected by the encounters.

We find that disc size is much more affected by encounters than disc mass. In the case in which disc spreading is fast, due to a high viscosity, only close encounters matter, as any mass redistribution in the disc caused by more distant encounters is quickly washed out. In this case, the close encounters simply truncate the disc at a given radius. If instead the spreading is not fast enough, we find a regime where distant encounters can have a significant impact on the discs, hardening their surface densities, and thus shrinking their radii. This also makes the discs more resistant to mass stripping by subsequent encounters. Therefore, we stress the importance of hydrodynamical numerical simulations of this kind to yield accurate predictions of the impact of stellar encounters on disc sizes.

Finally, we confirm that theoretically we expect to see a cut-off at stellar densities higher than $10^{3.5}\,\mathrm{pc}^{-2}$ in the disc sizes due to the effect of encounters. Further work is needed to probe the high stellar densities present in real stellar clusters.

\section*{Acknowledgements}
GR acknowledges the support of the International Max Planck Research School (IMPRS). This research was supported by the DFG cluster of excellence `Origin and Structure of the Universe' (JED, DH, BE). We would like to acknowledge the Nordita program on Photo-Evaporation in Astrophysical Systems (June 2013) where part of the work for this paper was carried out. We thank Cathie Clarke, Leonardo Testi, Carlo Manara, Antonella Natta, Henny Lamers and all the participants of the ESO Star Formation Coffee for stimulating discussions. We are grateful to the referee (Richard Alexander) for the constructive criticism. Our visualisations made use of the SPLASH software package \citep{2007PASA...24..159P}. JMDK acknowledges the hospitality of the Aspen Center for Physics, which is supported by the National Science Foundation Grant No.~PHY-1066293

\bibliography{BiblioRosotti}{}

\begin{thebibliography}{}

\bibitem[\protect\citeauthoryear{{Aarseth}, {Henon} \& {Wielen}}{{Aarseth}
  et~al.}{1974}]{AarsethPlummer}
{Aarseth} S.~J.,  {Henon} M.,    {Wielen} R.,  1974, \aap, 37, 183

\bibitem[\protect\citeauthoryear{{Adamo}, {{\"O}stlin} \& {Zackrisson}}{{Adamo}
  et~al.}{2011}]{adamo11}
{Adamo} A.,  {{\"O}stlin} G.,    {Zackrisson} E.,  2011, \mnras, 417, 1904

\bibitem[\protect\citeauthoryear{{Adams}}{{Adams}}{2010}]{2010ARA&A..48...47A}
{Adams} F.~C.,  2010, \araa, 48, 47

\bibitem[\protect\citeauthoryear{{Adams}, {Hollenbach}, {Laughlin} \&
  {Gorti}}{{Adams} et~al.}{2004}]{adams04}
{Adams} F.~C.,  {Hollenbach} D.,  {Laughlin} G.,    {Gorti} U.,  2004, \apj,
  611, 360

\bibitem[\protect\citeauthoryear{{Alexander}, {Clarke} \&
  {Pringle}}{{Alexander} et~al.}{2006}]{AlexanderEvol}
{Alexander} R.~D.,  {Clarke} C.~J.,    {Pringle} J.~E.,  2006, \mnras, 369, 229

\bibitem[\protect\citeauthoryear{{Andrews} \& {Williams}}{{Andrews} \&
  {Williams}}{2005}]{Andrews2005}
{Andrews} S.~M.,  {Williams} J.~P.,  2005, \apj, 631, 1134

\bibitem[\protect\citeauthoryear{{Armitage}}{{Armitage}}{2011}]{Armitage2011}
{Armitage} P.~J.,  2011, \araa, 49, 195

\bibitem[\protect\citeauthoryear{{Artymowicz} \& {Lubow}}{{Artymowicz} \&
  {Lubow}}{1994}]{artymowiczlubow1994}
{Artymowicz} P.,  {Lubow} S.~H.,  1994, \apj, 421, 651

\bibitem[\protect\citeauthoryear{{Balsara}}{{Balsara}}{1995}]{balsara1995}
{Balsara} D.~S.,  1995, Journal of Computational Physics, 121, 357

\bibitem[\protect\citeauthoryear{{Bastian}}{{Bastian}}{2008}]{bastian08}
{Bastian} N.,  2008, \mnras, 390, 759

\bibitem[\protect\citeauthoryear{{Bate}, {Bonnell} \& {Price}}{{Bate}
  et~al.}{1995}]{1995MNRAS.277..362B}
{Bate} M.~R.,  {Bonnell} I.~A.,    {Price} N.~M.,  1995, \mnras, 277, 362

\bibitem[\protect\citeauthoryear{{Bate} \& {Burkert}}{{Bate} \&
  {Burkert}}{1997}]{1997MNRAS.288.1060B}
{Bate} M.~R.,  {Burkert} A.,  1997, \mnras, 288, 1060

\bibitem[\protect\citeauthoryear{{Casertano} \& {Hut}}{{Casertano} \&
  {Hut}}{1985}]{1985ApJ...298...80C}
{Casertano} S.,  {Hut} P.,  1985, \apj, 298, 80

\bibitem[\protect\citeauthoryear{{Clarke}}{{Clarke}}{2007}]{clarke07}
{Clarke} C.~J.,  2007, \mnras, 376, 1350

\bibitem[\protect\citeauthoryear{{Clarke}, {Gendrin} \& {Sotomayor}}{{Clarke}
  et~al.}{2001}]{UVswitch}
{Clarke} C.~J.,  {Gendrin} A.,    {Sotomayor} M.,  2001, \mnras, 328, 485

\bibitem[\protect\citeauthoryear{{Clarke} \& {Pringle}}{{Clarke} \&
  {Pringle}}{1993}]{1993MNRAS.261..190C}
{Clarke} C.~J.,  {Pringle} J.~E.,  1993, \mnras, 261, 190

\bibitem[\protect\citeauthoryear{{Craig} \& {Krumholz}}{{Craig} \&
  {Krumholz}}{2013}]{2013ApJ...769..150C}
{Craig} J.,  {Krumholz} M.~R.,  2013, \apj, 769, 150

\bibitem[\protect\citeauthoryear{{Dale}, {Ercolano} \& {Bonnell}}{{Dale}
  et~al.}{2013}]{dale13}
{Dale} J.~E.,  {Ercolano} B.,    {Bonnell} I.~A.,  2013, \mnras, 430, 234

\bibitem[\protect\citeauthoryear{{de Juan Ovelar}, {Kruijssen}, {Bressert},
  {Testi}, {Bastian} \& {C{\'a}novas}}{{de Juan Ovelar}
  et~al.}{2012}]{2012A&A...546L...1D}
{de Juan Ovelar} M.,  {Kruijssen} J.~M.~D.,  {Bressert} E.,  {Testi} L.,
  {Bastian} N.,    {C{\'a}novas} H.,  2012, \aap, 546, L1

\bibitem[\protect\citeauthoryear{{Ercolano}, {Clarke} \& {Hall}}{{Ercolano}
  et~al.}{2011}]{2011MNRAS.410..671E}
{Ercolano} B.,  {Clarke} C.~J.,    {Hall} A.~C.,  2011, \mnras, 410, 671

\bibitem[\protect\citeauthoryear{{Fedele}, {van den Ancker}, {Henning},
  {Jayawardhana} \& {Oliveira}}{{Fedele} et~al.}{2010}]{Fedele2010}
{Fedele} D.,  {van den Ancker} M.~E.,  {Henning} T.,  {Jayawardhana} R.,
  {Oliveira} J.~M.,  2010, \aap, 510, A72

\bibitem[\protect\citeauthoryear{{Forgan} \& {Rice}}{{Forgan} \&
  {Rice}}{2009}]{2009MNRAS.400.2022F}
{Forgan} D.,  {Rice} K.,  2009, \mnras, 400, 2022

\bibitem[\protect\citeauthoryear{{Goddard}, {Bastian} \& {Kennicutt}}{{Goddard}
  et~al.}{2010}]{goddard10}
{Goddard} Q.~E.,  {Bastian} N.,    {Kennicutt} R.~C.,  2010, \mnras, 405, 857

\bibitem[\protect\citeauthoryear{{Gorti}, {Dullemond} \& {Hollenbach}}{{Gorti}
  et~al.}{2009}]{2009ApJ...705.1237G}
{Gorti} U.,  {Dullemond} C.~P.,    {Hollenbach} D.,  2009, \apj, 705, 1237

\bibitem[\protect\citeauthoryear{{Guilloteau}, {Dutrey}, {Pi{\'e}tu} \&
  {Boehler}}{{Guilloteau} et~al.}{2011}]{2011A&A...529A.105G}
{Guilloteau} S.,  {Dutrey} A.,  {Pi{\'e}tu} V.,    {Boehler} Y.,  2011, \aap,
  529, A105

\bibitem[\protect\citeauthoryear{{Gullbring}, {Hartmann}, {Briceno} \&
  {Calvet}}{{Gullbring} et~al.}{1998}]{Gullbring98}
{Gullbring} E.,  {Hartmann} L.,  {Briceno} C.,    {Calvet} N.,  1998, \apj,
  492, 323

\bibitem[\protect\citeauthoryear{{Hall}}{{Hall}}{1997}]{1997MNRAS.287..148H}
{Hall} S.~M.,  1997, \mnras, 287, 148

\bibitem[\protect\citeauthoryear{{Hartmann}, {Calvet}, {Gullbring} \&
  {D'Alessio}}{{Hartmann} et~al.}{1998}]{Hartmann98}
{Hartmann} L.,  {Calvet} N.,  {Gullbring} E.,    {D'Alessio} P.,  1998, \apj,
  495, 385

\bibitem[\protect\citeauthoryear{{Hayashi}}{{Hayashi}}{1981}]{MMSN}
{Hayashi} C.,  1981, Progress of Theoretical Physics Supplement, 70, 35

\bibitem[\protect\citeauthoryear{{Heller}}{{Heller}}{1995}]{1995ApJ...455..252H}
{Heller} C.~H.,  1995, \apj, 455, 252

\bibitem[\protect\citeauthoryear{{Herczeg} \& {Hillenbrand}}{{Herczeg} \&
  {Hillenbrand}}{2008}]{2008ApJ...681..594H}
{Herczeg} G.~J.,  {Hillenbrand} L.~A.,  2008, \apj, 681, 594

\bibitem[\protect\citeauthoryear{{Hubber}, {Allison}, {Smith} \&
  {Goodwin}}{{Hubber} et~al.}{2013}]{HybridNbodySPH}
{Hubber} D.~A.,  {Allison} R.~J.,  {Smith} R.,    {Goodwin} S.~P.,  2013,
  \mnras, 430, 1599

\bibitem[\protect\citeauthoryear{{Hubber}, {Batty}, {McLeod} \&
  {Whitworth}}{{Hubber} et~al.}{2011}]{SerenPaper}
{Hubber} D.~A.,  {Batty} C.~P.,  {McLeod} A.,    {Whitworth} A.~P.,  2011,
  \aap, 529, A27

\bibitem[\protect\citeauthoryear{{Hubber}, {Falle} \& {Goodwin}}{{Hubber}
  et~al.}{2013}]{2013MNRAS.432..711H}
{Hubber} D.~A.,  {Falle} S.~A.~E.~G.,    {Goodwin} S.~P.,  2013, \mnras, 432,
  711

\bibitem[\protect\citeauthoryear{{Isella}, {Carpenter} \& {Sargent}}{{Isella}
  et~al.}{2009}]{Isella2009}
{Isella} A.,  {Carpenter} J.~M.,    {Sargent} A.~I.,  2009, \apj, 701, 260

\bibitem[\protect\citeauthoryear{{Johnstone}, {Hollenbach} \&
  {Bally}}{{Johnstone} et~al.}{1998}]{Johnstone98}
{Johnstone} D.,  {Hollenbach} D.,    {Bally} J.,  1998, \apj, 499, 758

\bibitem[\protect\citeauthoryear{{Koepferl}, {Ercolano}, {Dale}, {Teixeira},
  {Ratzka} \& {Spezzi}}{{Koepferl} et~al.}{2013}]{2013MNRAS.428.3327K}
{Koepferl} C.~M.,  {Ercolano} B.,  {Dale} J.,  {Teixeira} P.~S.,  {Ratzka} T.,
    {Spezzi} L.,  2013, \mnras, 428, 3327

\bibitem[\protect\citeauthoryear{{Kraus}, {Ireland}, {Hillenbrand} \&
  {Martinache}}{{Kraus} et~al.}{2012}]{2012ApJ...745...19K}
{Kraus} A.~L.,  {Ireland} M.~J.,  {Hillenbrand} L.~A.,    {Martinache} F.,
  2012, \apj, 745, 19

\bibitem[\protect\citeauthoryear{{Kruijssen}}{{Kruijssen}}{2012}]{kruijssen12d}
{Kruijssen} J.~M.~D.,  2012, \mnras, 426, 3008

\bibitem[\protect\citeauthoryear{{Kruijssen}, {Maschberger}, {Moeckel},
  {Clarke}, {Bastian} \& {Bonnell}}{{Kruijssen} et~al.}{2012}]{kruijssen12}
{Kruijssen} J.~M.~D.,  {Maschberger} T.,  {Moeckel} N.,  {Clarke} C.~J.,
  {Bastian} N.,    {Bonnell} I.~A.,  2012, \mnras, 419, 841

\bibitem[\protect\citeauthoryear{{Lada} \& {Lada}}{{Lada} \&
  {Lada}}{2003}]{lada03}
{Lada} C.~J.,  {Lada} E.~A.,  2003, \araa, 41, 57

\bibitem[\protect\citeauthoryear{{Lada}, {Margulis} \& {Dearborn}}{{Lada}
  et~al.}{1984}]{1984ApJ...285..141L}
{Lada} C.~J.,  {Margulis} M.,    {Dearborn} D.,  1984, \apj, 285, 141

\bibitem[\protect\citeauthoryear{{Landau} \& {Lifshitz}}{{Landau} \&
  {Lifshitz}}{2010}]{LandauLifshitz}
{Landau} L.~D.,  {Lifshitz} E.~M.,  2010, {Mechanics}.
Elsevier Butterworth-Heinemann

\bibitem[\protect\citeauthoryear{{Lodato}, {Meru}, {Clarke} \& {Rice}}{{Lodato}
  et~al.}{2007}]{2007MNRAS.374..590L}
{Lodato} G.,  {Meru} F.,  {Clarke} C.~J.,    {Rice} W.~K.~M.,  2007, \mnras,
  374, 590

\bibitem[\protect\citeauthoryear{{Lodato} \& {Price}}{{Lodato} \&
  {Price}}{2010}]{LodatoViscosity}
{Lodato} G.,  {Price} D.~J.,  2010, \mnras, 405, 1212

\bibitem[\protect\citeauthoryear{{Lodato} \& {Pringle}}{{Lodato} \&
  {Pringle}}{2007}]{2007MNRAS.381.1287L}
{Lodato} G.,  {Pringle} J.~E.,  2007, \mnras, 381, 1287

\bibitem[\protect\citeauthoryear{{Longmore}, {Kruijssen}, {Bastian}, {Bally},
  {Rathborne}, {Testi}, {Stolte}, {Dale}, {Bressert} \& {Alves}}{{Longmore}
  et~al.}{2014}]{longmore14}
{Longmore} S.~N.,  {Kruijssen} J.~M.~D.,  {Bastian} N.,  {Bally} J.,
  {Rathborne} J.,  {Testi} L.,  {Stolte} A.,  {Dale} J.~E.,  {Bressert} E.,
  {Alves} J.,  2014, Protostars and Planets VI, p. submitted

\bibitem[\protect\citeauthoryear{{Luhman}, {Allen}, {Espaillat}, {Hartmann} \&
  {Calvet}}{{Luhman} et~al.}{2010}]{2010ApJS..186..111L}
{Luhman} K.~L.,  {Allen} P.~R.,  {Espaillat} C.,  {Hartmann} L.,    {Calvet}
  N.,  2010, \apjs, 186, 111

\bibitem[\protect\citeauthoryear{{Lynden-Bell} \& {Pringle}}{{Lynden-Bell} \&
  {Pringle}}{1974}]{lyndenbellpringle}
{Lynden-Bell} D.,  {Pringle} J.~E.,  1974, \mnras, 168, 603

\bibitem[\protect\citeauthoryear{{Malmberg}, {Davies} \& {Heggie}}{{Malmberg}
  et~al.}{2011}]{malmberg2011}
{Malmberg} D.,  {Davies} M.~B.,    {Heggie} D.~C.,  2011, \mnras, 411, 859

\bibitem[\protect\citeauthoryear{{Malmberg}, {de Angeli}, {Davies}, {Church},
  {Mackey} \& {Wilkinson}}{{Malmberg} et~al.}{2007}]{malmberg2007}
{Malmberg} D.,  {de Angeli} F.,  {Davies} M.~B.,  {Church} R.~P.,  {Mackey} D.,
     {Wilkinson} M.~I.,  2007, \mnras, 378, 1207

\bibitem[\protect\citeauthoryear{{Manara}, {Robberto}, {Da Rio}, {Lodato},
  {Hillenbrand}, {Stassun} \& {Soderblom}}{{Manara}
  et~al.}{2012}]{2012ApJ...755..154M}
{Manara} C.~F.,  {Robberto} M.,  {Da Rio} N.,  {Lodato} G.,  {Hillenbrand}
  L.~A.,  {Stassun} K.~G.,    {Soderblom} D.~R.,  2012, \apj, 755, 154

\bibitem[\protect\citeauthoryear{{Mann} \& {Williams}}{{Mann} \&
  {Williams}}{2010}]{2010ApJ...725..430M}
{Mann} R.~K.,  {Williams} J.~P.,  2010, \apj, 725, 430

\bibitem[\protect\citeauthoryear{{Miotello}, {Robberto}, {Potenza} \&
  {Ricci}}{{Miotello} et~al.}{2012}]{2012ApJ...757...78M}
{Miotello} A.,  {Robberto} M.,  {Potenza} M.~A.~C.,    {Ricci} L.,  2012, \apj,
  757, 78

\bibitem[\protect\citeauthoryear{{Moeckel} \& {Bally}}{{Moeckel} \&
  {Bally}}{2007}]{2007ApJ...656..275M}
{Moeckel} N.,  {Bally} J.,  2007, \apj, 656, 275

\bibitem[\protect\citeauthoryear{Monaghan}{Monaghan}{1997}]{Monaghan1997}
Monaghan J.~J.,  1997, Journal of Computational Physics, 136, 298

\bibitem[\protect\citeauthoryear{{Monaghan} \& {Lattanzio}}{{Monaghan} \&
  {Lattanzio}}{1985}]{1985A&A...149..135M}
{Monaghan} J.~J.,  {Lattanzio} J.~C.,  1985, \aap, 149, 135

\bibitem[\protect\citeauthoryear{Morris \& Monaghan}{Morris \&
  Monaghan}{1997}]{morris&monaghan1997}
Morris J.~P.,  Monaghan J.~J.,  1997, Journal of Computational Physics, 136, 41

\bibitem[\protect\citeauthoryear{{Murray}}{{Murray}}{1996}]{murray96}
{Murray} J.~R.,  1996, \mnras, 279, 402

\bibitem[\protect\citeauthoryear{{Natta}, {Testi}, {Muzerolle}, {Randich},
  {Comer{\'o}n} \& {Persi}}{{Natta} et~al.}{2004}]{Natta2004}
{Natta} A.,  {Testi} L.,  {Muzerolle} J.,  {Randich} S.,  {Comer{\'o}n} F.,
  {Persi} P.,  2004, \aap, 424, 603

\bibitem[\protect\citeauthoryear{{O'dell}}{{O'dell}}{1998}]{1998AJ....115..263O}
{O'dell} C.~R.,  1998, \aj, 115, 263

\bibitem[\protect\citeauthoryear{{Olczak}, {Kaczmarek}, {Harfst}, {Pfalzner} \&
  {Portegies Zwart}}{{Olczak} et~al.}{2012}]{2012ApJ...756..123O}
{Olczak} C.,  {Kaczmarek} T.,  {Harfst} S.,  {Pfalzner} S.,    {Portegies
  Zwart} S.,  2012, \apj, 756, 123

\bibitem[\protect\citeauthoryear{{Olczak}, {Pfalzner} \& {Spurzem}}{{Olczak}
  et~al.}{2006}]{2006ApJ...642.1140O}
{Olczak} C.,  {Pfalzner} S.,    {Spurzem} R.,  2006, \apj, 642, 1140

\bibitem[\protect\citeauthoryear{{Owen}, {Ercolano} \& {Clarke}}{{Owen}
  et~al.}{2011}]{Owen11Models}
{Owen} J.~E.,  {Ercolano} B.,    {Clarke} C.~J.,  2011, \mnras, 412, 13

\bibitem[\protect\citeauthoryear{{Owen}, {Ercolano}, {Clarke} \&
  {Alexander}}{{Owen} et~al.}{2010}]{Owen1st}
{Owen} J.~E.,  {Ercolano} B.,  {Clarke} C.~J.,    {Alexander} R.~D.,  2010,
  \mnras, 401, 1415

\bibitem[\protect\citeauthoryear{{Pelupessy} \& {Portegies Zwart}}{{Pelupessy}
  \& {Portegies Zwart}}{2012}]{pelupessy12}
{Pelupessy} F.~I.,  {Portegies Zwart} S.,  2012, \mnras, 420, 1503

\bibitem[\protect\citeauthoryear{{Pfalzner}}{{Pfalzner}}{2008}]{2008A&A...492..735P}
{Pfalzner} S.,  2008, \aap, 492, 735

\bibitem[\protect\citeauthoryear{{Pfalzner}, {Tackenberg} \&
  {Steinhausen}}{{Pfalzner} et~al.}{2008}]{2008A&A...487L..45P}
{Pfalzner} S.,  {Tackenberg} J.,    {Steinhausen} M.,  2008, \aap, 487, L45

\bibitem[\protect\citeauthoryear{{Pfalzner}, {Umbreit} \& {Henning}}{{Pfalzner}
  et~al.}{2005}]{2005ApJ...629..526P}
{Pfalzner} S.,  {Umbreit} S.,    {Henning} T.,  2005, \apj, 629, 526

\bibitem[\protect\citeauthoryear{{Pfalzner}, {Vogel}, {Scharw{\"a}chter} \&
  {Olczak}}{{Pfalzner} et~al.}{2005}]{2005A&A...437..967P}
{Pfalzner} S.,  {Vogel} P.,  {Scharw{\"a}chter} J.,    {Olczak} C.,  2005,
  \aap, 437, 967

\bibitem[\protect\citeauthoryear{{Price}}{{Price}}{2007}]{2007PASA...24..159P}
{Price} D.~J.,  2007, \pasa, 24, 159

\bibitem[\protect\citeauthoryear{{Price} \& {Monaghan}}{{Price} \&
  {Monaghan}}{2007}]{2007MNRAS.374.1347P}
{Price} D.~J.,  {Monaghan} J.~J.,  2007, \mnras, 374, 1347

\bibitem[\protect\citeauthoryear{{Ricci}, {Robberto} \& {Soderblom}}{{Ricci}
  et~al.}{2008}]{ricci2008}
{Ricci} L.,  {Robberto} M.,    {Soderblom} D.~R.,  2008, \aj, 136, 2136

\bibitem[\protect\citeauthoryear{{Robberto}, {Soderblom}, {Bergeron},
  {Kozhurina-Platais}, {Makidon}, {McCullough}, {McMaster}, {Panagia}, {Reid},
  {Levay}, {Frattare}, {Da Rio}, {Andersen}, {O'Dell}, {Stassun} \&
  {Simon}}{{Robberto} et~al.}{2013}]{robberto13}
{Robberto} M.,  {Soderblom} D.~R.,  {Bergeron} E.,  {Kozhurina-Platais} V.,
  {Makidon} R.~B.,  {McCullough} P.~R.,  {McMaster} M.,  {Panagia} N.,  {Reid}
  I.~N.,  {Levay} Z.,  {Frattare} L.,  {Da Rio} N.,  {Andersen} M.,  {O'Dell}
  C.~R.,  {Stassun} K.~G.,    {Simon} 2013, \apjs, 207, 10

\bibitem[\protect\citeauthoryear{{Scally} \& {Clarke}}{{Scally} \&
  {Clarke}}{2001}]{2001MNRAS.325..449S}
{Scally} A.,  {Clarke} C.,  2001, \mnras, 325, 449

\bibitem[\protect\citeauthoryear{{Shakura} \& {Sunyaev}}{{Shakura} \&
  {Sunyaev}}{1973}]{AlphaViscosity}
{Shakura} N.~I.,  {Sunyaev} R.~A.,  1973, \aap, 24, 337

\bibitem[\protect\citeauthoryear{{Sicilia-Aguilar}, {Henning}, {Kainulainen} \&
  {Roccatagliata}}{{Sicilia-Aguilar} et~al.}{2011}]{SiciliaAguilar11}
{Sicilia-Aguilar} A.,  {Henning} T.,  {Kainulainen} J.,    {Roccatagliata} V.,
  2011, \apj, 736, 137

\bibitem[\protect\citeauthoryear{{Sicilia-Aguilar}, {Kim}, {Sobolev}, {Getman},
  {Henning} \& {Fang}}{{Sicilia-Aguilar} et~al.}{2013}]{SiciliaAguilar13}
{Sicilia-Aguilar} A.,  {Kim} J.~S.,  {Sobolev} A.,  {Getman} K.,  {Henning} T.,
     {Fang} M.,  2013, ArXiv e-prints

\bibitem[\protect\citeauthoryear{{Silva-Villa}, {Adamo} \&
  {Bastian}}{{Silva-Villa} et~al.}{2013}]{silvavilla13}
{Silva-Villa} E.,  {Adamo} A.,    {Bastian} N.,  2013, ArXiv e-prints

\bibitem[\protect\citeauthoryear{{Springel} \& {Hernquist}}{{Springel} \&
  {Hernquist}}{2002}]{2002MNRAS.333..649S}
{Springel} V.,  {Hernquist} L.,  2002, \mnras, 333, 649

\bibitem[\protect\citeauthoryear{{Stolte}, {Morris}, {Ghez}, {Do}, {Lu},
  {Wright}, {Ballard}, {Mills} \& {Matthews}}{{Stolte}
  et~al.}{2010}]{2010ApJ...718..810S}
{Stolte} A.,  {Morris} M.~R.,  {Ghez} A.~M.,  {Do} T.,  {Lu} J.~R.,  {Wright}
  S.~A.,  {Ballard} C.,  {Mills} E.,    {Matthews} K.,  2010, \apj, 718, 810

\bibitem[\protect\citeauthoryear{{Toomre} \& {Toomre}}{{Toomre} \&
  {Toomre}}{1972}]{1972ApJ...178..623T}
{Toomre} A.,  {Toomre} J.,  1972, \apj, 178, 623

\bibitem[\protect\citeauthoryear{{Walch}, {Naab}, {Whitworth}, {Burkert} \&
  {Gritschneder}}{{Walch} et~al.}{2010}]{2010MNRAS.402.2253W}
{Walch} S.,  {Naab} T.,  {Whitworth} A.,  {Burkert} A.,    {Gritschneder} M.,
  2010, \mnras, 402, 2253

\bibitem[\protect\citeauthoryear{{Williams} \& {Cieza}}{{Williams} \&
  {Cieza}}{2011}]{WilliamsCieza2011}
{Williams} J.~P.,  {Cieza} L.~A.,  2011, \araa, 49, 67

\end{thebibliography}
\bibliographystyle{mn2e}

\appendix

\section{Resolution tests}
In order to quantify the effect of the limited resolution available on viscosity, we run resolution tests of the discs in isolation. In addition to the runs presented in the text, which employed a resolution of $10^4$ particles, we run simulations at resolutions of $3 \times 10^4$, $10^5$ and $3 \times 10^5$ particles. For the smallest disc (run R10) we did not run the highest resolution case; being the smallest one, it is the one that requires the highest number of integration steps and therefore the strongest computational constraint. We then fitted the evolution of the radius of the disc as done in section 3.2, obtaining the values reported in table \ref{tab_res_test}.  It can be noted that increasing the resolution (in terms of number of particles) of a factor of 30 does not lead to a substantial difference in the spreading rates of the discs: the variation in $t_\mathrm{spread}$ is always less than a factor of 2. This change in the viscous time scale can be explained by the change in resolution: as spatial resolution scales with $N_\mathrm{part}^{1/3}$, the expected increase in spatial resolution is around $~3$.

\begin{table}
\begin{tabular}{crccc}
\toprule
Run		&	Resolution (particles) & $t_\nu \ [\mathrm{yr}]$	&	$\gamma$	&	$t_\mathrm{spread} [ \mathrm{yr}]$\\
\midrule
R10		&	$10^4$	& 18891	&	1.11		& 16800\\
R10 & $3 \times 10^4$		&	13356	&	0.92	 &	14430\\
R10 	&	$10^5$	&  20584	&	1.02	 &	20080\\
\midrule[0.3pt]
R30		&	$10^4$	&	23218	 &	0.44	&	36220\\
R30 & $3 \times 10^4$		&	21400	 &	0.29	  &	36405\\
R30 &	$10^5$		&	14600	&	-0.02	&	29500\\
R30 &	$3 \times 10^5$		&	18800	&	0.08	 &	36039\\
\midrule[0.3pt]
R100	&	$10^4$	&		11760	&	-1.69	&	43400\\
R100 & $3 \times 10^4$	&		12054	&	-1.84	&	43600\\
R100 &	$10^5$	 &	12103	&	-1.93	&	47640\\
R100 &	$3 \times 10^5$	 &	19763	&	-1.57	&	70600\\
\midrule[0.3pt]
R300	&	$10^4$	&		25430	&	-3.19	&	132000\\
R300 & $3 \times 10^4$			&	29490	&	-3.17	&	152750\\
R300 &	$10^5$	 &	30965	&	-3.26	&	162950\\
R300 &	$3 \times 10^5$	 &	35710	&	-3.64	&	201700\\
\bottomrule
\end{tabular}
\caption{We report in the table the results of the resolution tests run.}
\label{tab_res_test}
\end{table}

This shows that, while it is certainly true that the spreading rates are enhanced by the limited resolution, the effect is rather small and is not likely to substantially affect our results and conclusions. In addition, the discs do not spread at a rate very incompatible with what is seen in observations. We note that higher viscosities cause the effect of encounters to be underestimated, since they would make viscous spreading more important. The good overall agreement of the semi-analytical model with the results of the simulations shows that, despite having rather large viscosities, the disc behaviour is still physical. The existence of a competition between disc spreading and encounters is therefore robust and has clear physical explanation.

\bsp

\label{lastpage}

\end{document}